\documentclass[singlecolumn,secnumarabic,amssymb,nobibnotes,aps,prd]{revtex4-1}
\pdfoutput=1
\usepackage{amsmath,amssymb,graphicx,subcaption,feynmp}
\numberwithin{equation}{section}
\newcommand{\ba}[1]{\begin{aligned}#1\end{aligned}}

\newcommand{\para}[1]{\left( #1 \right)}
\newcommand{\sqpara}[1]{\left [ #1 \right ]}

\newcommand{\slashed}[1]{\not\! #1}
\makeatletter
\def\endfmffile{%
  \fmfcmd{\p@rcent\space the end.^^J%
          end.^^J%
          endinput;}%
  \if@fmfio
    \immediate\closeout\@outfmf
  \fi
  \IfFileExists{\thefmffile.mp}{\immediate\write18{mpost \thefmffile}}{}
  \let\thefmffile\relax
}
\makeatother

\makeatletter
\def\@author#1{\g@addto@macro\elsauthors{\normalsize%
    \def\baselinestretch{1}%
    \upshape\authorsep#1\unskip\textsuperscript{%
      \ifx\@fnmark\@empty\else\unskip\sep\@fnmark\let\sep=,\fi
      \ifx\@corref\@empty\else\unskip\sep\@corref\let\sep=,\fi
      }%
    \def\authorsep{\space and\space}%
    \global\let\@fnmark\@empty
    \global\let\@corref\@empty
    \global\let\sep\@empty}%
    \@eadauthor={#1}
}

\def\ps@pprintTitle{%
 \let\@oddhead\@empty
 \let\@evenhead\@empty
 \def\@oddfoot{}%
 \let\@evenfoot\@oddfoot}
 
\long\def\MaketitleBox{%
  \resetTitleCounters
  \def\baselinestretch{1}%
  \begin{center}%
   \def\baselinestretch{1}%
    \Large\@title\par\vskip18pt
    \normalsize\elsauthors\par\vskip10pt
    \footnotesize\itshape\elsaddress\par\vskip36pt
    \rule{\textwidth}{1.5pt}\vskip12pt
    \ifvoid\absbox\else\unvbox\absbox\par\vskip10pt\fi
    \ifvoid\keybox\else\unvbox\keybox\par\vskip10pt\fi
    \rule{\textwidth}{1.5pt}\vskip12pt
    \end{center}%
}

\begin{document}

\title{{\bf \Large Matter Quantum Corrections to the \\[2mm]
 Graviton Self-Energy and the Newtonian Potential}\medskip}

\begin{flushright}
MAN/HEP/2014/18 \\
December 2014
\end{flushright}
\bigskip

\author{\large Daniel Burns}
\email{daniel.burns@hep.manchester.ac.uk}
\author{\large Apostolos Pilaftsis}
\email{apostolos.pilaftsis@manchester.ac.uk}

\affiliation{\vspace{2mm}
Consortium for Fundamental Physics, School of
  Physics and Astronomy, University of Manchester, Manchester M13
  9PL, United Kingdom}

\begin{abstract}
\vspace{3mm}
\centerline{\bf ABSTRACT}
\vspace{2mm}
\noindent 
We revisit the  calculation of matter quantum effects  on the graviton
self-energy on a flat Minkowski  background, with the aim to acquire a
deeper  understanding  of  the  mechanism that  renders  the  graviton
massless.  To this end, we  derive a low-energy theorem which directly
relates  the radiative  corrections  of the  cosmological constant  to
those of the  graviton mass to all orders  in perturbation theory.  As
an  illustrative example,  we  consider an  Abelian  Higgs model  with
minimal  coupling  to  gravity  and  show explicitly  how  a  suitable
renormalization of the cosmological constant leads to the vanishing of
the graviton  mass at the one-loop  level.  In the  same Abelian Higgs
model,  we  also  calculate  the  matter quantum  corrections  to  the
Newtonian  potential  and  present  analytical formulae  in  terms  of
modified Bessel  and Struve  functions of the  particle masses  in the
loop.  We~show that the correction to the Newtonian potential exhibits
an  exponential  fall-off dependence  on  the  distance~$r$, once  the
non-relativistic  limit with  respect  to the  non-zero  loop mass  is
carefully considered.  For massless scalars, fermions and gauge bosons
in  the loops,  we recover  the  well-known results  presented in  the
literature.
\end{abstract}

\maketitle

\makeatletter
\def\appendixname{Appendix}
\renewcommand\@makefntext[1]{\leftskip=0em\hskip1em\@makefnmark\space #1}

\makeatother

\section{Introduction}

Symmetries play an instrumental role in quantum field theory to ensure
that massless particles at the classical level remain massless against
quantum  loop  effects.   For  instance,  massless  vector  bosons  in
Yang--Mills  theories stay  massless, as  a consequence  of  the gauge
symmetry of  the effective  action. This fact  can be  understood more
easily within  the gauge-invariant  framework of the  background field
method  \cite{DeWitt1967a,DeWitt1967b}, in which  a non-zero  mass for
the background Yang--Mills vector boson  is forbidden to all orders in
perturbation  theory.  Likewise,  massless fermions  can  be protected
from     receiving     a    non-zero     mass     due    to     chiral
symmetry~\cite{GellMann:1968rz}.   Scalar   particles  can  also  stay
massless  to all  orders, as  a  result of  symmetries.  For  example,
massless scalar particles could  result from the spontaneous breakdown
of   a  global   Goldstone   symmetry~\cite{Goldstone:1961eq}.   Other
potential  symmetries   leading  to  massless   scalar  particles  are
supersymmetry~\cite{Volkov:1973ix,Wess:1974tw}  or  classical  scaling
(conformal)   symmetries~\cite{Coleman:1973jx,Gildener:1976ih}.   Such
symmetries  have been extensively  discussed within  the context  of a
related   problem  in   the   Standard  Model   (SM),  the   so-called
gauge-hierarchy problem~\cite{Haber:1984rc,Martin:1997ns,Ibrahim:2007fb}.

The aim  of the present paper is  to shed light on  the mechanism that
protects the spin-2 graviton from receiving a non-zero mass beyond the
tree  level.    In  this  context,  we  should   mention  that  matter
contributions to the graviton self-energy have already been studied in
the      past     to      a      great     extent~\cite{Capper:1973bk,
  Capper:1973mv,Capper:1974ed,Capper:1973pv,Capper:1984qq}.    However,
in our opinion, the actual mechanism that lies behind the masslessness
of  the   graviton  has  not  yet  been   adequately  elucidated.   In
particular,  a radiatively  generated  graviton mass  will affect  the
scattering  of  two  scalar   fields  beyond  the  tree  level.   Such
calculations  are  relevant to  the  study  of  the quantum  corrected
Newtonian  potential  and may  be  in  conflict with  well-established
observations.  It is therefore important  to state here that the gauge
or diffeomorphisms invariance of the effective action, even within the
linearized  framework of  perturbative quantum  gravity (PQG),  is not
sufficient by  itself to guarantee that the  graviton remains massless
against  quantum  loop  corrections.  Specifically,  the  cosmological
constant term  is invariant under diffeomorphisms and  contains a mass
term for  the graviton.  At the  tree level, this  problem is resolved
(see,  e.g.~\cite{Gabadadze2005})  after  imposing  the  equations  of
motion with respect to the  background graviton field, with the aid of
which  a would-be  graviton  mass  can be  removed.   Beyond the  tree
approximation,  however, the masslessness  of the  graviton is  not an
obvious property,  as this problem becomes  strongly interrelated with
the renormalization of the cosmological constant $\Lambda$.

In quantum  field theory, the  pole position of a  particle propagator
encodes all  the information  about the mass  of the particle.   As we
will show in this paper,  the cosmological constant $\Lambda$ plays an
important role, as it  receives radiative corrections independently of
the graviton propagator.  These  corrections are divergent and must be
renormalized,  or otherwise  naturally suppressed,  to give  the small
value    of   $\Lambda$    that    we   observe    in   the    present
epoch~\cite{Tegmark2004,PlanckCollaboration2013}.     Upon    suitable
renormalization  of  the cosmological  constant  $\Lambda$  to a  flat
(Minkowski)   background   metric,   the  generated   counterterm~(CT)
$\delta\Lambda$ enters the  graviton self-energy explicitly within our
linearized framework  of PQG.  We find that  the masslessness property
of the graviton  is protected by a shift symmetry  which is present in
any diffeomorphisms  invariant theory  described by a  flat background
metric.   The  absence  of   the  graviton  mass  will  be  explicitly
demonstrated at  the one-loop  level in PQG  within the context  of an
Abelian Higgs model.

Given    that   the   framework    of   PQG    is   non-renormalizable
\cite{'tHooft:1974bx,Deser1974a,Deser:1974cz,Deser1974}, we follow the
general lore and  treat General Relativity (GR) as  an effective field
theory~\cite{Donoghue1994}  with  a  characteristic ultra-violet  (UV)
scale equal  to the Planck mass~$M_{\rm  P}$. Much work  has been done
within  this   effective  field-theoretic  framework,   including  PQG
corrections  to  the Newtonian  and  Coulomb  potentials,  as well  as
one-loop calculations of  graviton-mediated scatterings between matter
fields          in         the          non-relativistic         limit
\cite{Donoghue1994,Donoghue1993,Hamber1995,Muzinich1995,Akhundov1997,
  Bjerrum-Bohr2002,Bjerrum-Bohr2002a,Khriplovich2002,Butt2006,Faller2008}.
Taking  into account  the  contributions from  the  graviton and  from
massless fields of different spin, the established analytic result for
the  Newtonian potential  $V(r)$, between  two masses  $m_1$ and~$m_2$
being   at    distance   $r$   apart,    may   be   cast    into   the
form~\cite{Hamber1995,Bjerrum-Bohr2002a,Duff:1974ud,Duff:2000mt}:
\begin{equation}
  \label{Vr}
V(r) = -\frac{G m_1 m_2}{r} \bigg[1 + 3\frac{G(m_1 + m_2)}{r c^2}+
  \frac{41 \hbar G}{10 \pi c^3 
    r^2} + \bigg( \frac{9}{4}N_0 + 3 N_{\frac{1}{2}} + 12
    N_1\bigg)\frac{\hbar G}{45 \pi c^3 r^2} + O(\hbar^2) \bigg]\;,
\end{equation}
where $G = \hbar c/M^2_{\rm P}$  is Newton's constant and $N_s$ is the
number  of fields  with spin  $s =  0$ (scalar) $,\, \frac{1}2$ (Weyl fermion)$,\,  1 $ (vector boson) in  units of
$\hbar$.  Note  that the first  two terms in~\eqref{Vr}  correspond to
the  classical and  quantum  graviton contributions  to the  Newtonian
potential $V(r)$, respectively.   The leading radiative corrections to
$V(r)$  come  from  the  so-called  {\it non-analytic}  parts  of  the
amplitude,  which diverge  in the  infra-red (IR)  limit  of vanishing
3-momenta for the  external gravitationally-scattered fields.  Using a
similar approach,  we compute the  general matter loop  corrections to
the graviton propagator, as well as the modifications to the Newtonian
potential $V(r)$.  The matter  contributions to $V(r)$ at the one-loop
level  effect only  the graviton  self-energy  in a  generic $2\to  2$
scattering process. Thus, we shall show that the contributions of {\it
  massive} matter fields to the resummed graviton self-energies become
relevant in the non-relativistic limit and therefore contribute to the
Newtonian potential.

The  layout  of the  paper  is  as  follows. After  this  introductory
section, Section  \ref{TF} presents a gauged Abelian  Higgs model with
minimal  coupling to gravity.   This model  serves as  an illustrative
example, which  will help us  to define our theoretical  PQG framework
that can  include scalars, fermions  and spin-1 fields. Based  on this
framework,   we   discuss   the   properties  of   the   corresponding
diffeomorphically invariant path integral for the gauged Abelian Higgs
model.      Given~that    the     model    has     no    gravitational
anomalies~\cite{AlvarezGaume:1983ig},   we  derive  the   master  Ward
identity~(WI)  associated with  the  invariance of  the path  integral
under diffeomorphisms.

In Section \ref{CC}, we study the minimization conditions pertinent to
the  one-loop  effective  action,  where the  re\-normalization  of  the
cosmological   constant~$\Lambda$   plays    a   key   role   to   the
renormalization of the graviton  tadpole graphs. To further illuminate
this  deep connection, we  derive a  low-energy theorem  that involves
graviton correlation  functions to all orders  in perturbation theory.
This Graviton Low Energy Theorem~(GLET) may also be utilized to obtain
a non-perturbative relation between  the tadpole contributions and the
graviton self-energy at zero external momentum.

In  Section \ref{MC},  we calculate  the matter  contributions  to the
graviton self-energy  for the gauged Abelian Higgs  model with minimal
coupling to gravity.  To deal  with UV infinities, we adopt the method
of  dimensional regularisation~\cite{'tHooft:1972fi}. We  then proceed
to  renormalize the  {\em massive}  matter-field contributions  to the
graviton  self-energy,  after   properly  including  the  cosmological
constant   CT  $\delta   \Lambda$,  as   well   as  higher-dimensional
Planck-suppressed operators  of the Riemann tensor. We  thus show that
the  graviton  field acquires  no  mass  at  the one-loop  level.   We
explicitly demonstrate  how this result  persists to all orders,  as a
consequence  of the  GLET and  the WI  due to  invariance of  the path
integral under diffeomorphisms.

In Section  \ref{NP}, we first  review the tree-level  calculation for
the  gravitationally mediated  scattering process  between  two scalar
fields, where the classical part  of the Newtonian potential $V(r)$ is
recovered. We  then incorporate  the self-energy contributions  to the
graviton  propagator, which is  used to  determine the  matter quantum
corrections  to the  Newtonian  potential.  Our  analytic results  are
expressed  in terms  of modified  Bessel and  Struve functions  of the
particle masses in the loop. In the massless limit of the loop masses,
we reproduce  the analytic  result given in~\eqref{Vr},  for particles
with spin $s = 0,\, \frac{1}2,\,  1$.  In the same section, we comment
on  the  independence  of  $V(r)$ on  the  gravitational  gauge-fixing
parameters, as well as on  gauge-fixing parameters due to gauge bosons
in the loop.  Section~\ref{Concl} summarizes our conclusions. Finally,
relevant Feynman rules and other technical details that were useful in
our computations have been presented in Appendix~\ref{FR}.

\section{Theoretical framework of Quantum Gravity\label{TF}}

In this section, we first outline our theoretical framework within the
context of an Abelian Higgs model with minimal coupling to gravity, by
making  use of the  background field  method. We  then write  down the
generating functional for this  model and discuss its invariance under
transformations  of  diffeomorphism.  From  the  latter,  we derive  a
master WI  for diffeomorpshims,  which gives rise  to an  important WI
that relates  the graviton self-energy to the  graviton tadpole graphs
to all orders in perturbation theory.

To begin with, we write down  the action $S$ of an Abelian Higgs model
minimally coupled to gravity as a sum of two terms:
\begin{equation}
  \label{Stot} 
S = S_G + S_{M} = \int d^4 x \sqrt{-g}\Big (\Lambda +
\frac{1}{\kappa^2}R  + \mathcal{L}_M \Big)\; ,
\end{equation}
where  $S_G$  is  the  Hilbert--Einstein  action  of  gravity  with  a
cosmological constant $\Lambda$ and  $S_M \equiv \int d^4x \sqrt{-g}\,
{\cal L}_M$  is the part of  the action that only  contains the matter
Lagrangian ${\cal  L}_M$.  In addition,  we denote with  $g_{\mu \nu}$
the global  metric of the space  and $g \equiv  {\rm det} g_{\mu\nu}$,
whilst  our convention  for  the Minkowski  metric $\eta_{\mu\nu}$  is
$\eta_{\mu  \nu} = {\rm  diag}(1,-1,-1,-1)$.  In~\eqref{Stot},  $R$ is
the Ricci scalar and $\kappa$ a gravitational coupling constant, which
is related to Newton's constant $G$ by $\kappa^2 = 16 \pi G$.

The matter action  $S_M$ describes a gauged Abelian  Higgs model based
on  the  gauge group  $U(1)_Y$,  which  realizes spontaneous  symmetry
breaking. In detail, the matter action $S_M$ is given by
\begin{equation}
  \label{eq:SM}
S_{M}\ =\ \int d^4 x \sqrt{-g}\bigg[-\frac{1}{4} g^{\mu \rho}g^{\nu
      \sigma}F_{\mu \nu} F_{\rho \sigma} + g^{\mu \nu}(\nabla_\mu
    \phi)^\dagger \nabla_\nu \phi - \lambda\,\bigg(\phi^{\dagger} \phi -
      \frac{\mu^2}{2\lambda}\bigg)^2\bigg]\ +\ S_f\; ,
\end{equation}
where $S_f$ is the fermionic  sector of the model, $F_{\mu \nu} \equiv
\partial_\mu  \mathcal{A}_\nu -  \partial_\nu \mathcal{A}_\mu$  is the
field    strength   tensor   associated    with   the    gauge   field
$\mathcal{A}_\mu$, and  $\phi = \frac{1}{\sqrt{2}}(v  +\mathcal{H} + i
\mathcal{G})$  is a complex  scalar field  with hypercharge  $Y_\phi =
1$. Moreover, $\nabla_\mu$ is the covariant derivative with respect to
both the gauge group and  the group of diffeomorphisms.  Thus, for the
scalar  field $\phi$,  the  covariant derivative  is  simply given  by
$\nabla_\mu  \phi =  \partial_\mu \phi  - i  e  \mathcal{A}_\mu \phi$.
Here, we  follow the  standard procedure of  general covariantization,
namely by  first writing  down the matter  Lagrangian ${\cal  L}_M$ in
flat space and then making the substitution $\eta_{\mu \nu} \to g_{\mu
  \nu}$ and  $\partial_\mu \to \nabla_\mu$.  In~\eqref{eq:SM}, we have
also  included an overall  factor $\sqrt{-g}$,  so as  to get  a fully
frame-independent action.

Adopting the  background field method  (BFM), we decompose  the fields
into background and quantum fields as follows:
\begin{equation}
\mathcal{H} =  \bar{H} + H^Q, \;\;\; \mathcal{G} = \bar{G} + G^Q,
  \;\;\; \mathcal{A}_\mu = \bar{A}_\mu + A^Q_\mu\; ,
\end{equation}
where an overbar denotes a  background field, whilst a superscript $Q$
denotes a  quantum field.   The Higgs mechanism  will generate  a mass
$m_A$ to the gauge field in the broken phase, given by $m_A = e v$, as
well  as a  mass for  the Higgs  field itself  determined  through the
relation: $m_H^2 = 2 \lambda v^2$.

The  fermionic part $S_f$  in~\eqref{eq:SM} of  the matter  action may
contain left-  and right-handed  chiral fermions.  For  simplicity, we
assume  one  Dirac  fermion  $\psi$ with  hypercharge  quantum  number
$Y_\psi  =  1$,  with  vector-like  couplings to  the  $U(1)_Y$  gauge
bosons.    This    simple   setup    is    also    free   of    chiral
anomalies~\cite{Adler:1969gk,Bell:1969ts}.    In   curved   spacetime,
spinors have non-trivial transformation  properties under the group of
diffeomorphisms,  which   is  usually   accounted  for  by   the  spin
connection.   Hence, with  the inclusion  of the  Dirac  fermion field
$\psi = \psi^Q$, the fermionic part of the action $S_f$ reads:
\begin{equation}
\mathcal{S}_f = \int d^4 x\sqrt{-g} \bigg [\frac{1}{2}
    \big(\nabla_\mu\bar{\psi}^Q\big) i e^\mu_a \gamma^a \psi^Q -
    \frac{1}{2}\bar{\psi}^Q i e^\mu_a \gamma^a \big(\nabla_\mu \psi^Q\big) -
    m_\psi \bar{\psi}^Q \psi^Q \bigg] \; ,
\end{equation}
where the covariant derivative acting on $\psi$ is given by
\begin{equation}
\nabla_\mu \psi^Q = \partial_\mu \psi^Q - \omega^{ab}_\mu
  \sigma_{ab}\psi^Q - i e \mathcal{A}^Q_\mu \psi^Q\; .
\end{equation}
In the above, $\sigma_{ab}  = \frac{1}{4}[\gamma_a, \gamma_b]$ are the
Lorentz-group    generators   in    the    spinorial   representation,
$\omega_\mu^{ab}$ is the spin connection, which is determined by means
of the vielbeins $e^a_\mu$ as follows:
\begin{equation}
\omega^{ab}_\mu = - g^{\nu \lambda}e^{a}_\lambda(\partial_\mu
  e^{b}_{\nu} - e^b_\sigma \Gamma^{\sigma}_{\mu \nu})\; .
\end{equation}
Note  that  the vielbein  fields  $e^a_\mu$  are  defined through  the
relations:
\begin{equation}
g_{\mu \nu} \equiv e^a_\mu e^b_\nu \eta_{ab}, \qquad e^a_\mu e^\mu_b
  = \delta^a_b, \qquad e^a_\mu e^\nu_a = \delta^\mu_\nu\; ,
\end{equation} 
where the  Latin indices $a,\, b$  etc.~refer to the  tangent space of
the curved spacetime which is locally flat.

To quantise gravity within the  BFM framework, we decompose the metric
$g_{\mu\nu}$   as  
\begin{equation}
  \label{eq:metricdecomposition}
g_{\mu   \nu}   =  \eta_{\mu   \nu}  +   \kappa
  (\bar{h}_{\mu  \nu} + h^Q_{\mu  \nu}) =  \bar{g}_{\mu \nu}  + \kappa
  h^{Q}_{\mu \nu}\; , 
\end{equation}
where  $h^Q_{\mu  \nu}$ is  the  quantum  fluctuation  of the  metric,
$\bar{h}_{\mu \nu}$  is the background field and  $\bar{g}_{\mu \nu} =
\eta_{\mu  \nu} +  \kappa  \bar{h}_{\mu  \nu}$. In  the  absence of  a
classical   gravitational   field   $\bar{h}_{\mu   \nu}$,   we   have
$\bar{g}_{\mu \nu} = \eta_{\mu \nu}$ and the curved space reduces to a
Minkowski flat space in this case.   In this paper, we will consider a
flat  background to  carry  out perturbative  calculations within  the
framework of linearized quantum gravity.

To eliminate the degeneracy in the  field space due to the symmetry of
diffeomorphisms, we use the gauge fixing condition
\begin{equation}
 G_a = (-\bar{g})^{\frac{1}{4}}\sqpara{\bar{g}^{\alpha
      \beta}\para{\overline{\nabla}_\alpha h^Q_{\beta \mu} - \sigma
      \overline{\nabla}_\mu h^Q_{\alpha \beta}}} \bar{e}^\mu_a =
  \omega_a\; ,
\end{equation} 
where $\omega_a(x)$ is an  arbitrary function and $\bar{e}^{\mu}_a$ is
the  background  vielbein  field.  Employing the  Faddeev-Popov  gauge
fixing procedure, we introduce the gauge-fixing action
\begin{equation}
S_{\rm GF, Diff} = - \frac{1}{2 \xi_D} \int d^4 x \sqrt{-\bar{g}}
  \bar{g}^{\mu \nu} \sqpara{\bar{g}^{\alpha
      \beta}\para{\overline{\nabla}_\alpha h^Q_{\beta \mu} - \sigma
      \overline{\nabla}_\mu h^Q_{\alpha
        \beta}}}\sqpara{\bar{g}^{\delta
      \gamma}\para{\overline{\nabla}_\delta h^Q_{\gamma \nu} - \sigma
      \overline{\nabla}_\nu h^Q_{\delta \gamma}}}\; ,
\end{equation}
which in turn induces the ghost action
\begin{equation}
S_{\rm Gh, Diff} = -\int d^4 x \sqrt{-\bar{g}} \,\bar{\eta}^\mu
  \Big( \bar{g}^{\alpha \beta}\overline{\nabla}_\alpha
  \overline{\nabla}_\beta \eta_\mu + \bar{g}^{\alpha \beta}
  \bar{R}_{\mu \alpha} \eta_\beta + (1-2\sigma)\bar{g}^{\alpha
    \beta}\overline{\nabla}_\mu \overline{\nabla}_\alpha \eta_\beta
  \Big)\; ,
\end{equation}
where  $\eta_\mu$ and  $\bar{\eta}_\nu$  are the  ghost vector  fields
associated with the graviton field $h_{\mu\nu}$.

In addition to  the diffeomorphisms group, we must  also gauge-fix the
$U(1)_Y$ gauge group. To this end, we consider the gauge fixing term
\begin{equation}
S_{\rm GF, U(1)} = - \frac{1}{2 \xi_G} \int d^4 x \sqrt{-g} \Big
  [g^{\mu \nu} \nabla_\mu A^Q_\nu + e \xi_G G^Q\big(v+H^Q\big) \Big]^2\; ,
\end{equation}
which  has  the  property  of  preserving  general  covariance  whilst
breaking the  invariance of  the gauge group.   It also  preserves the
Higgs-boson  low-energy theorem~(HLET)~\cite{Ellis:1975, Shifman:1978,
  Vainshtein:1979,   Dawson:1992,   Kniehl:1995}   in  its   canonical
form~\cite{Pilaftsis:1997fe}.   The  gauge-fixing  action $S_{\rm  GF,
  U(1)}$ also induces a Faddeev-Popov ghost action, which is given by
\begin{equation}
S_{\rm Gh, U(1)} = -\int d^4 x \sqrt{-g} \,\bar{c}\, \Big\{ g^{\mu
    \nu} \nabla_\mu \nabla_\nu + \frac{e^2}{2} \xi_G \big[(v +H^Q)^2 -
    (G^Q)^2\big] \Big\}\,c 
\end{equation}
where  $c,\, \bar{c}$  are the  $U(1)_Y$ Faddeev--Popov  ghosts.  Note
that the  scalar ghosts $c,\,  \bar{c}$ and their  vector counterparts
$\eta_\mu,\,  \bar{\eta}_\nu$  are  all anti-commuting  negative  norm
fields.

\subsection{The Diffeomorphically Invariant Path Integral}

To quantize the Abelian Higgs  model with minimal coupling to gravity,
we introduce the generating  functional 
\begin{equation}
  \label{eq:generatingfunctionaldef}
\ba{ Z[\bar{h}_{\mu \nu},
      \bar{H},      \bar{G},      \bar{A}_\mu,     J_h^{\mu      \nu},
      J_\psi,\bar{J}_\psi, J_H, J_G,  J_A^\mu] & = N \int\;\mathcal{D}
    \Phi\; \exp \bigg [ iS[\bar{h}_{\mu \nu},h^Q_{\mu \nu}, \mathcal{H},
      \mathcal{G},  \psi,\bar{\psi}, \mathcal{A}_\mu]  \\  & \;\;\;  +
    \int d^4 x \sqrt{- \bar{g}} \big(J_h^{\mu \nu}h^Q_{\mu \nu} +
    \bar{J}_{\psi} \psi^Q 
    +  \bar{\psi}^Q J_{\psi} +  J_H H^Q  + J_G  G^Q +  J_A^\mu A_\mu^Q
    \big) \bigg ]\;,} 
\end{equation}
where $N$ is an unphysical overall normalization constant and 
\begin{equation}
  \label{eq:DPhi}
\mathcal{D} \Phi \equiv \mathcal{D} h^Q_{\mu \nu}\, \mathcal{D}
  A^Q_\mu\, \mathcal{D} H^Q\, \mathcal{D} G^Q\, \mathcal{D}
  \bar{\psi}^Q\,\mathcal{D} \psi^Q
\end{equation}
is   a  short-hand   notation   for  the   integral  measure.    Under
infinitesimal  diffeomorphisms, $x^\mu  \to  x'^\mu =  x^\mu +  \kappa
\epsilon^\mu(x)$ with  $\epsilon^\mu(x) \ll 1$, the action  $S$ of the
theory remains invariant provided the fields transform as follows:
\begin{subequations}
\label{eq:diffeomorphisms}
\begin{align}
g_{\mu \nu}' &= g_{\mu \nu} + \kappa(g^\alpha_\nu \partial_\mu
\epsilon_\alpha + g^\alpha_\mu \partial_\nu \epsilon_\alpha +
\epsilon_\alpha \partial^\alpha g_{\mu \nu})\;,\\ 
\mathcal{H}' &= \mathcal{H} + \kappa \epsilon^\alpha \partial_\alpha
\mathcal{H}\;, \\ 
\mathcal{G}' &= \mathcal{G} + \kappa \epsilon^\alpha \partial_\alpha
\mathcal{G}\;, \\ 
\psi^{\prime Q} &= \psi^Q + \kappa \epsilon^\alpha \partial_\alpha \psi^Q\; ,\\
\bar{\psi}^{\prime Q} &= \bar{\psi}^Q + \kappa \epsilon^\alpha
\partial_\alpha \bar{\psi}^Q \;,\\ 
\mathcal{A}_\mu' & =  \mathcal{A}_\mu + \kappa \epsilon^\alpha
\partial_\alpha \mathcal{A}_\mu + \kappa (\partial_\mu
\epsilon^\alpha) \mathcal{A}_\alpha\; . 
\end{align}
\end{subequations}
There is now a degree  of arbitrariness in the way the transformations
are  attributed  separately for  the  background  and quantum  fields,
within  the context of  the BFM.  We choose  to distribute  the metric
transformation as
\begin{subequations}
\label{eq:gravitoninvariance}
\begin{align}
\bar{h}_{\mu \nu}' &=\bar{h}_{\mu \nu} + \partial_\mu \epsilon_\nu +
\partial_\nu \epsilon_\mu + \kappa(\bar{h}^\alpha_\nu \partial_\mu
\epsilon_\alpha + \bar{h}^\alpha_\mu \partial_\nu \epsilon_\alpha +
\epsilon_\alpha \partial^\alpha \bar{h}_{\mu \nu})\; ,\\
h_{\mu \nu}^{\prime Q} &= h^Q_{\mu \nu} + \kappa(h^Q_{\alpha\nu}
\partial_\mu \epsilon^\alpha + h^Q_{\alpha\mu} \partial_\nu
\epsilon^\alpha + \epsilon_\alpha \partial^\alpha h^Q_{\mu
  \nu}\label{eq:quantumgravitoninvariance})\; .
\end{align}
\end{subequations}
Similarly, we distribute the transformations of the $\mathcal{H},
\mathcal{G}$ and $\mathcal{A}_\mu$ fields as 
\begin{subequations}
\begin{align}
\bar{H}' &= \bar{H} + \kappa \epsilon^\alpha \partial_\alpha \bar{H},
&H^{\prime Q } &= H^{Q} + \kappa \epsilon^\alpha \partial_\alpha
H^{Q}, \\  
\bar{G}' &= \bar{G} + \kappa \epsilon^\alpha \partial_\alpha \bar{G},
&G^{\prime Q } &= G^{Q} + \kappa \epsilon^\alpha \partial_\alpha G^Q
\\  
\bar{A}_\mu' & =  \bar{A}_\mu + \kappa \epsilon^\alpha \partial_\alpha
\bar{A}_\mu + \kappa (\partial_\mu \epsilon^\alpha) \bar{A}_\alpha, &
A^{\prime Q }_\mu & =  A^Q_\mu + \kappa \epsilon^\alpha
\partial_\alpha A^Q_\mu + \kappa (\partial_\mu \epsilon^\alpha)
A^Q_\alpha.  
\end{align}
\end{subequations}

It  is  now crucial  to  check  whether  the symmetry  transformations
in~\eqref{eq:diffeomorphisms} for  the action $S$ of  the theory leave
the integral  measure $\mathcal{D} \Phi$ invariant as  well.  For this
purpose, we need to calculate the Jacobian determinant associated with
the  transformations of  diffeomorphism, i.e.   
\begin{equation}
J[\epsilon] \equiv
  \det \Bigg ( \frac{\delta \Phi'_i(x)}{\delta \Phi_j(y)}\Bigg )
\end{equation}
where  $\Phi_i \in  \{h^Q_{\mu \nu},  H^Q, G^Q,  \psi^Q, \bar{\psi}^Q,
A_\mu^Q \}$. Using the fact that
\begin{equation}
\det(I + A) = 1 + {\rm Tr}(A) + O(A^2)
\end{equation}
for small $A$, we obtain that 
\begin{subequations}
\begin{align}
\det \Bigg (\frac{\delta H^{\prime Q}(x)}{\delta H^Q(y)} \Bigg )
  & = 1 - \frac{1}{2} \kappa \delta(0) \int d^4 x \partial_\mu
  \epsilon^\mu(x)\; ,\\ 
\det \Bigg (\frac{\delta \psi^{\prime Q}(x)}{\delta \psi^Q(y)} \Bigg )
  & = 1 - \frac{1}{2} \kappa \delta(0) \int d^4 x \partial_\mu
  \epsilon^\mu(x)\; ,\\ 
\det \Bigg (\frac{\delta A^{\prime Q}_\mu(x)}{\delta A^Q_\nu(y)}
  \Bigg ) & = 1 - \kappa \delta(0) \int d^4 x \partial_\mu
  \epsilon^\mu(x)\; ,\\
\det \Bigg (\frac{\delta h^{\prime Q}_{\mu \nu}(x)}{\delta
    h^Q_{\rho \sigma}(y)} \Bigg ) & = 1\; .
\end{align}
\end{subequations}
Consequently, for  scalars, fermions and spin-1  fields, there seems
to be a deviation from $1$. However, one may observe that the integral
appearing in the measure's transformation actually vanishes,
\begin{equation}
\int d^4 x \: \partial_\mu \epsilon^\mu\ =\ 0\; ,
\end{equation}
since fields (as well as gauge transformed fields) are taken to vanish
sufficiently rapidly at the boundaries, i.e.,~$\epsilon (x) \to 0$, as
$x \to \pm \infty$.

\subsection{Master Ward Identity for Diffeomorphisms}

Given the diffeomorphisms invariance of the generating functional $Z$,
we may now  derive a master WI associated with  this symmetry. To this
end,  we require  that the  part of  $Z$ containing  the  source terms
remains        invariant         under        the        infinitesimal
diffeomorphisms~\eqref{eq:diffeomorphisms}.   To accomplish  this, the
sources need to transform as tensors of the relevant rank as follows:
\begin{subequations}
\begin{align}
 \label{eq:sourcetransformations} 
 J_h'^{\mu  \nu} &=  J_h^{\mu
      \nu}  + \kappa(\epsilon^\alpha  \partial_\alpha J_h^{\mu  \nu} -J_h^{\nu\alpha}
    \partial^\mu   \epsilon_\alpha   -  J_h^{\mu\alpha}   \partial^\nu
    \epsilon_\alpha)\;,\\   
J_H'  &=   J_H   +  \kappa \epsilon^\alpha\partial_\alpha  J_H\;,\\  
J_G' &=  J_G +  \kappa \epsilon^\alpha \partial_\alpha J_G\; ,\\   
J_\psi'    &=   J_\psi    +   \kappa \epsilon^\alpha \partial_\alpha J_\psi\; ,\\   
\bar{J}_\psi' &= \bar{J}_\psi + \kappa
                          \epsilon^\alpha\partial_\alpha \bar{J}_\psi\;,\\ 
J_A'^{\mu} &  = J_A^\mu + \kappa(\epsilon^\alpha
    \partial_\alpha J_A^\mu - \epsilon_\alpha \partial^\mu J_A^\alpha)\;.
\end{align}
\end{subequations}
Under   these   transformations,   along  with   the   diffeomorphisms
\eqref{eq:diffeomorphisms}   and   \eqref{eq:gravitoninvariance},  the
generating functional $Z$ remains  invariant. Therefore, writing $X' =
X  + \delta  X$ for  $X \in  \{ \bar{h}_{\mu  \nu},  \bar{H}, \bar{G},
\bar{A}_\mu, J_h^{\mu \nu}, J_\psi,\bar{J}_\psi, J_H, J_G, J_A^\mu\}$,
we obtain the identity 
\begin{equation}
\int d^4 x \sum_X \frac{\delta Z}{\delta X} \delta X\ =\ 0\; .
\end{equation}
Defining the generating functional of connected Green's functions $W$ by
\begin{equation} 
  \label{eq:connectedgeneratingfunctionaldef}
Z[\bar{h}_{\mu \nu}, \bar{H}, \bar{G}, \bar{A}_\mu, J_h^{\mu
      \nu}, J_\psi,\bar{J}_\psi, J_H, J_G, J_A^\mu]\ =\ \exp( i
  W[\bar{h}_{\mu \nu}, \bar{H}, \bar{G}, \bar{A}_\mu, J_h^{\mu \nu},
    J_\psi,\bar{J}_\psi, J_H, J_G, J_A^\mu])\; ,  
\end{equation}
we obtain
\begin{equation}\int d^4 x \sum_X \frac{\delta W}{\delta X} \delta X =
  0\; .
\end{equation}
Next, we define the one particle irreducible (1PI) effective action
$\Gamma$ by means of a Legendre transform of $W$: 
\begin{equation}
\ba{
 \Gamma[\bar{h}_{\mu \nu}, \psi, \bar{\psi},\bar{H},\bar{G},
   \bar{A}_\mu, h_{\mu \nu}, H, G, A_\mu]= &\;  W[\bar{h}_{\mu \nu},
   \bar{H}, \bar{G}, \bar{A}_\mu, J_h^{\mu \nu}, J_\psi,\bar{J}_\psi,
   J_H, J_G, J_A^\mu]\\& - \int d^4 x  \sqrt{- \bar{g}}\big(J_h^{\mu \nu}h_{\mu \nu}
 + \bar{J}_{\psi} \psi + \bar{\psi} J_{\psi} + J_H H + J_G G + J_A^\mu
 A_\mu \big)\; , \label{eq:effectiveactiondef} 
}
\end{equation}
where
\begin{equation}
h_{\mu \nu} \equiv \frac{\delta W }{\delta J_h^{\mu \nu}}\;, \quad \psi
  \equiv \frac{\delta W }{\delta \bar{J}_\psi}\;, \quad \bar{\psi} \equiv
  \frac{\delta W}{\delta J_\psi}\;, \quad H \equiv \frac{\delta W}{\delta
    J_H}\;, \quad G \equiv \frac{\delta W}{\delta J_G}\;, \quad A_\mu \equiv
  \frac{\delta W}{\delta J_A^\mu}\;.
\end{equation}
To have an invariant effective action, we must require that the source
terms  remain  invariant. As  a  consequence,  the Legendre  transform
variables   transform   like   their   quantum   field   counterparts,
i.e.~according to  the transformations (\ref{eq:diffeomorphisms}) with
the  identification  $X^Q \to  X$  for  $X  \in \{h_{\mu  \nu},  \psi,
\bar{\psi}, H, G, A_\mu\}$. This allows us to write
\begin{equation}
\Gamma[\bar{h}_{\mu \nu}', \psi', \bar{\psi}',\bar{H}', \bar{G}',
    \bar{A}_\mu', h_{\mu \nu}', H', G', A_\mu']\ =\ \Gamma[\bar{h}_{\mu
      \nu}, \psi, \bar{\psi},\bar{H},\bar{G}, \bar{A}_\mu, h_{\mu
      \nu}, H, G, A_\mu]\; .
\end{equation}
For vanishing  arguments of the  quantum fields $h_{\mu  \nu},\, H,\,
G,\, A_\mu$, we have
\begin{equation}
\bar{\Gamma}[\bar{h}_{\mu \nu}', \psi', \bar{\psi}',\bar{H}',\bar{G}',
    \bar{A}_\mu']\ =\ \bar{\Gamma}[\bar{h}_{\mu \nu}, \psi,
    \bar{\psi},\bar{H},\bar{G}, \bar{A}_\mu]\; ,
\end{equation}
which is a statement of invariance for the background field effective
action defined by 
\begin{equation}
\bar{\Gamma}[\bar{h}_{\mu \nu}, \psi, \bar{\psi}, \bar{H},
    \bar{G}, \bar{A}_\mu]\ \equiv\ \Gamma[\bar{h}_{\mu \nu}, \psi,
    \bar{\psi},\bar{H},\bar{G}, \bar{A}_\mu, 0,0,0,0]\; .
\end{equation}
An immediate consequence of this invariance is the master Ward identity
\begin{equation}
   \label{eq:matterwardidentity}
\ba{
& &\sqpara{\delta^\alpha_\mu \partial_\nu  + \kappa
    \para{\bar{h}^\alpha_{\nu} \partial_{\mu} + \partial_\mu
      \bar{h}^\alpha_\nu + \frac{1}{2} \partial^\alpha \bar{h}_{\mu
        \nu} }} \frac{\delta \bar{\Gamma}}{\delta \bar{h}_{\mu
      \nu}(x)} + \kappa \Big(\partial^\alpha \bar{A}_\mu -
  \partial_\mu \bar{A}^\alpha -  \bar{A}^\alpha \partial_\mu \Big
  )\frac{\delta \bar{\Gamma}}{\delta \bar{A}_{\mu}}  
\\& & + \; \kappa \partial^\alpha \bar{H} \frac{\delta
  \bar{\Gamma}}{\delta \bar{H}} + \kappa \partial^\alpha \bar{G}
\frac{\delta \bar{\Gamma}}{\delta \bar{G}}  + \frac{\delta
  \bar{\Gamma}}{\delta \psi} \kappa \partial^\alpha \psi  +  \kappa
\partial^\alpha \bar{\psi} \frac{\delta \bar{\Gamma}}{\delta
  \bar{\psi}}\ =\ 0\; ,
}
\end{equation}
where  $\alpha$  is  a  free  index.  By  appropriate  differentiation
of~\eqref{eq:matterwardidentity}  with respect  to the  fields  of the
theory, this master WI can be used to deduce further WIs and relations
between correlation functions of the background fields.

\begin{figure}
\centering
\begin{equation} 
\nonumber
p_\mu \Bigg
(\raisebox{-0.45\height}{\includegraphics[scale=1]{selfenergyblob.1}}\Bigg)^{\mu
  \nu, \rho \sigma} +\ \frac{\kappa}{2}\;\Bigg[\, \eta^{\nu \rho} p_\mu
\Bigg(\raisebox{-0.45\height}{\includegraphics[scale=1]{tadpoleblob.1}} \hspace{0.75em}\Bigg)^{\mu\sigma} 
+\ \eta^{\nu \sigma} p_\mu
\Bigg(\raisebox{-0.45\height}{\includegraphics[scale=1]{tadpoleblob.1}}\hspace{0.75em}\Bigg)^{\mu\rho} -\ p^\nu
\Bigg(\raisebox{-0.45\height}{\includegraphics[scale=1]{tadpoleblob.1}}\hspace{0.75em}\Bigg)^{\rho \sigma}\Bigg]\ =\ 0\; .
\end{equation}
\caption{Diagrammatic representation of the Ward 
                    Identity~(\ref{eq:secondwardidentitymomentum}), where the
zigzag lines denote gravitons.}
\label{fig:wardidentitygraphs}
\end{figure}

Since we  are interested here only in  graviton correlation functions,
we  may take the  matter field  arguments of  the effective  action to
zero. This yields a simpler version of the master WI:
\begin{equation}
\sqpara{\delta^\alpha_\mu \partial_\nu  + \kappa
    \para{\bar{h}^\alpha_{\nu} \partial_{\mu} + \partial_\mu
      \bar{h}^\alpha_\nu + \frac{1}{2} \partial^\alpha \bar{h}_{\mu
        \nu} }} \frac{\delta \bar{\Gamma}}{\delta \bar{h}_{\mu
      \nu}(x)}\ =\ 0\; .
\end{equation}
Differentiating   functionally   with   respect  to   $\bar{h}_{\rho
\sigma}(y)$  and converting  the result  into the  momentum  space, we
  obtain the Ward identity
\begin{equation}
  \label{eq:secondwardidentitymomentum} 
p_\mu \Pi^{\mu \nu, \rho \sigma}(p)\: +\: \frac{\kappa}{2}\,
  \Big(\,\eta^{\nu \rho} p_{\mu} T_h^{\sigma
      \mu}\: +\: \eta^{\nu \sigma} p_{\mu} T_h^{\rho \mu}\: -\: p^\nu T_h^{\rho
      \sigma}\,\Big)\ =\ 0\;, 
\end{equation}
where  $p^\mu$   is  the   graviton  momentum,  $\Pi^{\mu   \nu,  \rho
  \sigma}(p)$ is  the 1PI graviton self-energy and  $T^{\mu \nu}_h$ is
the   1-point  correlation   function  for   the   graviton  tadpoles.
Figure~\ref{fig:wardidentitygraphs}  gives a  graphical representation
of the Ward  identity (\ref{eq:secondwardidentitymomentum}), where the
zigzag lines indicate gravitons.

We conclude this section by  commenting on the appearance of the terms
depending  on  the  graviton  tadpoles  $T^{\mu \nu}_h$  in  the  Ward
identity~\eqref{eq:secondwardidentitymomentum}.      In~fact,    their
appearance is  where Yang-Mills theory and PQG  explicitly differ, as
tadpole graphs for Yang-Mills fields vanish identically due to Lorentz
covariance. On the other hand,  previous studies in PQG mostly focused
on  massless  particle   contributions  to  the  graviton  self-energy
\cite{Capper:1973mv,Capper:1974ed,Capper:1973pv,Capper:1984qq},     for
which  the   tadpole  contributions  were   unimportant,  since  these
contributions  vanish identically  in the  context of  DR.   Thus, the
self-energy becomes transverse  in this case, as a  consequence of the
WI~\eqref{eq:secondwardidentitymomentum}, with $T^{\mu\nu}_h = 0$.  In
the massive  case, however,  the tadpole graphs  do not vanish  in DR,
thus  signifying the presence  of longitudinal  modes in  the graviton
self-energy. In the next  two sections, we will explicitly demonstrate
how  these  longitudinal modes  disappear  after  minimisation of  the
effective action and renormalization of the cosmological constant.

\section{Minimisation Conditions and Cosmological Constant
  Renormalization\label{CC}}

In this section,  we discuss the minimization of  the effective action
$\Gamma$, in the presence of background graviton fields, and elucidate
its   connection  with   the  renormalization   of   the  cosmological
constant~$\Lambda$. We  also derive a low-energy  theorem that relates
graviton  tadpoles  with the  graviton  self-energy  at zero  external
momentum. As we will see, this  theorem plays a central role to ensure
the masslessness of gravitons.

In the  context of the BFM,  the minimisation of  the effective action
with  respect to  the background  fields translates  into  the generic
condition:
\begin{equation}
\left. \frac{\delta \Gamma}{\delta X} \right |_{X = 0}\ =\ 0\; ,
\end{equation}
for  $X   \in  \{h_{\mu  \nu},  \psi,  \bar{\psi},   H,  G,  A_\mu\}$.
Specifically, we  require that the  vacuum expectation value~(VEV)~$v$
of   the   Higgs  boson   be   translation   and  Lorentz   invariant,
i.e.,~$\partial_\mu v  = 0$.   If we define  $\Gamma =  \Gamma^{(0)} +
\Gamma^{(n  \geq  1)}$, where  $\Gamma^{(n  \geq  1)}$ represents  the
quantum   corrections,  we  obtain   the  following   equations:  
\begin{eqnarray}
   \label{eq:higgstadpolecondition}
\frac{\delta \Gamma}{\delta H}\ &=&\
  f_{\bar{H}}(\bar{H},\bar{G},\bar{A}_\mu,v_0,\mu^2_0,\lambda_0,e_0) - 
\lambda_0 v_0 \bigg( v_0^2  -  \frac{\mu^2_0}{\lambda_0}\bigg) + 
\frac{\delta  \Gamma^{(n \geq 1)}}{\delta H}\ =\ 0\;,\\ 
\frac{\delta \Gamma}{\delta
    G}\ &=&\   f_{\bar{G}}(\bar{H},\bar{G},\bar{A}_\mu,v_0,\mu^2_0,\lambda_0,e_0)+
  \frac{\delta  \Gamma^{(n \geq  1)}}{\delta  G}\ =\  0\; , \\  
\frac{\delta \Gamma}{\delta  \psi}\ &=&\  \frac{\delta\Gamma^{(n \geq  1)}}{\delta
    \psi}\  =\  0\;  ,  \\  
\frac{\delta  \Gamma}{\delta  \bar{\psi}}\ &=&\
\frac{\delta  \Gamma^{(n  \geq   1)}}{\delta  \bar{\psi}}\ =\ 0\;,\\
\frac{\delta          \Gamma}{\delta          A_\mu}\ &=&\
f^\mu_{\bar{A}}(\bar{H},\bar{G},\bar{A}_\mu,v_0,\mu^2_0,\lambda_0,e_0)          +
\frac{\delta   \Gamma^{(n   \geq   1)}}{\delta   A_\mu}\  =\   0\;,\\   
  \label{eq:gravitontadpolecondition} 
\frac{\delta   \Gamma}{\delta   h_{\mu   \nu}}\ &=&\   \frac{1}{2}
\bar{g}^{\mu \nu} \para{\frac{1}{\kappa}\bar{R} + \kappa(\Lambda_0 +
\Lambda^H_0)}   -    \frac{1}{\kappa}\bar{R}^{\mu    \nu}   -
\frac{\kappa}{2}  \bar{T}^{\mu \nu}  + \frac{\delta  
\Gamma^{(n \geq1)}}{\delta          h_{\mu         \nu}}\ =\ 0\; , 
\end{eqnarray}
where
\begin{align}
f_{\bar{H}}(\bar{H},\bar{G},\bar{A}_\mu,v_0,\mu^2_0,\lambda_0,e_0)\ =&\ \;
\bar{g}^{\mu \nu}\partial_\mu \partial_\nu \bar{H} + e_0^2 \bar{A}^\mu
\bar{A}_\mu (v_0+\bar{H})\\ &\ -\, \lambda_0\,\bigg[\,v_0
  \Big(2v_0\bar{H} + \bar{H}^2 + \bar{G}^2\Big)+ \bar{H}\bigg( (v_0 +
  \bar{H})^2 + \bar{G}^2 - \frac{\mu^2_0}{\lambda_0}\bigg) \bigg]\; ,\\ 
f_{\bar{G}}(\bar{H},\bar{G},\bar{A}_\mu,v_0,\mu^2_0,\lambda_0,e_0)\ =&\ \;
\bar{g}^{\mu \nu} \partial_\mu \partial_\nu \bar{G}+ e_0^2 \bar{A}^\mu
\bar{A}_\mu \bar{G} +\frac{\lambda_0}{2}\bar{G}\bigg[ (v_0 + \bar{H})^2 +
\bar{G}^2 - \frac{\mu^2_0}{\lambda_0} \bigg]\; ,\\ 
f^\mu_{\bar{A}}(\bar{H},\bar{G},\bar{A}_\mu,v_0,\mu^2_0,\lambda_0,e_0)\ =&\ \;
\bar{g}^{\mu \rho} \bar{g}^{\nu \sigma} \Big(\partial_\nu
\partial_\sigma \bar{A}_\rho- \partial_\rho \partial_\sigma
\bar{A}_{\nu}\Big)- e_0 \Big( (v_0+\bar{H})\partial^\mu \bar{G} -
\bar{G}\partial^\mu \bar{H}\Big)\\ 
&\ +\, e_0^2 \bar{A}^\mu\Big( (v_0 + \bar{H})^2 + \bar{G}^2 \Big)\; ,\\
  \label{eq:LH0}
\Lambda^H_0\ = &\ -\, \frac{\lambda_0}{4}\, 
\bigg(v_0^2 - \frac{\mu^2_0}{\lambda_0}\bigg)^2
\end{align}
and $\bar{e}^\mu_a$ represents the  background vielbein field. Here, a
bar on a field (other than $\bar{\psi}$) represents a background field
and  a   subscript~0  indicates  a   bare  (unrenormalized)  kinematic
parameter, such as the bare coupling constant $e_0$ and the bare Higgs
VEV $v_0$.  In  the BFM, the background fields are  not free, but obey
their  respective equations  of  motion with  some specified  boundary
conditions.  Thus,  we assume that  all the background  fields satisfy
these constraints without  determining their analytical form. Finally,
in  the  present model  under  study, only  the  Higgs  boson and  the
graviton can have non-zero tadpole contributions.

Let us  now turn our attention  to discussing quantum  loop effects on
the  cosmological  constant~$\Lambda$.   Observe that  the  generating
functional  $Z$ defined in~\eqref{eq:generatingfunctionaldef}  is well
specified,  except  of  an  overall  normalization  constant~$N$.   In
theories,  in  which gravitons  are  treated  as classical  background
fields, such a constant $N$  seems to be equivalent to renormalization
of  $\Lambda$.    However,  in   theories  of  quantum   gravity,  the
cosmological constant  is accompanied  by a factor  $\sqrt{-g}$, which
prevents the  factorization of $\Lambda$  from the path  integral.  To
deal   with   this   problem,   we   treat   the   cosmological   term
$\sqrt{-g}\,\Lambda$ as  an interaction in the  action and renormalize
$\Lambda$, by renormalizing the  effective action $\Gamma[0]$ by means
of a  gauge-invariant CT~$\delta \Lambda$.  This can be done  by first
writing $\Lambda_0 =  \Lambda + \delta \Lambda$ and  then imposing the
condition
\begin{equation}
  \label{eq:CCren}
\Gamma[0]\ =\ \Lambda_0\: +\: \Lambda^H_0\: +\: \Gamma^{(n \geq 1)}[0]\ =\
\Lambda\; .
\end{equation}
Assuming  a flat  Minkowski background  after renormalization,  we set
$\Lambda =  0$, such that $\eta_{\mu  \nu}$ remains a  solution of the
background  equations   of  motion.   Notice   that  $\Lambda^H_0$  is
renormalized only through the Higgs VEV $v_0$ and the quartic coupling
constant $\lambda_0$ [cf.~\eqref{eq:LH0}].  At the one-loop level, the
contribution  $\Gamma^{(1)}[0]$  to   the  cosmological  constant  may
graphically be represented as
\begin{equation}
\ba{
i \Gamma^{(1)}[0]\
&=\ \hspace{-1em}\raisebox{-0.45\height}{\begin{fmffile}{psibubble} 
\begin{fmfgraph*}(50,50)
\fmfleft{i}
\fmfright{o}
\fmf{phantom,tension=5}{i,v1}
\fmf{phantom,tension=5}{v2,o}
\fmf{fermion,left,tension=0.4}{v1,v2,v1}
\fmflabel{$\psi$}{v2}
\end{fmfgraph*}
\end{fmffile}}
\hspace{0.5em}+\hspace{-0.5em}
\raisebox{-0.45\height}{\begin{fmffile}{Hbubble}
\begin{fmfgraph*}(50,50)
\fmfleft{i}
\fmfright{o}
\fmf{phantom,tension=5}{i,v1}
\fmf{phantom,tension=5}{v2,o}
\fmf{dashes,left,tension=0.4}{v1,v2,v1}
\fmflabel{$H$}{v2}
\end{fmfgraph*}
\end{fmffile}}
\hspace{0.5em}+\hspace{-0.5em}
\raisebox{-0.45\height}{\begin{fmffile}{Gbubble}
\begin{fmfgraph*}(50,50)
\fmfleft{i}
\fmfright{o}
\fmf{phantom,tension=5}{i,v1}
\fmf{phantom,tension=5}{v2,o}
\fmf{dashes,left,tension=0.4}{v1,v2,v1}
\fmflabel{$G$}{v2}
\end{fmfgraph*}
\end{fmffile}}
\hspace{0.5em}+\hspace{-0.5em}
\raisebox{-0.45\height}{\begin{fmffile}{ghostbubble}
\begin{fmfgraph*}(50,50)
\fmfleft{i}
\fmfright{o}
\fmf{phantom,tension=5}{i,v1}
\fmf{phantom,tension=5}{v2,o}
\fmf{ghost,left,tension=0.4}{v1,v2,v1}
\fmflabel{$c$}{v2}
\end{fmfgraph*}
\end{fmffile}}
\\ &\hspace{3em}+\hspace{-0.5em}
\raisebox{-0.45\height}{\begin{fmffile}{photonbubble}
\begin{fmfgraph*}(50,50)
\fmfleft{i}
\fmfright{o}
\fmf{phantom,tension=5}{i,v1}
\fmf{phantom,tension=5}{v2,o}
\fmf{photon,left,tension=0.4}{v1,v2,v1}
\fmflabel{$A_\mu$}{v2}
\end{fmfgraph*}
\end{fmffile}}
\hspace{0.5em}+\hspace{-0.5em}
\raisebox{-0.45\height}{\begin{fmffile}{gravitonghostbubble}
\begin{fmfgraph*}(50,50)
\fmfleft{i}
\fmfright{o}
\fmf{phantom,tension=5}{i,v1}
\fmf{phantom,tension=5}{v2,o}
\fmf{ghost,left,tension=0.4}{v1,v2,v1}
\fmflabel{$\eta_\mu$}{v2}
\end{fmfgraph*}
\end{fmffile}}
\hspace{0.5em}+\hspace{-0.5em}
\raisebox{-0.45\height}{\begin{fmffile}{gravitonbubble}
\begin{fmfgraph*}(50,50)
\fmfleft{i}
\fmfright{o}
\fmf{phantom,tension=5}{i,v1}
\fmf{phantom,tension=5}{v2,o}
\fmf{zigzag,left,tension=0.4}{v1,v2,v1}
\fmflabel{$h_{\mu \nu}$}{v2}
\end{fmfgraph*}
\end{fmffile}} \;.
}
\end{equation}
Writing  $\Lambda^H_0  = \Lambda_H  +  \delta  \Lambda_H$,  it is  not
difficult  to  see  that  $\delta  \Lambda_H  =  0$  at  the  one-loop
level.   Therefore,  the   renormalization  condition~\eqref{eq:CCren}
simplifies to
\begin{equation}
\delta \Lambda\: +\: \Gamma^{(1)}[0]\ =\ 0\; .
\end{equation}
In   the   DR   scheme,   the  individual   graphs   contributing   to
$\Gamma^{(1)}[0]$ can be calculated explicitly. In this way, we obtain
\begin{eqnarray}
\raisebox{-0.45\height}{\begin{fmffile}{Hbubble}
\begin{fmfgraph*}(50,50)
\fmfleft{i}
\fmfright{o}
\fmf{phantom,tension=5}{i,v1}
\fmf{phantom,tension=5}{v2,o}
\fmf{dashes,left,tension=0.4}{v1,v2,v1}
\fmflabel{$H$}{v2}
\end{fmfgraph*}
\end{fmffile}}\hspace{1em} &=&\ \frac{1}{2} \int \frac{d^d k}{(2
  \pi)^d} \ln\para{-  k^2 + m_H^2}\ =\ \frac{i}{2(4 \pi)^2}
\para{\frac{m_H^2}{2} A_0(m_H^2) +
  \frac{m_H^4}{4}}\label{eq:higgsvacuumgraph}, \hspace{3.5cm}\\ 
\raisebox{-0.45\height}{\begin{fmffile}{psibubble}
\begin{fmfgraph*}(50,50)
\fmfleft{i}
\fmfright{o}
\fmf{phantom,tension=5}{i,v1}
\fmf{phantom,tension=5}{v2,o}
\fmf{fermion,left,tension=0.4}{v1,v2,v1}
\fmflabel{$\psi$}{v2}
\end{fmfgraph*}
\end{fmffile}} \hspace{1em} &=&\ - 2\int \frac{d^d k}{(2 \pi)^d}
\ln\para{- k^2 + m_\psi^2}\ =\ - \frac{2 i}{(4 \pi)^2}
\para{\frac{m_\psi^2}{2} A_0(m_\psi^2) + \frac{m_\psi^4}{4}},\\ 
\raisebox{-0.45\height}{\begin{fmffile}{photonbubble}
\begin{fmfgraph*}(50,50)
\fmfleft{i}
\fmfright{o}
\fmf{phantom,tension=5}{i,v1}
\fmf{phantom,tension=5}{v2,o}
\fmf{photon,left,tension=0.4}{v1,v2,v1}
\fmflabel{$A_\mu$}{v2}
\end{fmfgraph*}
\end{fmffile}}\hspace{1em} &=&\ \frac{d-1}{2} \int \frac{d^d k}{(2
  \pi)^d} \ln\para{-k^2 + m_A^2} + \frac{1}{2} \int \frac{d^d k}{(2
  \pi)^d} \ln\para{-k^2 + \xi_G m_A^2}\\ 
&=&\ \frac{3i}{2(4 \pi)^2} \para{\frac{m_A^2}{2} A_0(m_A^2) -
  \frac{m_A^4}{12}} + \frac{i}{2(4 \pi)^2} \para{\frac{\xi_G m_A^2}{2}
  A_0(\xi_G m_A^2) + \frac{ \xi_G^2 m_A^4}{4}}, \\ 
\raisebox{-0.45\height}{\begin{fmffile}{Gbubble}
\begin{fmfgraph*}(50,50)
\fmfleft{i}
\fmfright{o}
\fmf{phantom,tension=5}{i,v1}
\fmf{phantom,tension=5}{v2,o}
\fmf{dashes,left,tension=0.4}{v1,v2,v1}
\fmflabel{$G$}{v2}
\end{fmfgraph*}
\end{fmffile}} \hspace{1em} &=&\ \frac{1}{2} \int \frac{d^d k}{(2
  \pi)^d} \ln\para{- k^2 + \xi_G m_A^2}\ =\ \frac{i}{2(4 \pi)^2}
\para{\frac{\xi_G m_A^2}{2} A_0(\xi_G m_A^2) + \frac{\xi_G^2
    m_A^4}{4}}, \\ 
\raisebox{-0.45\height}{\begin{fmffile}{ghostbubble}
\begin{fmfgraph*}(50,50)
\fmfleft{i}
\fmfright{o}
\fmf{phantom,tension=5}{i,v1}
\fmf{phantom,tension=5}{v2,o}
\fmf{ghost,left,tension=0.4}{v1,v2,v1}
\fmflabel{$c$}{v2}
\end{fmfgraph*}
\end{fmffile}} \hspace{1em} &=&\ - \int \frac{d^d k}{(2 \pi)^d}
\ln\para{-k^2 + \xi_G m_A^2}\ =\ - \frac{i}{(4 \pi)^2} \para{\frac{\xi_G
    m_A^2}{2} A_0(\xi_G m_A^2) + \frac{\xi_G^2 m_A^4}{4}},\\ 
& &
\hspace{0.5cm}
\raisebox{-0.45\height}{\begin{fmffile}{gravitonghostbubble}
\begin{fmfgraph*}(50,50)
\fmfleft{i}
\fmfright{o}
\fmf{phantom,tension=5}{i,v1}
\fmf{phantom,tension=5}{v2,o}
\fmf{ghost,left,tension=0.4}{v1,v2,v1}
\fmflabel{$\eta_\mu$}{v2}
\end{fmfgraph*}
\end{fmffile}} \hspace{1em} =\ 0\;,
\hspace{2cm}
\raisebox{-0.45\height}{\begin{fmffile}{gravitonbubble}
\begin{fmfgraph*}(50,50)
\fmfleft{i}
\fmfright{o}
\fmf{phantom,tension=5}{i,v1}
\fmf{phantom,tension=5}{v2,o}
\fmf{zigzag,left,tension=0.4}{v1,v2,v1}
\fmflabel{$h_{\mu \nu}$}{v2}
\end{fmfgraph*}
\end{fmffile}} \hspace{1em} =\ 0\; .
\end{eqnarray}
Here,   $A_0(m^2)$   is  the   tadpole   loop   integral  defined   in
$d=4-2\epsilon$ as
\begin{equation}
  \label{eq:a0definition}
A_0(m^2)\ \equiv\ (2\pi\mu)^{4-d}\,
\int \frac{d^d k}{i\pi^2}\: \frac{1}{k^2 - m^2}\ =\ 
m^2\,\bigg[\ \frac{1}{\bar{\epsilon}}\ +1 -\: \ln
  \bigg(\frac{m^2}{\mu^2}\bigg)\, \bigg]\; ,  
\end{equation}
where  $1/\bar{\epsilon} =  1/\epsilon -  \gamma_E +  \ln  4\pi$, with
$\gamma_E$ being the Euler--Mascheroni constant and $\mu$ the 't Hooft
mass renormalization scale.  We note that the sum
\begin{equation}
\raisebox{-0.45\height}{\begin{fmffile}{photonbubble}
\begin{fmfgraph*}(50,50)
\fmfleft{i}
\fmfright{o}
\fmf{phantom,tension=5}{i,v1}
\fmf{phantom,tension=5}{v2,o}
\fmf{photon,left,tension=0.4}{v1,v2,v1}
\fmflabel{$A_\mu$}{v2}
\end{fmfgraph*}
\end{fmffile}}
\hspace{0.5em}+\hspace{-0.5em}
\raisebox{-0.45\height}{\begin{fmffile}{Gbubble}
\begin{fmfgraph*}(50,50)
\fmfleft{i}
\fmfright{o}
\fmf{phantom,tension=5}{i,v1}
\fmf{phantom,tension=5}{v2,o}
\fmf{dashes,left,tension=0.4}{v1,v2,v1}
\fmflabel{$G$}{v2}
\end{fmfgraph*}
\end{fmffile}}
\hspace{0.5em}+\hspace{-0.5em}
\raisebox{-0.45\height}{\begin{fmffile}{ghostbubble}
\begin{fmfgraph*}(50,50)
\fmfleft{i}
\fmfright{o}
\fmf{phantom,tension=5}{i,v1}
\fmf{phantom,tension=5}{v2,o}
\fmf{ghost,left,tension=0.4}{v1,v2,v1}
\fmflabel{$c$}{v2}
\end{fmfgraph*}
\end{fmffile}}
\ =\ \frac{3i}{2(4 \pi)^2} \para{\frac{m_A^2}{2} A_0(m_A^2) - \frac{m_A^4}{12}}
\end{equation}
is independent of the  $U(1)_Y$ gauge fixing parameter $\xi_G$.  Thus,
at the  one-loop level, the cosmological constant  CT $\delta \Lambda$
is found to be
\begin{equation}
  \label{eq:dLambda}
\delta \Lambda\ =\ \frac{2}{(4 \pi)^2} \para{\frac{m_\psi^2}{2}
  A_0(m_\psi^2) + \frac{m_\psi^4}{4}}  -\frac{1}{2(4 \pi)^2}
\para{\frac{m_H^2}{2} A_0(m_H^2) + \frac{m_H^4}{4}} - \frac{3}{2(4
  \pi)^2} \para{\frac{m_A^2}{2} A_0(m_A^2) - \frac{m_A^4}{12}}. 
\end{equation}
The  fact that  $\delta \Lambda$  is  independent of  $\xi_G$ and  the
diffeomorphisms-fixing  parameters $\xi_D$  and $\sigma$  reflects the
gauge   invariance   of   the   effective  action   at   its   minimum
\cite{Nielsen:1975}   and  provides  a   consistency  check   for  the
correctness of our analytic results.

Let     us     now     analyze     the     minimisation     conditions
(\ref{eq:higgstadpolecondition})                                    and
(\ref{eq:gravitontadpolecondition})  related  to  the  Higgs  and  the
graviton tadpoles,  respectively. For the Higgs  tadpole condition, we
have
\begin{equation}
 \begin{aligned}
\frac{\delta \Gamma^{(1)}}{\delta H}\ \equiv T_H\ =&
\raisebox{-0.45\height}{\begin{fmffile}{HHtadpole} 
  \begin{fmfgraph*}(80,50)
   \fmfleft{i}
    \fmfright{o}
\fmf{dashes,tension=0.8,label=$H$,l.side=left}{i,v1}
\fmf{phantom,tension=5}{v2,o}
\fmf{dashes,left,tension=0.4}{v1,v2,v1}
\fmflabel{$H$}{v2}
\end{fmfgraph*}
\end{fmffile}}\hspace{0.5em}+\raisebox{-0.45\height}{\begin{fmffile}{HGtadpole}
\begin{fmfgraph*}(80,50)
\fmfleft{i}
\fmfright{o}
\fmf{dashes,tension=0.8,label=$H$,l.side=left}{i,v1}
\fmf{phantom,tension=5}{v2,o}
\fmf{dashes,left,tension=0.4}{v1,v2,v1}
\fmflabel{$G$}{v2}
\end{fmfgraph*}
\end{fmffile}}\hspace{0.5em}+\raisebox{-0.45\height}{\begin{fmffile}{HAtadpole}
\begin{fmfgraph*}(80,50)
\fmfleft{i}
\fmfright{o}
\fmf{dashes,tension=0.8,label=$H$,l.side=left}{i,v1}
\fmf{phantom,tension=5}{v2,o}
\fmf{photon,left,tension=0.4}{v1,v2,v1}
\fmflabel{$A_\mu$}{v2}
\end{fmfgraph*}
\end{fmffile}}\hspace{0.5em}
\\&+\raisebox{-0.45\height}{\begin{fmffile}{HCtadpole}
\begin{fmfgraph*}(80,50)
\fmfleft{i}
\fmfright{o}
\fmf{dashes,tension=0.8,label=$H$,l.side=left}{i,v1}
\fmf{phantom,tension=5}{v2,o}
\fmf{ghost,left,tension=0.4}{v1,v2,v1}
\fmflabel{$c$}{v2}
   \end{fmfgraph*}
  \end{fmffile}}
 \end{aligned}
\end{equation}
Expressing the bare Higgs VEV $v_0$ as $v_0 = v + \delta v$,
(\ref{eq:higgstadpolecondition}) reads 
\begin{equation}
- 2 \lambda v^2 \delta v + T_H =0
\end{equation}
at the one-loop level, from which we deduce the Higgs VEV CT 
\begin{equation}
\delta v\ =\ \frac{T_H}{2 \lambda v^2}\ .
\end{equation}
To      deal     with      the     graviton      tadpole     condition
(\ref{eq:gravitontadpolecondition}), we first define
\begin{equation}
 \begin{aligned}
\frac{\delta \Gamma^{(1)}}{\delta h_{\mu \nu}}\ & =\
\raisebox{-0.45\height}{\begin{fmffile}{gravitonhiggstadpole} 
\begin{fmfgraph*}(80,50)
\fmfleft{i}
\fmfright{o}
\fmf{zigzag,tension=0.8,label=$h_{\mu \nu}$,l.side=left}{i,v1}
\fmf{phantom,tension=5}{v2,o}
\fmf{dashes,left,tension=0.4}{v1,v2,v1}
\fmflabel{$H$}{v2}
\end{fmfgraph*}
\end{fmffile}}
\hspace{0.5em}+
\raisebox{-0.45\height}{\begin{fmffile}{gravitongoldstonetadpole}
\begin{fmfgraph*}(80,50)
\fmfleft{i}
\fmfright{o}
\fmf{zigzag,tension=0.8,label=$h_{\mu \nu}$,l.side=left}{i,v1}
\fmf{phantom,tension=5}{v2,o}
\fmf{dashes,left,tension=0.4}{v1,v2,v1}
\fmflabel{$G$}{v2}
\end{fmfgraph*}
\end{fmffile}} 
\hspace{0.5em}+
\raisebox{-0.45\height}{\begin{fmffile}{gravitonfermiontadpole}
\begin{fmfgraph*}(80,50)
\fmfleft{i}
\fmfright{o}
\fmf{zigzag,tension=0.8,label=$h_{\mu \nu}$,l.side=left}{i,v1}
\fmf{phantom,tension=5}{v2,o}
\fmf{fermion,left,tension=0.4}{v1,v2,v1}
\fmflabel{$\psi$}{v2}
\end{fmfgraph*}
\end{fmffile}} 
\\ & \hspace{3em}+
\raisebox{-0.45\height}{\begin{fmffile}{gravitonphotontadpole}
\begin{fmfgraph*}(80,50)
\fmfleft{i}
\fmfright{o}
\fmf{zigzag,tension=0.8,label=$h_{\mu \nu}$,l.side=left}{i,v1}
\fmf{phantom,tension=5}{v2,o}
\fmf{photon,left,tension=0.4}{v1,v2,v1}
\fmflabel{$A_\mu$}{v2}
\end{fmfgraph*}
\end{fmffile}}
\hspace{0.5em}+
\raisebox{-0.45\height}{\begin{fmffile}{gravitonghosttadpole}
\begin{fmfgraph*}(80,50)
\fmfleft{i}
\fmfright{o}
\fmf{zigzag,tension=0.8,label=$h_{\mu \nu}$,l.side=left}{i,v1}
\fmf{phantom,tension=5}{v2,o}
\fmf{ghost,left,tension=0.4}{v1,v2,v1}
\fmflabel{$c$}{v2}
\end{fmfgraph*}
\end{fmffile}}
\\ &\ \equiv\ i T_h^{\mu \nu}.
 \end{aligned}
\end{equation}
As done with the Higgs field,  we allow for the quantum graviton field
$h_{\mu \nu}$ to  develop a VEV, by replacing  $h_{\mu \nu} \to h_{\mu
  \nu} + \delta  h_{\mu \nu}$. In this way,  the one-loop minimization
condition for the graviton field becomes:
\begin{equation}
  \label{eq:Tadgr}
\int d^4 y \sqpara{ \left . \frac{\delta^2 S}{\delta h_{\mu
        \nu}(x) \delta h_{\rho \sigma}(y)} \right |_{g_{\mu \nu} =
      \eta_{\mu \nu}}\delta h_{\rho \sigma}(y)}z\:
  +\: \frac{\kappa}{2}\eta^{\mu \nu}\delta \Lambda\: +\: T_h^{\mu
    \nu}\ =\ 0\; .
\end{equation}
From this  last equation, we  easily see that the  second functional
derivative with respect  to the quantum graviton field  is the inverse
graviton propagator in the flat space, i.e.
\begin{equation}
\left . \frac{\delta^2 S}{\delta h_{\mu \nu}(x) \delta h_{\rho
      \sigma}(y)} \right|_{g_{\mu \nu} = \eta_{\mu \nu}} =\ \Delta^{-1
    \mu \nu \rho \sigma}(x - y)\; .
\end{equation} 
By virtue of the latter, \eqref{eq:Tadgr} may be recast into the form:
\begin{equation}
  \label{eq:TgrII}
\int d^4 y \Big[ \Delta^{-1  \mu \nu \rho \sigma}(x - y)\delta
    h_{\rho \sigma}(y)\Big]\: +\: \frac{\kappa}{2}\eta^{\mu \nu}\delta
  \Lambda\: +\: T_h^{\mu \nu}(x)\ =\ 0\; .
\end{equation}
Solving equation~\eqref{eq:TgrII} for $\delta h_{\rho \sigma}$ yields 
\begin{equation}
  \label{eq:gravitonvevct}
\delta h_{\rho \sigma}(x)\ =\ -\, \int d^4 y \; \Delta_{\mu \nu \rho
    \sigma}(x - y)\: \Big( T_h^{\rho \sigma}(y)\: +\:
  \frac{\kappa}{2}\eta^{\rho \sigma}\delta \Lambda \Big)\; .
\end{equation}

It  is now  instructive  to calculate  the  one-loop graviton  tadpole
$T_h^{\mu \nu}$ resulting from our  Abelian Higgs model.  With the aid
of the Feynman rules given in Appendix A, the individual contributions
to $T_h^{\mu \nu}$ are given by
\begin{eqnarray}
\raisebox{-0.45\height}{\begin{fmffile}{gravitonhiggstadpole}
\begin{fmfgraph*}(80,50)
\fmfleft{i}
\fmfright{o}
\fmf{zigzag,tension=0.8,label=$h_{\mu \nu}$,l.side=left}{i,v1}
\fmf{phantom,tension=5}{v2,o}
\fmf{dashes,left,tension=0.4}{v1,v2,v1}
\fmflabel{$H$}{v2}
\end{fmfgraph*}
\end{fmffile}} \hspace{1em} &=&\ \frac{i \kappa}{4(4 \pi)^2}
\eta^{\mu \nu} \para{\frac{m_H^2}{2} A_0(m_H^2) + \frac{m_H^4}{4}}\;, \\ 
\raisebox{-0.45\height}{\begin{fmffile}{gravitongoldstonetadpole}
\begin{fmfgraph*}(80,50)
\fmfleft{i}
\fmfright{o}
\fmf{zigzag,tension=0.8,label=$h_{\mu \nu}$,l.side=left}{i,v1}
\fmf{phantom,tension=5}{v2,o}
\fmf{dashes,left,tension=0.4}{v1,v2,v1}
\fmflabel{$G$}{v2}
\end{fmfgraph*}
\end{fmffile}} \hspace{1em} &=&\ \frac{i \kappa}{4(4 \pi)^2}
\eta^{\mu \nu} \para{\frac{\xi_G m_A^2}{2} A_0(\xi_G m_A^2)+
  \frac{\xi_G^2 m_A^4}{4}}\; , 
\\
\raisebox{-0.45\height}{\begin{fmffile}{gravitonfermiontadpole}
\begin{fmfgraph*}(80,50)
\fmfleft{i}
\fmfright{o}
\fmf{zigzag,tension=0.8,label=$h_{\mu \nu}$,l.side=left}{i,v1}
\fmf{phantom,tension=5}{v2,o}
\fmf{fermion,left,tension=0.4}{v1,v2,v1}
\fmflabel{$\psi$}{v2}
\end{fmfgraph*}
\end{fmffile}} \hspace{1em} &=& - \frac{i \kappa}{2(4 \pi)^2}
\eta^{\mu \nu} \para{\frac{m_\psi^2}{2} A_0(m_\psi^2) +
  \frac{m_\psi^4}{4}}\;, \\ 
\raisebox{-0.45\height}{\begin{fmffile}{gravitonphotontadpole}
\begin{fmfgraph*}(80,50)
\fmfleft{i}
\fmfright{o}
\fmf{zigzag,tension=0.8,label=$h_{\mu \nu}$,l.side=left}{i,v1}
\fmf{phantom,tension=5}{v2,o}
\fmf{photon,left,tension=0.4}{v1,v2,v1}
\fmflabel{$A_\mu$}{v2}
\end{fmfgraph*}
\end{fmffile}}
\hspace{1em} & = &\ \frac{3 i \kappa}{4 (4 \pi)^2} \eta^{\mu \nu}
\para{\frac{m_A^2}{2} A_0(m_A^2) - \frac{m_A^4}{12}} \nonumber \\ & &
+\, \frac{i \kappa}{4(4 \pi)^2} \eta^{\mu \nu} \para{\frac{\xi_G
    m_A^2}{2} A_0(\xi_G m_A^2) - \frac{\xi_G^2 m_A^4}{12}}\; ,\\ 
\raisebox{-0.45\height}{\begin{fmffile}{gravitonghosttadpole}
\begin{fmfgraph*}(80,50)
\fmfleft{i}
\fmfright{o}
\fmf{zigzag,tension=0.8,label=$h_{\mu \nu}$,l.side=left}{i,v1}
\fmf{phantom,tension=5}{v2,o}
\fmf{ghost,left,tension=0.4}{v1,v2,v1}
\fmflabel{$c$}{v2}
\end{fmfgraph*}
\end{fmffile}}  &=&\ - \frac{i \kappa}{(4 \pi)^2} \eta^{\mu \nu}
\para{\frac{\xi_G m_A^2}{2} A_0(\xi_G m_A^2) + \frac{\xi_G^2
    m_A^4}{4}} \; . 
\end{eqnarray}
Interestingly enough, we observe that the sum
\begin{equation}
\raisebox{-0.45\height}{\begin{fmffile}{gravitongoldstonetadpole}
\begin{fmfgraph*}(80,50)
\fmfleft{i}
\fmfright{o}
\fmf{zigzag,tension=0.8,label=$h_{\mu \nu}$,l.side=left}{i,v1}
\fmf{phantom,tension=5}{v2,o}
\fmf{dashes,left,tension=0.4}{v1,v2,v1}
\fmflabel{$G$}{v2}
\end{fmfgraph*}
\end{fmffile}} \hspace{1em} +\ 
\raisebox{-0.45\height}{\begin{fmffile}{gravitonphotontadpole}
\begin{fmfgraph*}(80,50)
\fmfleft{i}
\fmfright{o}
\fmf{zigzag,tension=0.8,label=$h_{\mu \nu}$,l.side=left}{i,v1}
\fmf{phantom,tension=5}{v2,o}
\fmf{photon,left,tension=0.4}{v1,v2,v1}
\fmflabel{$A_\mu$}{v2}
\end{fmfgraph*}
\end{fmffile}}\hspace{1em} +
\raisebox{-0.45\height}{\begin{fmffile}{gravitonghosttadpole}
\begin{fmfgraph*}(80,50)
\fmfleft{i}
\fmfright{o}
\fmf{zigzag,tension=0.8,label=$h_{\mu \nu}$,l.side=left}{i,v1}
\fmf{phantom,tension=5}{v2,o}
\fmf{ghost,left,tension=0.4}{v1,v2,v1}
\fmflabel{$c$}{v2}
\end{fmfgraph*}
\end{fmffile}}\ =\ \frac{3 i \kappa}{4 (4 \pi)^2} \eta^{\mu \nu}
\para{\frac{m_A^2}{2} A_0(m_A^2) - \frac{m_A^4}{12}} 
\end{equation}
is independent  of the gauge  fixing parameter $\xi_G$,  implying that
the graviton tadpoles form a gauge-invariant set of graphs. This is in
stark contrast with the Higgs tadpole and its VEV CT $\delta v$, which
are  known   to  be  both  gauge-{\em   dependent}  quantities  (e.g.,
see~\cite{Pilaftsis:1997fe}). 

Our  effort  to  gain  a  better  understanding  of  the  gauge-fixing
parameter  independence  of $T_h^{\mu  \nu}$  led  us  to observe  the
following relation:
\begin{equation}
T_h^{\mu \nu}\: +\: \frac{\kappa}{2} \eta^{\mu \nu} \delta \Lambda\ =\ 0\;. 
  \label{eq:tadpolecosmocancellation}
\end{equation}
Remarkably, \eqref{eq:tadpolecosmocancellation} holds {\em separately}
for each of  the quantum field circulating in the  loop. Hence, at the
one-loop  level,   tadpole  graphs   are  directly  linked   with  the
gauge-invariant   renormalization   CT   $\delta   \Lambda$   of   the
cosmological constant, so $T^{\mu\nu}_h$ is a gauge-invariant quantity
as  well.  Moreover,  graviton tadpole  graphs cancel  against  the CT
$\delta \Lambda$ in the  one-loop effective action, which implies that
there is  no VEV renormalization for the  graviton field, i.e.~$\delta
h_{\mu\nu} = 0$.

It is important to stress  here that our approach to renormalizing the
graviton field  differs significantly from the one  outlined, e.g., in
\cite{Capper:1973bk}, where a  cosmological constant was introduced in
an  {\it ad  hoc}  manner, in  order  to cancel  the graviton  tadpole
effects.  In our case, such a  cancellation is a result of an explicit
computation, without the need  to impose an additional constraint.  In
the   next    subsection,   we    will   show   that    the   relation
\eqref{eq:tadpolecosmocancellation} leading to the non-renormalization
of the graviton VEV, with $\delta  h_{\mu\nu} = 0$, is not an one-loop
coincidence, but a result that holds to all orders in perturbation for
a gravitational theory renormalized to a Minkowski flat background.

\subsection{The Graviton Low Energy Theorem}

Here      we     will      explicitly     demonstrate      how     the
relation~(\ref{eq:tadpolecosmocancellation}) holds  true to all orders
in perturbation.  As we will  see, this non-perturbative relation is a
direct consequence of a Graviton Low Energy Theorem (GLET).

Given  the  conceptual  similarity  of  the GLET  with  the  so-called
Higgs-boson Low  Energy Theorem (HLET)~\cite{Ellis:1975, Shifman:1978,
  Vainshtein:1979,  Dawson:1992,  Kniehl:1995,  Pilaftsis:1997fe},  we
begin  our  demonstration  by  briefly  reminding the  reader  of  the
latter. The HLET may be stated by the following defining equation:
\begin{equation}
  \label{HLET}
\frac{\partial}{\partial v} \Gamma\ =\ \frac{\delta \Gamma}{\delta
  \bar{H}(0)}\; ,
\end{equation}
where   $\bar{H}(0)$   denotes   a  zero-momentum   background   Higgs
field. This  result may be derived  from a global  shift symmetry that
exists  between the  Higgs  VEV  $v$ and  the  background Higgs  field
$\bar{H}$ of the form:
\begin{equation}
v'\ =\ v + s\; , \qquad  \bar{H}'\ =\ \bar{H} - s\; ,
\end{equation}
for   some   infinitesimal  constant   $s$,   provided  a   compatible
gauge-fixing  condition is  chosen~\cite{Pilaftsis:1997fe}.   Taking a
functional  derivative with  respect to  $\bar{H}$,  invoking momentum
conservation and writing $\Gamma = \Gamma^{(0)} + \Gamma^{(n\geq 1)}$,
where  $\Gamma^{(n  \geq  1)}$  represents  the  part  of  the  action
containing one- and higher-order quantum loop effects, we obtain
\begin{equation}
\frac{\partial}{\partial v}\para{\frac{\delta \Gamma^{(n\geq
      1)}}{\delta \bar{H}(0)}}\ =\ 
\frac{\delta^2 \Gamma^{(n\geq 1)}}{\delta \bar{H}(0) \delta
  \bar{H}(0)}\ .
\end{equation}
Therefore, one consequence of the HLET relevant to our discussion here
is the relation  of the Higgs-boson tadpole to  quantum effects on the
Higgs-boson mass.

We may now try to extend the basic idea of HLET to theories of quantum
gravity. As  discussed in Section~\ref{TF}, the  full spacetime metric
$g_{\mu\nu}$ may be decomposed in the BFM framework of quantum gravity
as follows:
\begin{equation}
g_{\mu \nu}\ =\ \eta_{\mu \nu} + \kappa(\bar{h}_{\mu \nu} + h_{\mu
  \nu})\; ,
\end{equation}
where $\bar{g}_{\mu\nu} =  \eta_{\mu\nu} + \kappa \bar{h}_{\mu\nu}$ is
the background  metric [cf.~\eqref{eq:metricdecomposition}].  In close
analogy  to HLET,  it is  not  difficult to  observe that  there is  a
similar  symmetry for the  effective action  $\Gamma$ of  the complete
matter-gravity theory. In particular, the  effective action  $\Gamma$
remains invariant under the shift transformations:
\begin{equation}
 \label{eq:shiftsymmetry}
\eta_{\mu \nu}'\ =\ \eta_{\mu \nu}\: +\: s_{\mu \nu}\;, \qquad 
\bar{h}_{\mu \nu}'\ =\ \bar{h}_{\mu \nu}\: -\: \frac{1}{\kappa} s_{\mu\nu}\;, 
\end{equation}
where  $s_{\mu \nu}$  is an  arbitrary tensor.   Since  the generating
functional (\ref{eq:generatingfunctionaldef})  remains invariant under
the shift symmetry~\eqref{eq:shiftsymmetry}, we can derive the shift
Ward identity:
\begin{equation}
\frac{\partial Z}{\partial \eta_{\mu \nu}}\: -\:
\frac{1}{\kappa} \int d^4 x \frac{\delta Z}{\delta \bar{h}_{\mu
    \nu}(x)}\  =\ 0\; ,
\end{equation}
which implies
\begin{equation}
\frac{\partial W}{\partial \eta_{\mu \nu}}\: -\:
  \frac{1}{\kappa} \int d^4 x \frac{\delta W}{\delta \bar{h}_{\mu
      \nu}(x)} \  =\ 0\; ,
\end{equation}
by virtue of~(\ref{eq:connectedgeneratingfunctionaldef}). With the aid
of~(\ref{eq:effectiveactiondef}), we may translate the last result
into the shift WI for the effective action:
\begin{equation}
\frac{\partial \Gamma}{\partial \eta_{\mu \nu}}\: -\:
\frac{1}{\kappa} \int d^4 x \frac{\delta \Gamma}{\delta \bar{h}_{\mu
    \nu}(x)}\  =\ 0\; , 
\end{equation}
or equivalently in momentum space:
\begin{equation}
  \label{eq:GLET}
\kappa \frac{\partial}{\partial \eta_{\mu \nu}} \Gamma\ =\
  \frac{\delta \Gamma}{\delta \bar{h}_{\mu \nu}(0)}\ .
\end{equation}
Equation~\eqref{eq:GLET} is the defining  equation for the GLET, where
$\bar{h}_{\mu \nu}(0)$ is a  zero-momentum graviton field.  Now, if we
consider the counterterm in the effective action, 
\begin{equation}
  \label{eq:cccounterterm}
\Delta S\ =\  \int d^4 x \sqrt{-g}\,\delta \Lambda\; ,
\end{equation}
in  order   to  cancel  $\Gamma^{(n   \geq  1)}[0]$,  we   obtain  the
relation~(\ref{eq:tadpolecosmocancellation}):
\begin{equation*}
T_h^{\mu \nu}\: +\: \frac{\kappa}{2} \eta^{\mu \nu}\delta
\Lambda\ =\ 0\; ,
\end{equation*}
which holds  true to  all orders in  perturbation theory.   Hence, the
one-loop relation (\ref{eq:tadpolecosmocancellation}) is a consequence
of the~GLET.

In  addition to  relating  the graviton  tadpole  to the  cosmological
constant,  the  GLET can  also  relate  the  graviton tadpole  to  the
graviton self-energy at zero external momentum:
\begin{equation}
   \label{eq:Tgr0}
\kappa\frac{\partial}{\partial \eta_{\mu \nu}}\para{\frac{\delta
    \Gamma^{(n\geq 1)}}{\delta \bar{h}_{\rho \sigma}(0)}}\ =\
\frac{\delta^2 \Gamma^{(n\geq 1)}}{\delta \bar{h}_{\mu \nu}(0) \delta
  \bar{h}_{\rho \sigma}(0)}\; .
\end{equation}
Since graviton tadpoles vanish  identically for massless fields in the
loop  in the  DR scheme,  the  graviton self-energy  at zero  external
momentum   will   vanish  as   well,   by  means   of~\eqref{eq:Tgr0}.
Consequently,   the  GLET~\eqref{eq:GLET}   can  also   guarantee  the
masslessness  of the graviton  field in  DR, if  all particles  in the
quantum  loops are  massless.  As  we will  see in  the  next section,
however, this  is not in general  true, if massive  particles occur in
the graviton self-energy.  In this case, both the GLET~\eqref{eq:GLET}
and the  diffeomorphisms WI~\eqref{eq:secondwardidentitymomentum} will
be  needed  to   render  the  graviton  massless  to   all  orders  in
perturbation, assuming a flat Minkowski background.

\section{Matter Contributions to the Graviton Self-Energy \label{MC}}

In  this  section,  we  will  first demonstrate  explicitly  how  upon
renormalization,  the  graviton  self-energy  obeys  the  property  of
transversality entailing in  a massless graviton field.  Subsequently,
we will  compute the matter contributions to  the graviton self-energy
tensor  resulting  from  massive  scalar, pseudo-scalar,  fermion  and
vector-boson particles in the loops.

\subsection{Transversality of the Graviton Self-energy}

The   graviton   self-energy   transition   $\bar{h}_{\mu\nu}(p)   \to
\bar{h}_{\rho\sigma}(p)$,    which   we    denote    as   $\Pi^{\mu\nu
  ,\rho\sigma}(p)$, receives  two renormalizations: (i)  from the bare
cosmological constant  $\Lambda_0$ which induces a  CT proportional to
$\delta  \Lambda$  for  the  graviton  mass in  the  effective  action
[cf.~\eqref{eq:cccounterterm}]; (ii) from the Ricci scalar $R$ and the
higher-dimensional  operators $R^2$  and  $R^{\mu\nu}R_{\mu\nu}$.  The
latter  contributions~(ii),  which  we  denote  as  $\Delta\Pi^{\mu\nu
  ,\rho\sigma}(p)$, are  transverse in the  minimal subtraction scheme
($\overline{\rm MS}$)  of renormalization  and they have  therefore no
effect on the graviton~mass.

Taking  into  account  the  two  contributions  mentioned  above,  the
renormalized   graviton  self-energy   $\Pi^{\mu\nu  ,\rho\sigma}_{\rm
  R}(p)$ may then be written down as follows:
\begin{equation}
  \label{eq:PiR}
\Pi^{\mu\nu   ,\rho\sigma}_{\rm   R}(p)\ =\ \Pi^{\mu \nu, \rho \sigma}(p)\: -\:
\frac{\kappa^2}{4} P^{\mu \nu \rho \sigma} \delta \Lambda\: 
+\: \Delta\Pi^{\mu\nu   ,\rho\sigma}(p)\; .
\end{equation}
where we have defined the tensor
\begin{equation}
P^{\mu \nu \rho \sigma}\ \equiv\ \eta^{\mu \rho}\eta^{\nu \sigma}\: +\: \eta^{\mu
  \rho}\eta^{\nu \sigma}\: -\:  \eta^{\mu \nu}\eta^{\rho \sigma}\; ,
\end{equation}
for                brevity.                Employing               the
identity~\eqref{eq:tadpolecosmocancellation} deduced from the GLET, we
may readily obtain the relation
\begin{equation}
  \label{eq:LtoT}
\frac{\kappa}{2}\, P^{\mu \nu \rho \sigma}\,\delta\Lambda\ =\ -
\eta^{\nu\rho}\, T^{\sigma\mu}_h \: -\: \eta^{\nu\sigma}\,
T^{\rho\mu}_h\: +\: \eta^{\mu\nu}\, T^{\rho\sigma}_h\; . 
\end{equation}
Substituting this last expression back in~\eqref{eq:PiR} gives
\begin{equation}
\Pi^{\mu\nu   ,\rho\sigma}_{\rm   R}(p)\ =\ \Pi^{\mu \nu, \rho \sigma}(p)\: +\:
\frac{\kappa}{2}\,\Big(\,\eta^{\nu\rho}\, T^{\sigma\mu}_h \: +\: \eta^{\nu\sigma}\,
T^{\rho\mu}_h\: -\: \eta^{\mu\nu}\, T^{\rho\sigma}_h\Big)
\:  +\: \Delta\Pi^{\mu\nu   ,\rho\sigma}(p)\; .
\end{equation}
Based     on    the     WI~(\ref{eq:secondwardidentitymomentum})    of
diffeomorphisms            depicted           graphically           in
Fig.~\ref{fig:wardidentitygraphs}   and   the   fact   that   $p_\mu\,
\Delta\Pi^{\mu\nu ,\rho\sigma}(p)  = 0$, it  is not difficult  to show
that the renormalized graviton self-energy is transverse, i.e.
\begin{equation}
  \label{eq:wardidentitycts}
p_\mu\, \Pi^{\mu \nu, \rho \sigma}_{\rm R}(p)\ =\ 0\; . 
\end{equation}
Hence,  the  longitudinal  modes   of  the  graviton  self-energy  are
successfully  removed after  renormalizing the  cosmological constant.
We shall use the transversality identity~\eqref{eq:wardidentitycts} to
check the consistency of our analytic results.

We  may now  decompose  the renormalized  graviton self-energy  tensor
$\Pi^{\mu \nu, \rho \sigma}_{\rm R}(p)$ in terms of independent rank-4
Lorentz  tensors that depend  on $\eta^{\mu  \nu}$ and  $p^\mu p^\nu$.
More  explicitly,  $\Pi^{\mu  \nu,  \rho \sigma}_{\rm  R}(p)$  may  be
expressed as follows:
\begin{eqnarray}
  \label{eq:gravitonselfenergyform} 
\Pi^{\mu \nu,  \rho \sigma}_{\rm R}(p)\ &=&\ 
p^\mu p^\nu p^\rho p^\sigma F_1(p^2)
\: +\: \eta^{\mu \nu}\eta^{\rho \sigma}F_2(p^2) 
\: +\: \Big(\eta^{\mu \rho}\eta^{\nu \sigma} 
\: +\: \eta^{\nu \rho}\eta^{\mu \sigma}\Big)F_3(p^2)\nonumber\\ 
&&+\: \Big(\eta^{\mu \nu}p^\rho p^\sigma + \eta^{\rho \sigma}
p^\mu p^\nu \Big)F_4(p^2)\: +\: \Big(\eta^{\mu \rho}p^\nu p^\sigma +
\eta^{\nu \rho} p^\mu p^\sigma\: +\: \eta^{\mu \sigma} p^\nu p^\rho +
\eta^{\nu \sigma} p^\mu p^\rho\Big) F_5(p^2)\; ,\qquad
\end{eqnarray}
where $F_i$ (with  $i = 1,2,\dots, 5$) is a set  of form factors. Note
that the form factors $F_i$ are not independent of each other, as they
have          to          satisfy          the          transversality
condition~\eqref{eq:wardidentitycts},  which gives  rise  to following
set of relations:
\begin{eqnarray}
   \label{eq:FFWI}
 p^2 F_1\: +\:  F_4\: +\: 2 F_5\ &=&\ 0\;,\nonumber\\
 F_2\: +\: p^2 F_4\ &=&\ 0\;, \\
 F_3\: +\: p^2 F_5\ &=&\ 0\; .\nonumber
\end{eqnarray}
Finally,  it  is  important   to  remark  here  that  the  UV-infinite
contributions  of  $\Delta\Pi^{\mu\nu  ,\rho\sigma}(p)$  to  the  form
factors  $F_i$   satisfy  independently  the   three  relations  given
in~\eqref{eq:FFWI}.

\subsection{Massive Scalar Loops}

\begin{figure}
\centering
	\begin{subfigure}[a]{0.25\textwidth}
		\raisebox{0.1\height}{\begin{fmffile}{higgsselfenergy1}
		\begin{fmfgraph*}(100,50)
		\fmfleft{i}
		\fmfright{o}
		\fmf{zigzag,tension=1}{i,v1}
		\fmf{zigzag,tension=1}{v2,o}
		\fmf{dashes,left,tension=0.4,label=$H$}{v1,v2,v1}
		\end{fmfgraph*}
		\end{fmffile}
		}
		\caption{}
	\end{subfigure}%
		~
	\begin{subfigure}[d]{0.25\textwidth}
		\begin{fmffile}{higgsselfenergy2}
		\begin{fmfgraph*}(100,50)
		\fmfleftn{i}{5}
		\fmfrightn{o}{5}
		\fmf{phantom,tension=1}{i1,v1,o1}
		\fmf{phantom,tension=1}{i2,v2,o2}
		\fmf{phantom,tension=1}{i3,v3,o3}
		\fmf{phantom,tension=1}{i4,v4,o4}
		\fmf{phantom,tension=1}{i5,v5,o5}
		\fmffreeze
		\fmf{zigzag,tension=1}{i1,v2,o1}
		\fmf{dashes,tension=1,left}{v2,v5}
		\fmf{dashes,tension=1,left}{v5,v2}
		\fmflabel{$H$}{v5}
		\end{fmfgraph*}
		\end{fmffile}
		\caption{}
	\end{subfigure}
	
	\caption{The Higgs contribution to the graviton self-energy.
\label{fig:scalar} }
\end{figure}
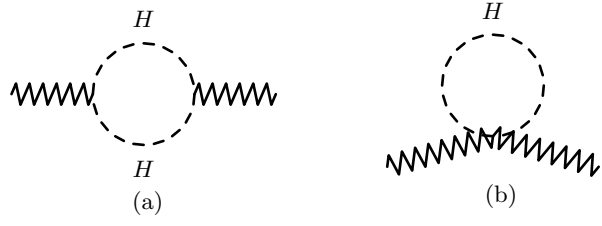

First,  we   consider  the   Higgs-scalar  effects  on   the  graviton
self-energy,  as   described  by  the  two   diagrams~(a)  and~(b)  in
Fig.~\ref{fig:scalar}. These are given by the loop integrals
\begin{eqnarray}
i \Pi^{\mu \nu, \rho \sigma}_{2(a)}(p)\ &=&\ \frac{1}{2}\int \frac{d^d
  k}{(2 \pi)^d} V^{\mu \nu}_{HHh}(k, -(p + 
k), m_H) V^{\rho \sigma}_{HHh}(-k, p + k, m_H) \sqpara{\frac{i}{k^2 -
    m_H^2}} \sqpara{\frac{i}{(p+k)^2 - m_H^2}}\ ,\\ 
i \Pi^{\mu \nu, \rho \sigma}_{2(b)}(p)\ &=&\  \frac{1}{2}\int \frac{d^d
  k}{(2 \pi)^d} V^{\mu \nu \rho \sigma}_{HHhh}(k, -k, m_H)
\sqpara{\frac{i}{k^2 - m_H^2}}\ .
\end{eqnarray}
Note  that the  contribution of  the would-be  Goldstone boson  $G$ is
obtained  by  replacing $m_H^2  \to  \xi_G  m_A^2$  in the  above  two
expressions.

Without   including   the    CTs   contained   in   $\Delta\Pi^{\mu\nu
  ,\rho\sigma}(p)$, the  Higgs contributions  to the form  factors are
given by
\begin{subequations}
  \label{eq:FFscalar}
\begin{align}
F_1(p^2)\ =&\ \frac{\kappa^2}{3600(4 \pi)^2 (p^2)^2}\Big[\Big( \alpha_1 +
  \alpha_4\Big) B_0(p^2,m_H^2,m_H^2) + \Big(\alpha_2 + \alpha_5\Big)
  A_0(m_H^2)+\Big(\alpha_3 + \alpha_6\Big)\Big]\; , \\ 
F_2(p^2)\ =&\
\frac{\kappa^2}{3600(4 \pi)^2} \Big(\alpha_1  B_0(p^2,m_H^2,m_H^2)+
\alpha_2 A_0(m_H^2)+ \alpha_3\Big)\;,\\ 
F_3(p^2)\ =&\
\frac{\kappa^2}{7200(4 \pi)^2} \Big(\alpha_4  B_0(p^2,m_H^2,m_H^2) +
\alpha_5 A_0(m_H^2)+ \alpha_6\Big)\; ,\\ 
F_4(p^2)\ =&\
-\frac{\kappa^2}{3600(4 \pi)^2 p^2} \Big(\alpha_1
B_0(p^2,m_H^2,m_H^2)+ \alpha_2  A_0(m_H^2)+ \alpha_3\Big)\; ,\\ 
F_5(p^2)\ =&\
-\frac{\kappa^2}{7200(4 \pi)^2 p^2} \Big(\alpha_4 B_0(p^2,m_H^2,m_H^2)
+ \alpha_5 A_0(m_H^2)+\alpha_6\Big)\; , 
\end{align}
\end{subequations}
where
\begin{subequations}
\begin{align}
\alpha_1 \; =&\ 15 \Big[8 m_H^4+16 m_H^2 p^2+3(p^2)^2\Big]\;,\\ 
\alpha_2 \; =&\ -30 \Big(4 m_H^2+3 p^2\Big)\;,\\ 
\alpha_3 \; =&\ 120 m_H^4+220 m_H^2 p^2-42(p^2)^2\;,\\ 
\alpha_4 \; =&\ 15 \Big(p^2-4 m_H^2\Big)^2 \;,\\ 
\alpha_5 \; =&\ -30 \Big(8 m_H^2+p^2\Big)\;,\\ 
\alpha_6 \; =&\ 16 \Big[15 m_H^4-10 m_H^2 p^2+(p^2)^2\Big]\; .
\end{align}
\end{subequations}
In addition, $B_0$ is the one-loop scalar self-energy integral defined
in $d= 4 - 2 \epsilon$ as
\begin{eqnarray}
B_0(p^2,m_1^2,m_2^2)\ &\equiv &\ (2\pi\mu)^{4-d}\,
\int \frac{d^d k}{i\pi^2}\:
\frac{1}{k^2 - m_1^2}\: \frac{1}{(k+p)^2 - m_2^2}\ =\ 
\frac{1}{\bar{\epsilon}}\ +\: 2\: -\: \ln\bigg(\frac{m_1m_2}{\mu^2}\bigg)
\nonumber\\ 
&&\ +\: \frac{1}{p^2}\, \bigg[\, (m^2_2 - m^2_1)\,
  \ln\bigg(\frac{m_1}{m_2}\bigg)\ +\
\lambda^{1/2}(p^2,m^2_1,m^2_2)\; \cosh^{-1}\bigg( \frac{m^2_1 +
  m^2_2 - p^2}{2 m_1 m_2}\bigg)\,\bigg]\; ,
\end{eqnarray}
with $\lambda  (x,y,z) \equiv (x-y-z)^2  - 4 yz$.   For $p^2 =  0$ and
$m_1 = m_2 = m$, the loop function $B_0 (0,m^2,m^2)$ is related to the
tadpole loop integral $A_0 (m^2)$ as follows:
\begin{equation}
A_0 (m^2)\ =\ m^2\, \Big(\, 1\: +\: B_0 (0, m^2, m^2)\,\Big)\; .
\end{equation}

We  may   now  calculate  the  CTs   described  by  $\Delta\Pi^{\mu\nu
  ,\rho\sigma}(p)$    in   the    $\overline{\rm   MS}$    scheme   of
renormalization~\cite{Bardeen:1978yd}.   For the  Higgs  and Goldstone
effects,  these  CTs  may  be  represented by  the  following  set  of
diffeomorphisms invariant operators:
\begin{eqnarray}
\Delta S_H\ &=&\ -\int d^4 x \frac{\sqrt{-g}}{2(4 \pi)^2 \bar{\epsilon} }
\sqpara{\frac{m_H^2}{6} R + \frac{1}{120}R^2+ \frac{1}{60} R^{\mu
    \nu}R_{\mu \nu}}\; ,\\
\Delta S_G\ &=&\ -\int d^4 x \frac{\sqrt{-g}}{2(4 \pi)^2 \bar{\epsilon} }
\sqpara{\frac{\xi_G m_A^2}{6} R + \frac{1}{120}R^2+ \frac{1}{60}
  R^{\mu \nu}R_{\mu \nu}}\; .
\end{eqnarray}
We note that the CTs  in $\Delta S_H$ agree with \cite{Capper:1973bk},
after    making    the    obvious   replacement:    $1/\epsilon    \to
1/\bar{\epsilon}$.  The inclusion of the CTs given by $\Delta S_H$ and
$\Delta S_G$  has the effect  to remove simply  the $1/\bar{\epsilon}$
poles that  occur through  the loop integrals  $A_0$ and $B_0$  in the
form   factors    $F_{1,2,\dots,5}$   listed   in~\eqref{eq:FFscalar}.
Finally, observe that the  five form factors $F_{1,2,\dots,5}$ satisfy
the transversality relations given in~\eqref{eq:FFWI}.

\subsection{Massive Fermion Loops}

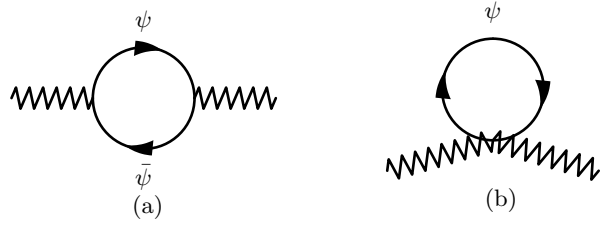
\begin{figure}
\centering
	\begin{subfigure}[a]{0.25\textwidth}
		\raisebox{0.1\height}{\begin{fmffile}{fermionselfenergy1}
		\begin{fmfgraph*}(100,50)
		\fmfleft{i}
		\fmfright{o}
		\fmf{zigzag,tension=1}{i,v1}
		\fmf{zigzag,tension=1}{v2,o}
		\fmf{fermion,left,tension=0.4,label=$\psi$}{v1,v2}
		\fmf{fermion,left,tension=0.4,label=$\bar{\psi}$}{v2,v1}
		\end{fmfgraph*}
		\end{fmffile}
		}
		\caption{}
	\end{subfigure}%
	~
\begin{subfigure}[d]{0.25\textwidth}
		\begin{fmffile}{fermionselfenergy2}
		\begin{fmfgraph*}(100,50)
		\fmfleftn{i}{5}
		\fmfrightn{o}{5}
		\fmf{phantom,tension=1}{i1,v1,o1}
		\fmf{phantom,tension=1}{i2,v2,o2}
		\fmf{phantom,tension=1}{i3,v3,o3}
		\fmf{phantom,tension=1}{i4,v4,o4}
		\fmf{phantom,tension=1}{i5,v5,o5}
		\fmffreeze
		\fmf{zigzag,tension=1}{i1,v2,o1}
		\fmf{fermion,tension=1,left}{v2,v5}
		\fmf{fermion,tension=1,left}{v5,v2}
		\fmflabel{$\psi$}{v5}
		\end{fmfgraph*}
		\end{fmffile}
		\caption{}
	\end{subfigure}
	
	\caption{The fermion contribution to the graviton self-energy.
\label{fig:fermion}}
\end{figure}

Quantum loops  due to  a Dirac fermion  $\psi$ contribute also  to the
graviton  self-energy  by  the   two  diagrams~(a)  and~(b)  shown  in
Fig.~\ref{fig:fermion}. These two diagrams may be calculated by
\begin{eqnarray}
i\Pi^{\mu \nu, \rho \sigma}_{3(a)}(p)\ &=&\ 
	 -\int \frac{d^d k}{(2 \pi)^d}\; {\rm Tr}\para{ V^{\mu
             \nu}_{\bar{\psi} \psi h}(k, -(p + k), m_\psi)
           \sqpara{\frac{i(\slashed{k} + m)}{k^2 - m_\psi^2}}  V^{\rho
             \sigma}_{\bar{\psi} \psi h}(-k, p + k, m_\psi)
           \sqpara{\frac{i(\slashed{p} + \slashed{k} + m)}{(p+k)^2 -
               m_\psi^2}}}\, ,\qquad\\
i\Pi^{\mu \nu, \rho \sigma}_{3(b)}(p)\ &=&\  - \int \frac{d^d k}{(2
  \pi)^d}\; {\rm Tr} \para{V^{\mu \nu \rho \sigma}_{\bar{\psi} \psi
    hh}(k,-k, m_H) \sqpara{\frac{i (\slashed{k} + m)}{k^2 -
      m_\psi^2}}}\, .
\end{eqnarray}

Upon including only the  cosmological constant CT $\delta \Lambda$, we
arrive at the following analytic expressions for the form factors:
\begin{subequations}
  \label{eq:FFfermion}
\begin{align}
F_1(p^2) \ =&\
\frac{\kappa^2}{1800(4 \pi)^2 (p^2)^2}\Big[\Big( \alpha_1 +
  \alpha_4\Big) B_0(p^2,m_H^2,m_H^2) + \Big(\alpha_2 + \alpha_5\Big)
  A_0(m_H^2)+\Big(\alpha_3 + \alpha_6\Big)\Big]\;, \\ 
F_2(p^2) \ =&\
\frac{\kappa^2}{1800(4 \pi)^2} \Big(\alpha_1
B_0(p^2,m_\psi^2,m_\psi^2)+ \alpha_2 A_0(m_\psi^2)+ \alpha_3\Big)\; ,\\
F_3(p^2) \ =&\
\frac{\kappa^2}{3600(4 \pi)^2} \Big(\alpha_4
B_0(p^2,m_\psi^2,m_\psi^2) + \alpha_5 A_0(m_\psi^2)+
\alpha_6\Big)\;,\\ 
F_4(p^2) \ =&\
-\frac{\kappa^2}{1800(4 \pi)^2 p^2} \Big(\alpha_1
B_0(p^2,m_\psi^2,m_\psi^2)+ \alpha_2  A_0(m_\psi^2)+
\alpha_3\Big)\;,\\ 
F_5(p^2) \ =&\
-\frac{\kappa^2}{3600(4 \pi)^2 p^2} \Big(\alpha_4
B_0(p^2,m_\psi^2,m_\psi^2) + \alpha_5 A_0(m_\psi^2)+\alpha_6\Big)\;, 
\end{align}
\end{subequations}
with
\begin{subequations}
\begin{align}
\alpha_1 \; =&\ - 15 \Big(p^2-4 m_\psi^2\Big)^2\, ,\\ 
\alpha_2 \ =&\ 30 \Big(8 m_\psi^2+ p^2\Big)\; ,\\ 
\alpha_3 \ =&\ -16\Big[15 m_\psi^4-10 m_\psi^2 p^2+(p^2)^2\Big]\;,\\ 
\alpha_4 \ =&\ -15\Big[32 m_\psi^4 + 4 m_\psi^2 p^2 - 3 (p^2)^2\Big]\;,\\ 
\alpha_5 \ =&\ -30 \Big(3 p^2 - 16 m_\psi^2\Big)\;,\\ 
\alpha_6 \ =&\ -480 m_\psi^4+20 m_\psi^2 p^2+18(p^2)^2\;.
\end{align}
\end{subequations}
The UV  poles proportional to  $1/\bar{\epsilon}$ that enter  the form
factors $F_{1,2,\dots,5}$  through the loop integrals  $A_0$ and $B_0$
may be  renormalized after taking into consideration  the CT effective
action
\begin{equation}
\Delta S_\psi\ =\ \int d^4 x \frac{\sqrt{-g}}{(4 \pi)^2 \bar{\epsilon} }
\sqpara{\frac{m_\psi^2}{6} R - \frac{1}{60}R^2 + \frac{1}{20} R^{\mu
    \nu}R_{\mu \nu}}\; .
\end{equation}
This last result is  in agreement with
\cite{DeMeyer:1974}.   As  with the  scalar case,  it is  not
difficult to  check that the form  factors $F_{1,2,\dots,5}$ exhibited
in~\eqref{eq:FFfermion}    satisfy   the    transversality   relations
in~\eqref{eq:FFWI}.

\subsection{Massive Gauge and Ghost Loops}

\begin{figure}
\centering
	\subcaptionbox{}[.24\textwidth]{
		\raisebox{0.1\height}{		
		\begin{fmffile}{gaugeselfenergy1}
		\begin{fmfgraph*}(100,50)
		\fmfleft{i}
		\fmfright{o}
		\fmf{zigzag,tension=1}{i,v1}
		\fmf{zigzag,tension=1}{v2,o}
		\fmf{photon,left,tension=0.4,label=$A_\mu$}{v1,v2}
		\fmf{photon,left,tension=0.4,label=$A_\nu$}{v2,v1}
		\end{fmfgraph*}
		\end{fmffile}
		}
}
\subcaptionbox{}[.24\textwidth]{
		\begin{fmffile}{gaugeselfenergy2}
		\begin{fmfgraph*}(100,50)
		\fmfleftn{i}{5}
		\fmfrightn{o}{5}
		\fmf{phantom,tension=1}{i1,v1,o1}
		\fmf{phantom,tension=1}{i2,v2,o2}
		\fmf{phantom,tension=1}{i3,v3,o3}
		\fmf{phantom,tension=1}{i4,v4,o4}
		\fmf{phantom,tension=1}{i5,v5,o5}
		\fmffreeze
		\fmf{zigzag,tension=1}{i1,v2,o1}
		\fmf{photon,tension=1,left,label=$A_\mu$}{v2,v5}
		\fmf{photon,tension=1,left,label=$A_\nu$}{v5,v2}
		\end{fmfgraph*}
		\end{fmffile}
}
\subcaptionbox{}[.24\textwidth]{
		\raisebox{0.1\height}{		
		\begin{fmffile}{ghostselfenergy1}
		\begin{fmfgraph*}(100,50)
		\fmfleft{i}
		\fmfright{o}
		\fmf{zigzag,tension=1}{i,v1}
		\fmf{zigzag,tension=1}{v2,o}
		\fmf{ghost,left,tension=0.4,label=$c$}{v1,v2}
		\fmf{ghost,left,tension=0.4,label=$\bar{c}$}{v2,v1}
		\end{fmfgraph*}
		\end{fmffile}
		}
}
\subcaptionbox{}[.24\textwidth]{
		\begin{fmffile}{ghostselfenergy2}
		\begin{fmfgraph*}(100,50)
		\fmfleftn{i}{5}
		\fmfrightn{o}{5}
		\fmf{phantom,tension=1}{i1,v1,o1}
		\fmf{phantom,tension=1}{i2,v2,o2}
		\fmf{phantom,tension=1}{i3,v3,o3}
		\fmf{phantom,tension=1}{i4,v4,o4}
		\fmf{phantom,tension=1}{i5,v5,o5}
		\fmffreeze
		\fmf{zigzag,tension=1}{i1,v2,o1}
		\fmf{ghost,tension=1,left}{v2,v5}
		\fmf{ghost,tension=1,left}{v5,v2}
		\fmflabel{$c$}{v5}
		\end{fmfgraph*}
		\end{fmffile}
}
\caption{Gauge- and ghost-field contributions to the
          graviton self-energy.\label{fig:gauge}} 
\end{figure}
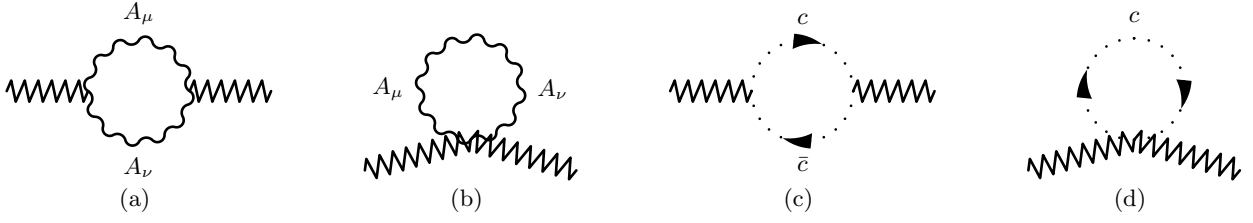

Finally, we  consider quantum  loop effects of  a massive  gauge boson
$A_\mu$ and  their respective  ghost fields $c$  and $\bar{c}$  on the
graviton  self-energy.   As  displayed in  Fig.~\ref{fig:gauge},  four
diagrams~(a), (b),  (c) and~(d)  contribute.  In the  Feynman-'t Hooft
gauge $\xi_G =  1$, these four diagrams may  respectively be evaluated
by the following integrals:
\begin{eqnarray}
i\Pi^{\mu \nu, \rho \sigma}_{4(a)}(p)\ &=&\ \frac{1}{2}\int \frac{d^d
  k}{(2 \pi)^d} V^{\mu \nu,\lambda, \delta}_{AAh}(k, -(p + k), m_A)
V^{\rho \sigma,\alpha,\beta}_{AAh}(-k, p + k, m_A) \sqpara{\frac{i
    \eta_{\alpha \lambda}}{k^2 - m_A^2}} \sqpara{\frac{i \eta_{\beta
      \gamma}}{(p+k)^2 - m_A^2}}\,,\qquad\\
i\Pi^{\mu \nu, \rho \sigma}_{4(b)}(p)\ &=&\  \frac{1}{2}\int \frac{d^d
  k}{(2 \pi)^d} V^{\mu \nu, \rho \sigma,\lambda,\delta}_{AAhh}(k, -k,
m_A^2) \sqpara{\frac{i \eta_{\lambda \delta}}{k^2 - m_A^2}}\,,\\
i \Pi^{\mu \nu, \rho \sigma}_{4(c)}(p)\ &=&\ -\int \frac{d^d k}{(2
  \pi)^d} V^{\mu \nu}_{\bar{c}c h}(k, -(p + k), m_A) V^{\rho
  \sigma}_{\bar{c}ch}(-k, p + k, m_A) \sqpara{\frac{i}{k^2 - m_A^2}}
\sqpara{\frac{i}{(p+k)^2 - m_A^2}}\,, \\
i \Pi^{\mu \nu, \rho \sigma}_{4(d)}(p)\ &=&\ -\int \frac{d^d k}{(2 \pi)^d}
V^{\mu \nu \rho \sigma}_{HHhh}(k, -k, m_A) \sqpara{\frac{i}{k^2 -
    m_A^2}}\,.
\end{eqnarray}

As done before,  we proceed by including only  the CT $\delta \Lambda$
of the cosmological constant. Then, the form factors are given~by
\begin{subequations}
  \label{eq:FFgauge}
\begin{align}
F_1(p^2)\ =&\ \frac{\kappa^2}{1800(4 \pi)^2 (p^2)^2}\;\Big[\Big(
  \alpha_1 + \alpha_4\Big) B_0(p^2,m_A^2,m_A^2) + \Big(\alpha_2 +
  \alpha_5\Big) A_0(m_H^2)+\Big(\alpha_3 + \alpha_6\Big)\Big]\,, \\ 
F_2(p^2)\ =&\
\frac{\kappa^2}{1800(4 \pi)^2}\; \Big(\alpha_1  B_0(p^2,m_A^2,m_A^2)+
\alpha_2 A_0(m_A^2)+ \alpha_3\Big)\,,\\ 
F_3(p^2)\ =&\
\frac{\kappa^2}{3600(4 \pi)^2}\; \Big(\alpha_4  B_0(p^2,m_A^2,m_A^2) +
\alpha_5 A_0(m_A^2)+ \alpha_6\Big)\;,\\ 
F_4(p^2)\ =&\
-\frac{\kappa^2}{1800(4 \pi)^2 p^2}\; \Big(\alpha_1  B_0(p^2,m_A^2,m_A^2)+
\alpha_2  A_0(m_A^2)+ \alpha_3\Big)\;,\\ 
F_5(p^2)\ =&\
-\frac{\kappa^2}{3600(4 \pi)^2 p^2}\; \Big(\alpha_4 B_0(p^2,m_A^2,m_A^2)
+ \alpha_5 A_0(m_A^2)+\alpha_6\Big)\; ,
\end{align}
\end{subequations}
with
\begin{subequations}
\begin{align}
\alpha_1 \; =&\ 30 \left(4 m_A^4-12 m_A^2 p^2-\left(p^2\right)^2\right)\;,\\ 
\alpha_2 \; =&\ 60 \Big(p^2 - 2m_A^2\Big)\;,\\ 
\alpha_3 \; =&\ 120 m^4-530 m^2 p^2+13 \big(p^2\big)^2\;,\\ 
\alpha_4 \; =&\ 30\Big[8 m_A^4 + 16 m_A^2 p^2 + 3\big(p^2\big)^2\Big]\;,\\ 
\alpha_5 \; =&\ -60 \Big( 4 m_A^2 + 3 p^2\Big)\;,\\ 
\alpha_6 \; =&\ 240 m_A^4+590 m_A^2 p^2-99\big(p^2\big)^2\;.
\end{align}
\end{subequations}
The form factors become UV  finite, after considering the CT effective
action
\begin{equation}
  \label{eq:DSgauge}
\Delta S_A\ =\ -\int d^4 x \frac{\sqrt{-g}}{(4 \pi)^2 \bar{\epsilon} }
\sqpara{\frac{m_A^2}{3} R - \frac{1}{30}R^2 + \frac{1}{10} R^{\mu
    \nu}R_{\mu \nu}}\; .
\end{equation}
Our result  in~\eqref{eq:DSgauge} agrees with  \cite{Capper:1974ed} in
the limit $m_A \to 0$. As with the scalar and fermion cases, gauge and
ghost  field  contributions  to  the  form  factors  $F_{1,2,\dots,5}$
satisfy the transversality relations stated in~\eqref{eq:FFWI}.

\subsection{Summary of Results}

Even  though our  calculations  pertain to  the  gauged Abelian  Higgs
model, the results we presented here for the graviton self-energy have
a general applicability. At the  one-loop order, only the kinetic part
of the matter Lagrangian contributes, whereas the part associated with
matter interactions  only enters at  two loops. Thus, at  the one-loop
level, we only need to know  the matter field content of the theory in
terms  of  scalar, fermionic  and  gauge  degrees  of freedom.   As  a
consequence, the total  {\em renormalized} graviton self-energy tensor
in a given theory may be summarized as follows:
\begin{equation}
  \label{eq:generalselfenergy}
\Pi^{\mu \nu, \rho \sigma}_{\rm R}(p)\ =\ \sum_{i = 1}^{N_0} \Pi^{\mu \nu, \rho
  \sigma}_0(p, m_{0, i})\: +\: \sum_{i = 1}^{N_\frac{1}{2}} \Pi^{\mu \nu,
  \rho \sigma}_{\frac{1}{2}}(p, m_{\frac{1}{2}, i})\: +\: \sum_{i =
  1}^{N_1} \Pi^{\mu \nu, \rho \sigma}_{1}(p, m_{1, i})\; ,
\end{equation}
where   $\Pi^{\mu  \nu,  \rho   \sigma}_0(p,m),  \Pi^{\mu   \nu,  \rho
  \sigma}_{\frac{1}{2}}(p,m)$, and $\Pi^{\mu \nu, \rho \sigma}_1(p,m)$
denote the  scalar, fermion, and gauge-  and ghost-field contributions
to   the    graviton   self-energy   with   a    generic   mass   $m$,
respectively. Correspondingly, $N_0$, $N_\frac{1}{2}$ and $N_1$ denote
the  number of  scalars, fermions,  and  gauge bosons.   As the  total
self-energy is  the sum of individual contributions  which satisfy the
transversality  condition~\eqref{eq:wardidentitycts},  it  is  evident
that the graviton  remains massless at the one-loop  level. Beyond the
one   loop,  the   validity  of   the  GLET~\eqref{eq:GLET}   and  the
diffeomorphisms    WI~\eqref{eq:secondwardidentitymomentum}   play   a
central role to preserve the  property of masslessness of the graviton
field to all orders in perturbation theory.

\section{Matter Quantum Corrections to the Newtonian Potential \label{NP}}

Having computed the matter effects on the graviton self-energy, we can
now proceed to study the one-loop quantum corrections to the Newtonian
potential  $V(r)$. As  we will  see in  this section,  the radiatively
corrected  Newtonian  potential can  be  derived  from the  $S$-matrix
element  describing the  elastic scattering  of two  massive particles
$\varphi_1$ and  $\varphi_2$ in  the non-relativistic limit.   We will
use  these results to  determine the  long and  short range  limits of
$V(r)$ and comment  on their relevance. In the  massless limit of loop
particles,  the known results  for $V(r)$  stated in  the introduction
are reproduced.

For definiteness, we consider  the scattering process $\varphi_1 (p_1)
\varphi_2   (p_2)  \to   \varphi_1  (k_1)   \varphi_2   (k_2)$,  where
$\varphi_1$ and  $\varphi_2$ are two  different gauge-singlet scalars.
The action describing the  interaction of $\varphi_{1,2}$ with gravity
is given by
\begin{equation}
S_{\varphi}\ =\ \frac{1}{2}\int d^4 x \sqrt{-g}\, \Big( g^{\mu \nu}
\partial_\mu \varphi_1 \partial_\nu \varphi_1\: -\: m_1^2 \varphi_1^2\ +
\ g^{\mu \nu} \partial_\mu \varphi_2 \partial_\nu \varphi_2\: -\: 
m_2^2 \varphi_2^2\,\Big)\; , 
\end{equation}
where $m_1$  and $m_2$  are the masses  of the fields  $\varphi_1$ and
$\varphi_2$,  respectively. We  shall use  this action  to  derive the
Feynman rules  for the interactions of $\varphi_1$  and $\varphi_2$ to
the graviton  field $h_{\mu \nu}$. The analytical  expressions for the
relevant vertices are given in Appendix \ref{FR}.

\subsection{Tree-Level Newtonian Potential}

\begin{figure}
\centering
		\raisebox{0.1\height}{
		\begin{fmffile}{treediagram}
		\begin{fmfgraph*}(150,100)
		\fmfleftn{i}{7}
		\fmfrightn{o}{7}
		\fmf{phantom,tension=1}{i1,v1,o1}
		\fmf{phantom,tension=1}{i2,v2,o2}
		\fmf{phantom,tension=1}{i3,v3,o3}
		\fmf{phantom,tension=1}{i4,v4,o4}
		\fmf{phantom,tension=1}{i5,v5,o5}
		\fmf{phantom,tension=1}{i6,v6,o6}
		\fmf{phantom,tension=1}{i7,v7,o7}
		\fmffreeze
\fmf{dashes,tension=1,label=$\varphi_1(p_1)$,l.side=right}{i1,v2}  
\fmf{dashes,tension=1,label=$\varphi_1(k_1)$,l.side=right}{v2,o1}
\fmf{dashes,tension=1,label=$\varphi_2(p_2)$,l.side=left}{i7,v6}
\fmf{dashes,tension=1,label=$\varphi_2(k_2)$,l.side=left}{v6,o7}
		\fmf{zigzag,label=$h_{\mu \nu}(q)$}{v2,v6}
		\end{fmfgraph*}
		\end{fmffile}
		}
	\caption{The tree level scattering diagram.}
	\label{fig:treediagram}
\end{figure}
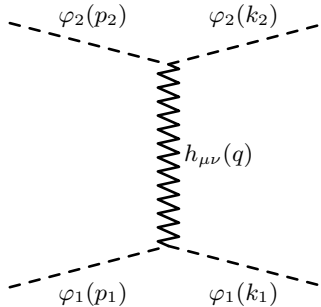

It  is  instructive to  briefly  review  how  the classical  Newtonian
potential can be inferred from  the tree-level $S$-matrix element of a
given  process $\varphi_1  (p_1) \varphi_2  (p_2) \to  \varphi_1 (k_1)
\varphi_2 (k_2)$. As illustrated in Fig.~\ref{fig:treediagram}, such a
process  proceeds at  the  tree level  via  the exchange  of a  single
graviton   in   the  $t$-channel.    The   momentum  space   amplitude
$\mathcal{M}_{\rm tree}$ is given by
\begin{equation}
   \label{eq:treeamplitude}
i\mathcal{M}_{\rm tree} = i V^{\mu \nu}_{\varphi_1 \varphi_1 h}(p_1, -
k_1) \Delta_{0,\mu \nu, \rho \sigma}(q) V^{\rho \sigma}_{\varphi_2
  \varphi_2 h}(p_2, - k_2)  
\end{equation}
where $q_\mu = (p_1 - k_1)_\mu = (p_2 - k_2)_\mu$ is the four-momentum
of the  graviton, $V^{\mu  \nu}_{\varphi_1 \varphi_1 h}$  and $V^{\rho
  \sigma}_{\varphi_2 \varphi_2 h}$ are the tree-level vertex functions
for the $\varphi_1 \varphi_1  h$ and $\varphi_2 \varphi_2 h$ vertices,
respectively,  and $\Delta_{0,\mu  \nu, \rho  \sigma}(q)$ is  the 
tree-level graviton propagator.

Let us briefly discuss the gauge independence of the tree-level graph.
The graviton propagator depends on the gauge fixing parameters $\xi_D$
and  $\sigma$. However,  as  a  consequence of  the  Ward identity  of
diffeomorphisms,   the   vertex   functions  $V^{\mu   \nu}_{\varphi_1
  \varphi_1   h}(p_1,-k_1)$  and  $V^{\mu   \nu}_{\varphi_2  \varphi_2
  h}(p_2,-k_2)$ satisfy:
\begin{equation}
  \label{eq:scalarwardidentity} 
(p_1 - k_1)_\mu V^{\mu \nu}_{\varphi_1 \varphi_1 h}(p_1,-k_1)\ =\ 0\;,\qquad
(p_2 - k_2)_\mu V^{\mu \nu}_{\varphi_2 \varphi_2 h}(p_2,-k_2)\ =\ 0\; , 
\end{equation}
when the scalar fields are taken  to be on-shell. As all dependence on
the gauge-fixing  parameters $\xi_D$ and $\sigma$ is  carried by terms
proportional   to   the   longitudinal  four-momentum   $q_\mu$   (see
Appendix~\ref{FR} for an expression  for the graviton propagator), all
$\xi_D$-     and     $\sigma$-dependent     terms    vanish     thanks
to~\eqref{eq:scalarwardidentity},  thus   yielding  a  gauge-invariant
result under the group of diffeomorphisms. In fact, this is equivalent
to replacing the graviton propagator  with the propagator in the gauge
$\xi_D = \frac{1}{2}, \sigma = 1$  (known as the harmonic or de Donder
gauge):
\begin{equation}
  \label{eq:propagatorreplacement}
\Delta_{0,\mu \nu, \rho \sigma}(q)\ \to\ 
\frac{P_{\mu \nu \rho \sigma}}{q^2 + i \epsilon}\ . 
\end{equation}

The non-relativistic  limit of the  amplitude $\mathcal{M}_{\rm tree}$
is obtained by  expanding in the three-momenta of  the external fields
and considering only  terms that diverge in the  IR limit of vanishing
3-momenta.   These terms  have  been called  {\it non-analytic}  terms
in~\cite{Bjerrum-Bohr2002a}.    Expanding    the   tree-level   vertex
functions,  $V_{\varphi_1 \varphi_1  h}$  and $V_{\varphi_2  \varphi_2
  h}$, about  the three-momenta of  the external particles,  we obtain
the leading terms of the expansion:
\begin{equation}
  \label{eq:nrvertices} 
V^{\mu \nu}_{\varphi_1 \varphi_1 h}\ =\ i \kappa m_1^2\delta^\mu_0
\delta^\nu_0\;;\qquad
V^{\mu \nu}_{\varphi_2 \varphi_2 h}\ =\ i \kappa m_2^2\delta^\mu_0\delta^\nu_0\; .
\end{equation}
Employing        the         elementary        identities        given
in~\eqref{eq:scalarwardidentity},     the     tree-level     amplitude
$\mathcal{M}_{\rm  tree}$  in~\eqref{eq:treeamplitude}  takes  on  the
simple form in the non-relativistic limit:
\begin{equation}
\mathcal{M}_{\rm tree} = -\frac{\kappa^2 m_1^2 m_2^2}{|\vec{q}|^2}
\end{equation}
where $|\vec{q}| = | \vec{p}_1  - \vec{k}_1|$ is the 3-momentum of the
exchange graviton.

To derive the Newtonian potential from the scattering amplitude ${\cal
  M}_{\rm tree}$, we use the relation~\cite{Bjerrum-Bohr2002a}:
\begin{equation}
  \label{eq:potentialdefinition}
V(\vec{r})\ =\ \frac{1}{2 m_1} \frac{1}{2 m_2} \int \frac{d^3 q}{(2
  \pi)^3} e^{i \vec{q} \cdot \vec{r}} \mathcal{M}_{\rm
  tree}(\vec{q})\; . 
\end{equation}
Note  that   the  factors  $1/2m_1$  and  $1/2m_2$   result  from  the
normalization  of   single  particle  states.   Using  the  definition
$\kappa^2 = 16 \pi G$ and the well-known result for the integral
\begin{equation}
\int \frac{d^3 q}{(2 \pi)^3} e^{i \vec{q} \cdot
  \vec{r}}\frac{1}{|\vec{q}|^2}\ =\ \frac{1}{4 \pi r}\; , 
\end{equation}
we obtain the scattering potential
\begin{equation}
V(r)\ =\ - \frac{G m_1 m_2}{r}\ ,
\end{equation}
which is  the classical Newtonian  potential.  Notice that  $V(r)$ has
been obtained by pure quantum field-theoretic means, and is manifestly
gauge invariant  and process  independent, i.e.~the same  result would
have been  obtained, if we  had considered fermions or  vector bosons,
instead of scalars, as external particles.

\subsection{Matter Quantum Corrections}

We shall  now compute the  one-loop matter quantum corrections  to the
scattering  process  $\varphi_1  \varphi_2 \to  \varphi_1  \varphi_2$,
shown  in  Fig.~\ref{fig:treediagram}.   Given  that  $\varphi_1$  and
$\varphi_2$ are gauge singlets, only self-energy effects contribute to
this process, as  illustrated in Fig.~\ref{fig:selfenergydiagram}.  If
these scalar fields were charged  under a $U(1)$ gauge group, one must
also include  vertex and  box contributions.  The  case of  an elastic
scattering       with      charged      scalars       was      studied
in~\cite{Bjerrum-Bohr2002},   whilst  the   scattering   process  with
external    charged    fermions     under    $U(1)$    was    analyzed
in~\cite{Butt2006}.   However, we  note  that quantum  effects on  the
Newtonian  potential do  not  depend  on the  specific  nature of  the
external  scattered particles,  i.e.~the quantum  effects  are process
independent.

We should  remark here that the  use of an  one-loop resummed graviton
propagator proves necessary.  A conventional perturbative expansion in
terms of graviton self-energies  produces corrections to the potential
which are  linear in  the separation i.e.~$\propto  r$, when  the loop
mass  is  {\em  non}-zero.   This  contribution  diverges  as  $r  \to
\infty$. We shall show that  only a calculation of the potential based
on the {\em resummed} graviton propagator gives the correct asymptotic
behaviour.

\subsubsection{The amplitude and its non-relativistic limit}

\begin{figure}
	\centering
		\raisebox{0.1\height}{		
		\begin{fmffile}{selfenergydiagram}
		\begin{fmfgraph*}(150,100)
		\fmfleftn{i}{7}
		\fmfrightn{o}{7}
		\fmf{phantom,tension=1}{i1,v1,o1}
		\fmf{phantom,tension=1}{i2,v2,o2}
		\fmf{phantom,tension=1}{i3,v3,o3}
		\fmf{phantom,tension=1}{i4,v4,o4}
		\fmf{phantom,tension=1}{i5,v5,o5}
		\fmf{phantom,tension=1}{i6,v6,o6}
		\fmf{phantom,tension=1}{i7,v7,o7}
		\fmffreeze
\fmf{dashes,tension=1,label=$\varphi_1(p_1)$,l.side=right}{i1,v2}
\fmf{dashes,tension=1,label=$\varphi_1(k_1)$,l.side=right}{v2,o1}
\fmf{dashes,tension=1,label=$\varphi_2(p_2)$,l.side=left}{i7,v6}
\fmf{dashes,tension=1,label=$\varphi_2(k_2)$,l.side=left}{v6,o7}
		\fmf{zigzag}{v2,v4}
		\fmfblob{.14w}{v4}
		\fmf{zigzag}{v4,v6}
		\end{fmfgraph*}
		\end{fmffile}
		}
	\caption{The class of diagrams corresponding to matter effects.}
	\label{fig:selfenergydiagram}
\end{figure}
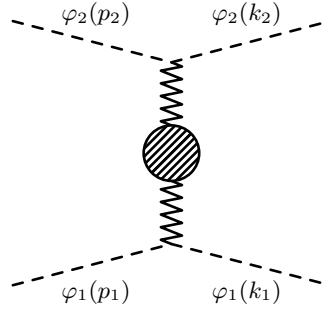

The       one-loop      transition       amplitude       shown      in
Fig.~\ref{fig:selfenergydiagram} is given by
\begin{equation}
   \label{eq:M1loop}
i \mathcal{M}_{\rm 1-loop} = i V^{\mu \nu}_{\varphi_1 \varphi_1
  h}(p_1, - k_1) \Delta_{\mu \nu, \rho \sigma}(q) V^{\rho
  \sigma}_{\varphi_2 \varphi_2 h}(p_2, - k_2). 
\end{equation}
where  $\Delta_{\mu \nu,  \rho  \sigma}(q)$ is  the resummed  graviton
propagator.  To  achieve this resummation  at one-loop order,  we must
resum the  Dyson series of  the one-loop graviton  self-energy graphs.
Specifically, the resummed  graviton propagator $\Delta_{\mu \nu, \rho
  \sigma}(q) $ is defined by the equation
\begin{equation} 
  \label{eq:dressedpropagator}
\Big(\Delta_0^{-1 \,\mu \nu, \alpha \beta}(q)\: +\: \Pi^{\mu \nu,
  \alpha \beta}_{\rm R}(q)\Big)\: 
\Delta_{\alpha \beta, \rho \sigma}(q)\ =\ 
\frac{1}{2} \,\Big( \delta_\mu^\rho \delta_\nu^\sigma +
\delta_\mu^\sigma \delta_\nu^\rho \Big)\; .
\end{equation}
Here,  $\Delta_0^{-1  \mu  \nu,  \rho \sigma}(q)$  is  the  tree-level
inverse propagator  and $\Pi_{\rm R}^{\mu  \nu \rho \sigma}(q)$  is the
renormalized graviton self-energy which has been calculated explicitly
in Section~\ref{MC}.

In order to invert the relation~\eqref{eq:dressedpropagator}, we first
write the resummed  graviton propagator in terms of  its possible form
factors:
\begin{align}
\Delta_{\mu \nu, \rho \sigma}(q)\ &=\  q_\mu q_\nu q_\rho q_\sigma
\Delta_1(q^2) + \eta_{\mu \nu}\eta_{\rho \sigma}\Delta_2(q^2) +
\Big(\eta_{\mu \rho}\eta_{\nu \sigma} + \eta_{\nu \rho}\eta_{\mu
  \sigma}\Big)\Delta_3(q^2)\nonumber\\ 
\ &\quad + \Big(\eta_{\mu \nu}q_\rho q_\sigma + \eta_{\rho \sigma} q_\mu
q_\nu \Big)\Delta_4(p^2) + \Big(\eta_{\mu \rho}q_\nu q_\sigma +
\eta_{\nu \rho} q_\mu q_\sigma + \eta_{\mu \sigma} q_\nu q_\rho +
\eta_{\nu \sigma} q_\mu q_\rho \Big)\Delta_5(q^2)\;. 
\end{align}
Employing the method of orthogonal projectors, we find
\begin{align}
\Delta_1(q^2)\ &=\ \frac{4}{3 (q^2)^2\big(q^2 - 4
  F_3(q^2)\big)}-\frac{4}{3(q^2)^2\big(q^2 + 3 F_2(q^2) + 2
  F_3(q^2)\big)}\ , \\
\Delta_2(q^2)\ &=\ -\frac{2}{3\big(q^2 - 4
  F_3(q^2)\big)}-\frac{1}{3\big(q^2 + 3 F_2(q^2) + 2 F_3(q^2)\big)}\ , \\
\Delta_3(q^2)\ &=\ \frac{1}{\big(q^2 - 4 F_3(q^2)\big)}\ , \\
\Delta_4(q^2)\ &=\ \frac{2}{3q^2\big(q^2 - 4
  F_3(q^2)\big)}-\frac{2}{3q^2\big(q^2 + 3 F_2(q^2) + 2
  F_3(q^2)\big)}\ , \\
\Delta_5(q^2)\ &=\ \frac{1}{(q^2)^2}-\frac{1}{q^2\big(q^2 - 4
  F_3(q^2)\big)}\ ,
\end{align}
where $F_2, F_3$ are the graviton self-energy form factors defined in
\eqref{eq:gravitonselfenergyform}.

Let us now discuss the gauge dependence of this amplitude. Writing out
the full Dyson series for the resummed propagator, we obtain 
\begin{equation}
\Delta_{\mu \nu, \rho \sigma}\ =\ \Delta_{0 \,\mu \nu, \rho \sigma}\: -\:
\Delta_{0\,\mu \nu, \alpha \beta} \Pi_{\rm R}^{\alpha \beta, \gamma
  \delta}\Delta_{0\,\gamma \delta, \rho \sigma}\: +\: \Delta_{0\,\mu \nu,
  \alpha \beta} \Pi_{\rm R}^{\alpha \beta, \gamma
  \delta}\Delta_{0\,\gamma \delta, \lambda \kappa} \Pi_{\rm
  R}^{\lambda \kappa, \epsilon \zeta}\Delta_{0\,\epsilon \zeta, \rho
  \sigma}\ +\ \cdots 
\end{equation}
Given that  the tree level  propagators must contract with  either the
tree level  vertex functions $V^{\mu \nu}_{\varphi_1  \varphi_1 h}$ or
$V^{\mu \nu}_{\varphi_2 \varphi_2 h}$, where the scalars are on-shell,
or the  renormalized graviton self-energy $\Pi_{\rm  R}^{\mu \nu, \rho
  \sigma}$,  any term in  the propagator  which explicitly  depends on
components of the longitudinal four-momenta $q_\mu$ will vanish due to
the                            identities~\eqref{eq:scalarwardidentity}
and~\eqref{eq:wardidentitycts}.     As~a~consequence,   the   one-loop
transition amplitude $\mathcal{M}_{\rm 1-loop}$ becomes independent of
the   gauge-fixing   parameters    $\xi_D$   and   $\sigma$   of   the
diffeomorphisms.  Like  the tree-level case,  we can use  the harmonic
gauge          for           the          graviton          propagator
[cf.~(\ref{eq:propagatorreplacement})] to simplify the calculation.

In the non-relativistic limit, the one-loop amplitude becomes
\begin{equation}
\mathcal{M}_{\rm 1-loop}\ =\ - \kappa^2 m_1^2 m_2^2\sqpara{\frac{4}{3}
  \para{\frac{1}{|\vec{q}|^2 + 4 F_3(-|\vec{q}|^2)}}\: +\:
  \frac{1}{3}\para{\frac{1}{3  F_2(-|\vec{q}|^2) + 2 F_3(-|\vec{q}|^2)
      - |\vec{q}|^2}}}\, .  
\label{eq:1loopamp}
\end{equation}
This  amplitude diverges  as $|\vec{q}|  \to 0$,  since both  the form
factors   $F_2$    and   $F_3$   vanish   in    this   limit,   thanks
to~\eqref{eq:FFWI}.   This  singularity  of the  transition  amplitude
$\mathcal{M}_{\rm  1-loop}$   as  $|\vec{q}|   \to  0$  is   a  simple
manifestation of  the masslessness of the graviton  field.  When going
to the  non-relativistic limit, the  presence of a particle  with mass
$m$ in the loop requires  special care, as $m$ is another dimensionful
parameter entering  the calculation of  the amplitude.  In  this case,
one needs to proceed carefully  and compare the size of $|\vec{q}|$ to
$m$, rather  than simply  taking the IR  limit $|\vec{q}| \ll  1$.  In
fact, one has  to distinguish between three possible  cases for a loop
particle  with mass  $m$: $|\vec{q}|  \gg m$,  $|\vec{q}| \sim  m$ and
$|\vec{q}| \ll m$.  In the  calculation that follows, we first compute
the   potential  in   the   general  case,   before  translating   the
aforementioned three limits into position space.

\subsubsection{Computation of the scattering potential}

Our aim  is now to compute  the Newtonian potential  from the one-loop
transition  amplitude.   As  before,   we  may  define  the  Newtonian
potential  in close  analogy  to~\eqref{eq:potentialdefinition}, which
may  be represented  by the  one-dimensional integral  of  the Fourier
transform:
\begin{equation}
  \label{eq:1DFourier}
V(r)\ =\ -\frac{i}{(2 \pi)^2} \int_{-\infty}^\infty dq\; \bigg(\frac{q}{r}
e^{i q r} \widehat{\mathcal{M}}_{\rm 1-loop}(q)\bigg)\;, 
\end{equation}
where  $q\equiv  |\vec{q}|$  and  $\widehat{\mathcal{M}}_{\rm  1-loop}
\equiv  \frac{1}{2 m_1}\frac{1}{2  m_2}\mathcal{M}_{\rm  1-loop}$. The
above   expression~\eqref{eq:1DFourier}    includes   the   tree-level
contribution to the potential, as  well as the one-loop matter quantum
corrections, through the resummed graviton propagator.

In order to perform the integration, we analytically continue $q$ to a
complex variable  and integrate over  a closed contour in  the complex
plane  which includes  the integral  of interest~\eqref{eq:1DFourier}.
Given that the  value of the closed contour  integral depends upon the
residue of the  poles within the contour, we  begin by identifying the
poles of the integrand. Explicitly,  we find that there are three real
poles for the  resummed graviton propagator: the standard  one at $q =
0$ and  two others that  occur in  the Planck mass  range at $q  = \pm
q_0$,  where  $q_0 \sim  M_{\rm  P}$.  The  latter poles  signify  the
breakdown of  perturbative quantum gravity and therefore  we call them
Landau poles.

An analytic expression for the  square $q_0^2$ of the Landau poles may
be  determined by  searching for  non-zero roots  of  the denominators
in~\eqref{eq:1loopamp}. Assuming the loop masses are small compared to
$1/\kappa^2  = M_{\rm  P}^2$,  we may  expand  the root  in powers  of
$\kappa^2$ and its inverse.  It can then be shown that the Landau pole
diverges as $\kappa^2 \to 0$ and that the pole is a simple pole. Thus,
the  leading  term  in  the  expansion is  the  term  proportional  to
$1/\kappa^2$. Hence, we obtain the approximate analytic expression for
$q_0^2$:
\begin{equation}
   \label{eq:landaupoles}
q_0^2\ =\ \frac{1920 \pi^2}{\kappa^2 \beta} \sqpara{W\para{\frac{1920
      \pi^2 \exp \para{-\gamma/\beta}}{\kappa^2 \mu^2 \beta}}}^{-1},
\end{equation}
with  $\beta  = N_0  +  3  N_{\frac{1}{2}} +  14  N_1$  and $\gamma  =
\frac{2}{15}(23 N_0 +  59 N_{\frac{1}{2}} + 142 N_1)$,  where $N_s$ is
the number  of fields  of spin $s  = 0$~(scalar),  $\frac{1}{2}$~(Weyl
fermion),  1~(vector boson)  and  $W(z)$ is  the Lambert  $W$-function
defined by the inverse relation: $z = W(z)e^{W(z)}$.

\begin{figure}[t]
\centering
\includegraphics[scale=1]{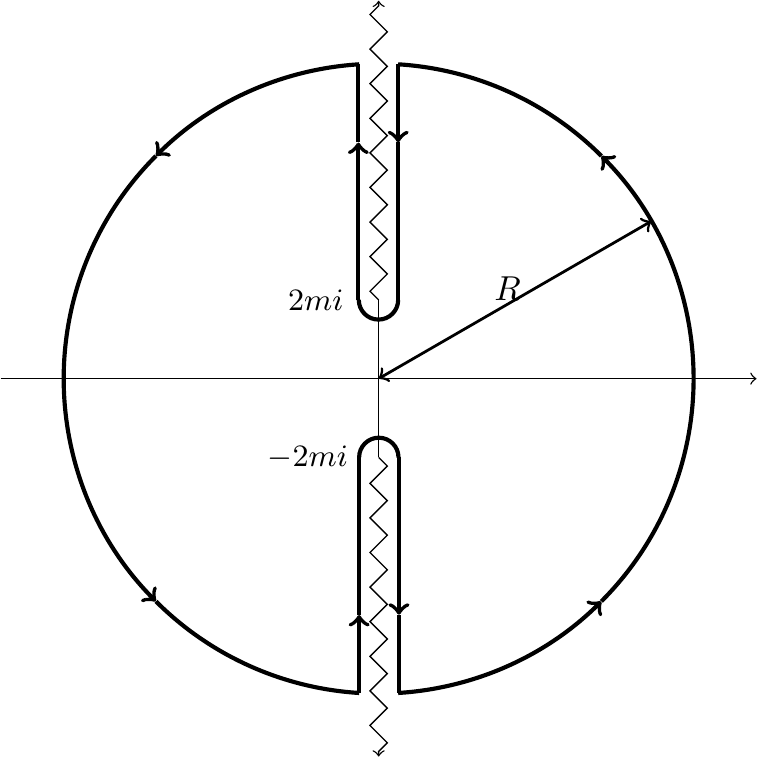}
	\caption{The contour used for the complex integral
          in~\eqref{eq:NP} to compute the number of poles and roots
          for a single loop mass $m$. This contour covers the whole
          complex plane as $R \to \infty$, whilst excluding the two
          branch cuts indicated by the zigzag lines.
	\label{fig:argumentprinciplecontour}}
\end{figure}

One may wonder  whether there are other complex  poles, in addition to
the three real poles mentioned above. To address this question, we use
the argument  principle, which states that, for  some complex function
$f(z)$, it holds
\begin{equation}
  \label{eq:NP}
\frac{1}{2 \pi i}\oint_\gamma \frac{f'(z)}{f(z)} dz\ =\ N\: -\: P\;,
\end{equation}
where $N$ is the number of roots of $f(z)$, $P$ is the number of poles
of $f(z)$, and $\gamma$ is a closed contour which contains the entire
complex plane whilst excluding the branch cuts of the function.  The
integrand $q e^{-qr} \widehat{\mathcal{M}}_{\rm 1-loop}(q)$ may be
split into two parts $f_1(q)$ and $f_2(q)$:
\begin{equation}
f_1(q)\ =\ -\: \frac{\kappa^2 m_1 m_2}{3}\para{\frac{q e^{-qr}}{-q^2 + 4
    F_3(q^2)}}, \qquad f_2(q)\ =\ -\: \frac{\kappa^2 m_1
  m_2}{12}\para{\frac{q e^{-qr}}{ 3  F_2(q^2) + 2 F_3(q^2) + q^2}}\;, 
\end{equation}
as there  are two terms  in~\eqref{eq:1loopamp}. We now  observe that,
for every matter field in the loop with mass $m$, there are two branch
cuts in the  complex plane for $f_1(q)$ and  $f_2(q)$. The first branch
cut is along the positive imaginary interval $[2mi, +i\infty)$, whilst
  the  second one  is along  the negative  imaginary  interval $[-2mi,
    -i\infty )$.   Taking these two  branch cuts into account,  we may
    determine  $N-P$  for  both  functions  independently,  using  the
    contour              $\gamma$              depicted             in
    Fig.~\ref{fig:argumentprinciplecontour}.  In  {\it both} cases, we
    obtain
\begin{equation}
N - P = - 3 \; .
\end{equation}
Since the form factors $F_2(q^2),  F_3(q^2)$ do not diverge for finite
values of $q$, $f_1(q)$ and $f_2(q)$ have no roots. This gives $P = 3$
for  both  functions.   Substituting  the expression  for  the  Landau
poles~\eqref{eq:landaupoles} into the denominator of each function, we
obtain zero in  both cases when the loop masses  are small compared to
$M_P$. Therefore, both functions diverge  at the same points: the real
pole at $q = 0$ and  the two Landau poles $q^2 = q_0^2$. Consequently,
the  resummed graviton  propagator and  so $\widehat{\mathcal{M}}_{\rm
  1-loop}(q)$ has no other complex poles that we need to worry about.

\begin{figure}[t]
\centering
\includegraphics[scale=1]{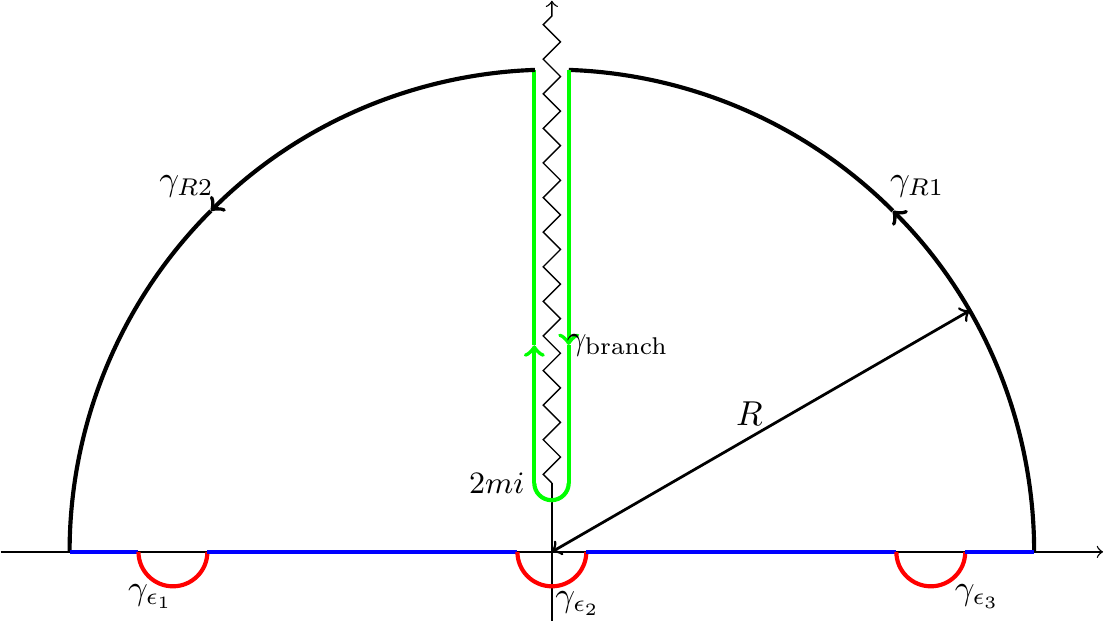}
	\caption{The  contour used  to compute  the  Fourier transform
          in~\eqref{eq:1DFourier}.  For  a generic non-zero  loop mass
          $m$, there is a branch  cut that starts at $2mi$ and extends
          to $i\infty$ as illustrated by the zigzag line.}
	\label{fig:ftcontour}
\end{figure}

Knowing the location  the three real poles, we  may construct a closed
contour to  compute the Fourier  transform~\eqref{eq:1DFourier}, which
is illustrated in Fig.~\ref{fig:ftcontour}.  By means of this contour,
we may evaluate the potential as follows:
\begin{equation}
V(r)\ =\ V_{\rm res}(r)\: +\: V_{\rm branch}(r)\; ,
\end{equation}
where
\begin{eqnarray}
  \label{eq:vres}
V_{\rm res}(r)\ & =&\ \frac{1}{2 \pi r} \sum_{n} {\rm Res}(q e^{i q r}
\widehat{\mathcal{M}}_{\rm 1-loop}(q), q_n)\: -\: \frac{i}{(2 \pi)^2 r}
\sum_{i = 1}^3 \lim_{\epsilon_i \to 0^+}\int_{\gamma_{\epsilon_i}} dq
\Big( q e^{iqr} \widehat{\mathcal{M}}_{\rm 1-loop}(q)\Big)\;,\\ 
V_{\rm branch}(r) \ & =&\ -\frac{1}{2 \pi^2 r} \lim_{\epsilon \to
  0^+}\int^\infty_{2m} dq \, q e^{-qr} \,{\rm
  Im}\big(\widehat{\mathcal{M}}_{\rm 1-loop}(iq + \epsilon)\big)\;, 
\end{eqnarray}
and  ${\rm  Res}(q  e^{i  q  r}  ${\small  $\widehat{\mathcal{M}}_{\rm
    1-loop}$}$(q),q_n)$ stands for the residue of a given complex pole
$q_n$.  The summation  in the first term of  $V_{\rm res}(r)$ is taken
over   all  complex   poles,  $q_n$,   of  $q   e^{i  q   r}  ${\small
  $\widehat{\mathcal{M}}_{\rm    1-loop}$}$(q)$.     There   are    no
contributions  from the $\gamma_{R1}$  and $\gamma_{R2}$  contours, as
they vanish as the radius of the contour $R$ goes to infinity. We note
that for a  radius $R$ bigger than the size of  the Landau pole $q_0$,
the contributions $\gamma_{\epsilon_1}$ and $\gamma_{\epsilon_3}$ must
be included.

Let us  first analyze the residue  at the physical pole  $q=0$ for the
resummed one-loop amplitude. This is given by
\begin{equation}
  \label{eq:zeroresidue}
{\rm Res}(q e^{i q r} \widehat{\mathcal{M}}_{\rm 1-loop}(q),0)\ =\ -\,
\alpha\: \frac{\kappa^2 m_1 m_2}{4}\ , 
\end{equation}
where
\begin{equation}
  \label{eq:alpha}
\alpha\ =\ \frac{1}{4}\bigg [ \frac{4}{3} \bigg(1 - 4 \sum_{i = 1}^n a_i
  \bigg)^{-1} - \frac{1}{3} \bigg(1 + 2 \sum_{i=1}^n a_i + 3 \sum_{i =
    1}^n b_i \bigg)^{-1} \bigg]\;, 
\end{equation}
with
\begin{equation}
  \label{eq:abi}
a_i\ =\ \left . \frac{\partial F_{2,i}(q^2)}{\partial q^2} \right |_{q^2
  = 0}, \qquad b_i\ =\ \left . \frac{\partial F_{3,i}(q^2)}{\partial
  q^2} \right |_{q^2 = 0}\;. 
\end{equation}
In the  above, $F_{j,i}$ is the  $j$th form factor of  the $i$th field
and $n$  is the  number of  fields. The formulae  for $a_i$  and $b_i$
derived  from~\eqref{eq:abi}  only  hold  if  the  form  factors  have
non-zero loop mass (or are  analytic in $q^2$). If all particle masses
in the loops vanish, we have $a_i  = b_i = 0$, implying that $\alpha =
1$.

As for the residues of the  Landau poles, we shall not include them in
the calculation, as these are related with the potential UV completion
of the theory of quantum gravity.  The simplest way to achieve this is
to introduce  a UV  cut-off just  below the Landau  pole $q_0$  in the
Fourier   transform   \eqref{eq:1DFourier}.    In   this   case,   the
contributions from  the $\gamma_{R1}$ and  $\gamma_{R2}$ contours will
not  vanish,  but  the  cut-off  integral  will  differ  by  terms  of
$O(m/q_0)$ in comparison to the other $O(1)$ terms.  Therefore, we may
safely ignore these cut-off  suppressed contributions in favour of the
other leading terms of order one.

We may now compute $V_{\rm  res}(r)$, using the result for the residue
in~\eqref{eq:zeroresidue}.   Computing the remaining  contour integral
$\gamma_{\epsilon_2}$ in~\eqref{eq:vres} gives rise to the potential
\begin{equation}
V_{\rm res}(r)\ =\  -\, \alpha\: \frac{G m_1 m_2}{r}\ .
\end{equation}
Evidently,   this   is   a   rescaled   version   of   the   Newtonian
potential. Specifically, for a scalar field of mass $m_H$, we have
\begin{equation}
a_H\ =\ \left. \frac{\partial F_{2,H}(p^2)}{\partial p^2} \right |_{p^2 = 0} =\
\frac{\kappa^2 m_H^2}{384\pi^2}\bigg[ \ln \para{\frac{m_H^2}{\mu^2}} -
  1 \bigg]\,, \qquad \left. b_H\ =\ \frac{\partial F_{3,H}(p^2)}{\partial p^2}
\right |_{p^2 = 0} =\ -\: \frac{\kappa^2 m_H^2}{768\pi^2}\bigg[ \ln
  \para{\frac{m_H^2}{\mu^2}} - 1 \bigg]\,. 
\end{equation}
For a fermion of mass $m_\psi$, we obtain
\begin{equation}
a_\psi\ =\ \left. \frac{\partial F_{2,\psi}(p^2)}{\partial p^2} \right
|_{p^2 = 0} =\ \frac{\kappa^2 m_\psi^2}{192\pi^2}\bigg[ \ln
  \para{\frac{m_\psi^2}{\mu^2}} - 1 \bigg]\,, \qquad
b_\psi\ =\ \left. \frac{\partial F_{3,\psi}(p^2)}{\partial p^2} \right
|_{p^2 = 0} =\ -\: \frac{\kappa^2 m_\psi^2}{384\pi^2}\bigg[ \ln
  \para{\frac{m_\psi^2}{\mu^2}} - 1 \bigg]\,. 
\end{equation}
Finally, for a massive gauge field of mass $m_A$ (without ghosts), we find
\begin{equation}
a_A \ =\ \left. \frac{\partial F_{2,A}(p^2)}{\partial p^2} \right |_{p^2 = 0} =\
-\frac{\kappa^2 m_A^2}{192\pi^2}\bigg[ \ln \para{\frac{m_A^2}{\mu^2}}
  - 2 \bigg]\,, \qquad 
b_A\ =\ \left. \frac{\partial F_{3,A}(p^2)}{\partial p^2}
\right |_{p^2 = 0} =\  \frac{\kappa^2 m_A^2}{384\pi^2}\bigg[ \ln
  \para{\frac{m_A^2}{\mu^2}} - 2 \bigg]\,. 
\end{equation}
Astronomical observations can only measure the combination $\alpha G$,
rather  than $G$  alone,  thus  leading to  a  renormalization of  the
Newtonian  constant $G$.  However,  we should  note that  the quantity
$\alpha$  differs significantly  from  $1$ when  the  loop masses  are
comparable to  the Planck mass  $M_{\rm P}$, which  is a case  that we
will not be considering here.

\subsubsection{The branch cut contribution}

Our  next task  is  to  compute the  branch  cut contribution  $V_{\rm
  branch}(r)$.  To  deal with  the  complexity  of  the integrand,  we
rewrite the one-loop corrected Newtonian potential as follows:
\begin{equation}
V(r)\ =\ -\: \frac{G m_1 m_2}{r}\; \Big(\alpha\: +\: \Delta V(r)\Big)\; ,
\end{equation}
where the  coefficient $\alpha$ given  by~\eqref{eq:alpha} pertains to
the residue contributions and the dimensionless quantity $\Delta V(r)$
refers to  the part  of the potential  resulting from the  branch cut,
i.e.
\begin{equation}
  \label{eq:Vbranch}
V_{\rm branch}(r)\  =\ -\: \frac{G m_1 m_2}{r}\;  \Delta V(r)\; .
\end{equation}
We  observe that  the  integral $V_{\rm  branch}(r)$  can be  computed
accurately by taking the first  order term in a perturbative expansion
in $\kappa^2$.   To leading order in $\kappa^2$,  the contributions to
$V_{\rm  branch}(r)$ from  scalar  ($H$), fermion  ($\psi$) and  gauge
boson~($A^\mu$)  loops  may  be  calculated  individually,  such  that
$\Delta V(r)$ is given by the sum:
\begin{equation}
\Delta V (r)\ =\ \Delta V_H (r)\: +\: \Delta  V_\psi (r)\:  +\:  
                 \Delta V_A  (r)\; .
\end{equation}   

We will  first present the calculation  for the scalar  loops and then
simply state the results of  the fermion and gauge fields.  The branch
cut effect due to a massive Higgs boson $H$ is given by the integral
\begin{equation}
\Delta V_H(r)\ =\ \frac{G}{60 \pi}\:\int^\infty_{2m} dq \,e^{-qr}\left(3 -
\frac{4m_H^2}{q^2} + \frac{28m_H^4}{q^4}\right)\,\sqrt{q^2 -
  4m_H^2}\, ,
\end{equation}
which is  analytically calculable.  Using  the substitution $q  = 2m_H
\cosh x$, $\Delta V_H(r)$ may be rewritten as
\begin{equation}
  \label{eq:deltavhyperbolic} 
\Delta V_H (r)\ =\ \frac{G m_H^2}{15 \pi}\int^\infty_{0} dx \,e^{-2mr\cosh
  x}\Big(\cosh^2 x - 1\Big)\bigg(3 -\,\text{sech}^2 \,x + \frac{7}{4}
\,\text{sech}^4\, x\bigg)\; .
\end{equation}

To proceed further, we first remind ourselves that the modified Bessel
functions  of the  second kind  $K_\alpha(\hat{r})$ have  the integral
representation
\begin{equation}
K_\alpha(\hat{r})\ =\ \int^\infty_0 dx\,e^{-\hat{r} \cosh x}
\cosh(\alpha x)\; ,
\end{equation}
and so it is
\begin{equation}
K_0(\hat{r}) = \int^\infty_0 dx \, e^{- \hat{r} \cosh x}\; .
\end{equation}
Moreover,  we  can  use  the  the  hyperbolic  trigonometric  identity
$\cosh^2 x = (1 + \cosh 2x)/2 $ to calculate the following integral:
\begin{equation}
\int^\infty_0 dx \, e^{- \hat{r} \cosh x} \cosh^2 x\ =\
\frac{1}{2}\, \Big(K_0(\hat{r}) + K_2(\hat{r}) \Big)\; . 
\end{equation}
Apart from  integrals containing $\cosh  x$, there are  also integrals
involving sech$\,x$, defined as
\begin{equation}
I_n (r)\ =\ \int^\infty_0 dx\, e^{- r \cosh x} \,\text{sech}^n x\; .
\end{equation}
We  may  compute  the  functions  $I_n(r)$ recursively,  by  means  of
integration by parts.  The relevant integrals of interest are
\begin{eqnarray}
I_1(r)\ &=&\ \int^\infty_r dx K_0(x) = -\frac{1}{2} \pi  \Big(r
\pmb{L}_{-1}(r) K_0(r)+r \pmb{L}_0(r) K_1(r)-1\Big)\;,\\ 
I_2(r)\ &=&\ r \Big(K_1(r) - I_1(r)\Big)\;,\\
I_3(r)\ &=&\ \frac{1}{2}\Big( r K_0(r) - r I_2(r) + I_1(r)\Big)\;, \\
I_4(r)\ &=&\ \frac{2}{3} I_2(r) + \frac{r}{3} \Big( I_1(r) -
I_3(r)\Big)\; ,
\end{eqnarray}
where $\pmb{L}_\alpha(r)$ is the modified Struve function which has
the integral representation 
\begin{equation}
\pmb{L}_\alpha(r)\ =\ \frac{2^{1-\alpha} r^\alpha}{\sqrt{\pi}\,
  \Gamma\big(\alpha + \frac{1}{2}\big)} \int^\frac{\pi}{2}_0 dx
\sinh(r \cos x)\sin^{2\alpha} x\; , 
\end{equation}
for ${\rm Re}(\alpha) >  -\frac{1}{2}$.  The latter representation may
be  analytically  continued  to  include  other values  of  the  index
$\alpha$ of the modified Struve function $\pmb{L}_\alpha(r)$.

\begin{figure}
\centering
\includegraphics[scale=0.4]{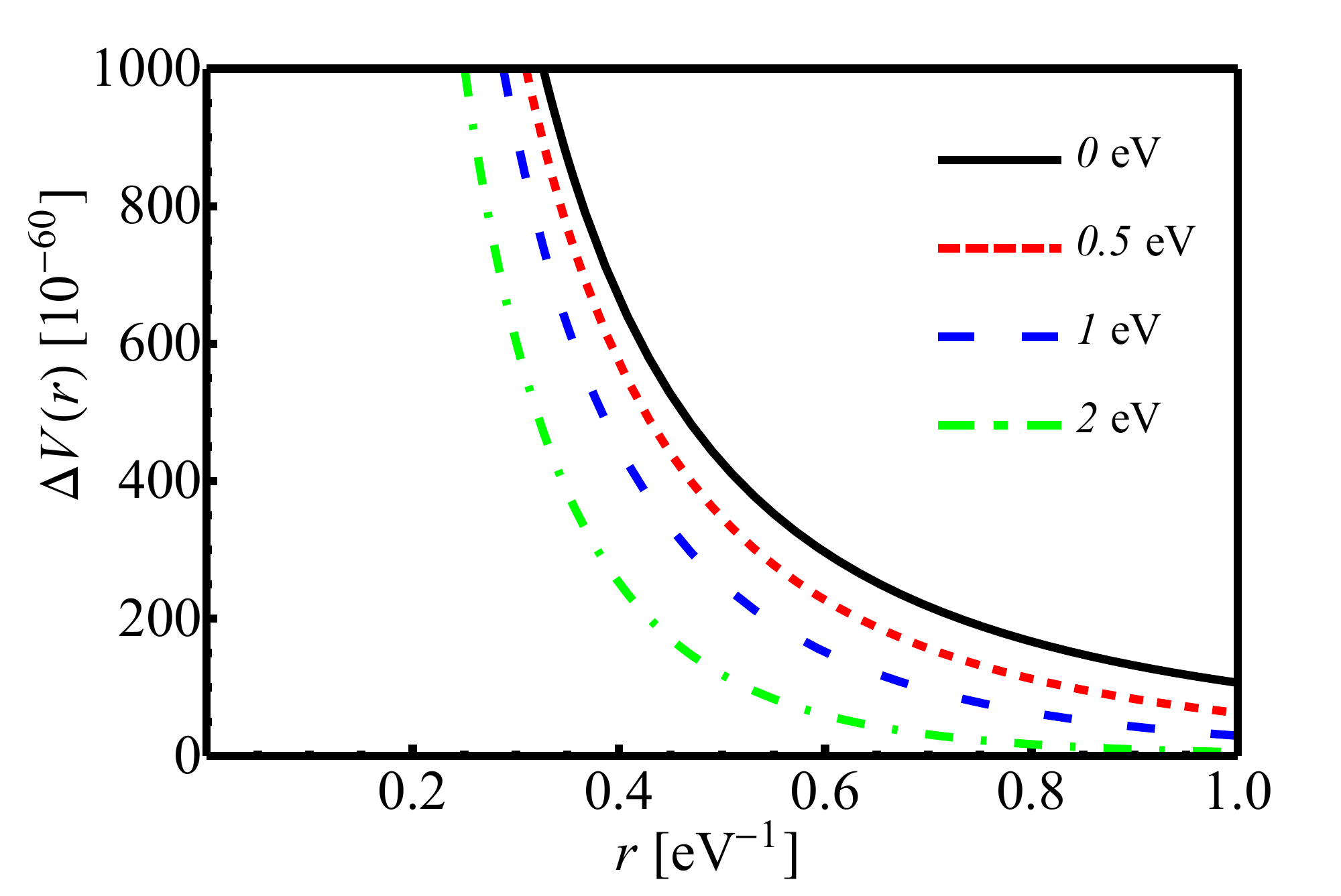}
~
\includegraphics[scale=0.4]{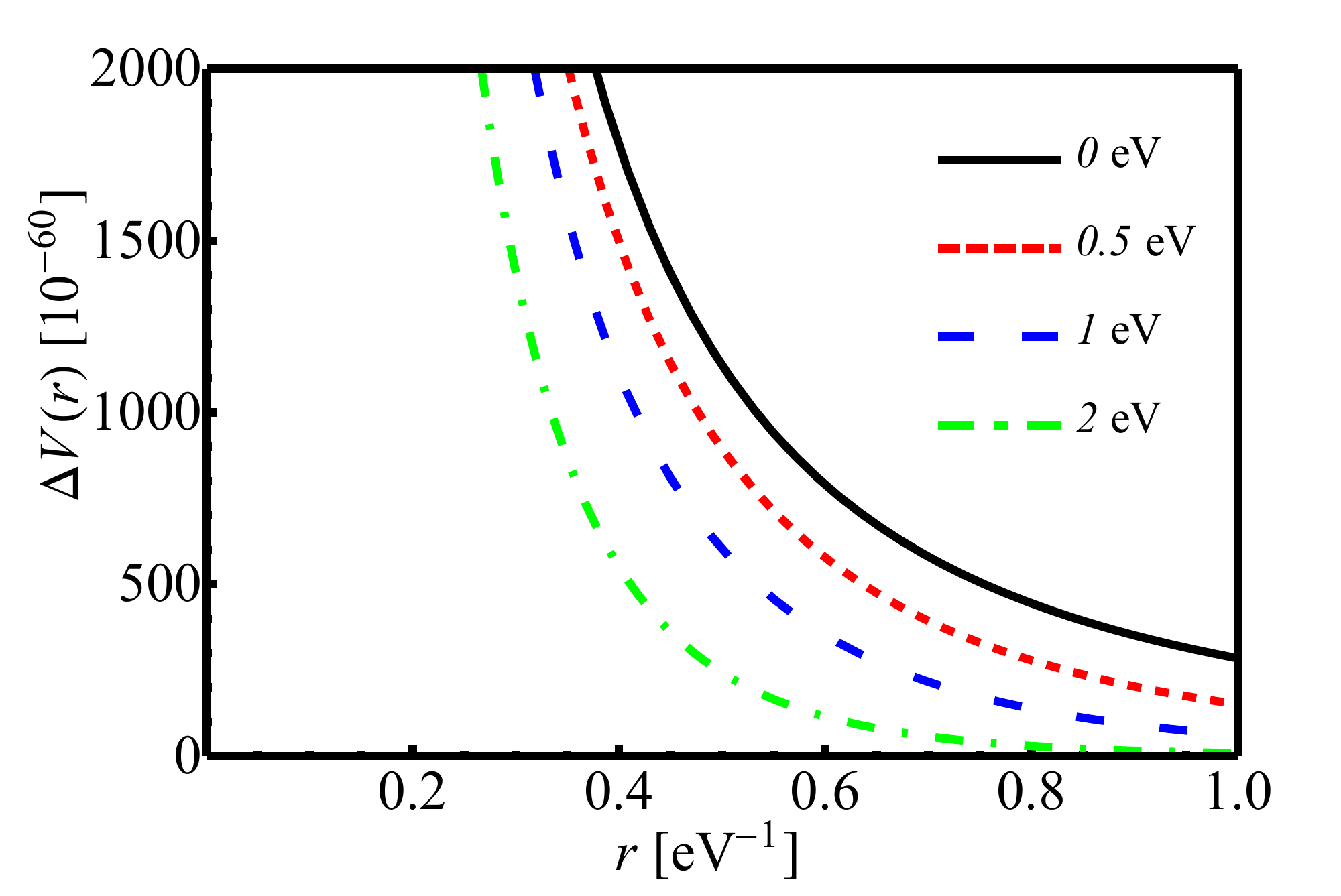}
\\
\includegraphics[scale=0.35]{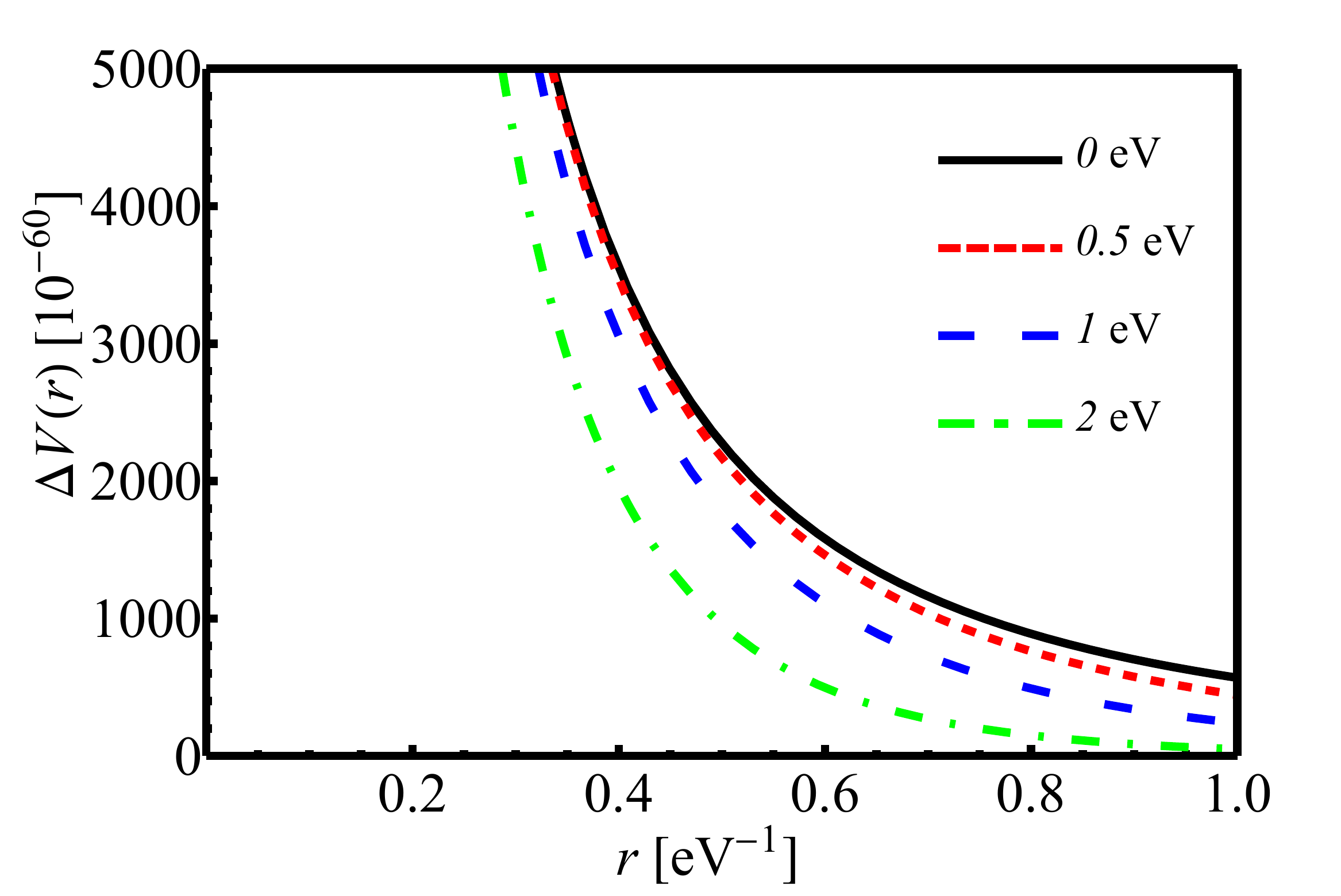}

\caption{Estimates  of the branch-cut  terms $\Delta  V_H(r)$, $\Delta
  V_\psi (r)$ and $\Delta V_A  (r)$ resulting from scalar, fermion and
  gauge fields,  as functions  of the distance~$r$,  are shown  in the
  upper left, upper right and lower pannels, respectively.
\label{fig:plots}} 

\end{figure}

We are now  in a position to analytically compute  the branch cut term
$\Delta V_H (r)$  in terms of the modified  Bessel and Struve functions.
Defining the dimensionless parameter $\hat{r}_H  = 2 m_H r$, we obtain
for the Higgs-scalar contribution:
\begin{align}
\Delta V_H(r)\ & =\ -\frac{G m_H^2}{360 \pi}\; \bigg[\,\frac{1}{2} \pi
\left(7 \hat{r}_H^2-45\right) \hat{r}_H^2 \Big(\pmb{L}_{-1}(\hat{r}_H)
K_0(\hat{r}_H)+ \pmb{L}_0(\hat{r}_H) K_1(\hat{r}_H)\Big)-\frac{7 \pi
  \hat{r}_H^3}{2}+\frac{45 \pi  \hat{r}_H}{2}\\&\hspace{6em}+7 \hat{r}_H^3
K_1(\hat{r}_H)-7 \hat{r}_H^2 K_0(\hat{r}_H)-38 \hat{r}_H K_1(\hat{r}_H)+60
K_0(\hat{r}_H)-36 K_2(\hat{r}_H)\,\bigg]\; . 
\end{align}
Similarly,  for a Dirac  fermion $\psi$,  we define  the dimensionless
parameter  $\hat{r}_\psi  =  2  m_\psi  r$,  in  terms  of  which  the
branch-cut contribution reads:
\begin{align}
\Delta V_\psi (r)\ & =\ -\frac{G m_\psi^2}{180 \pi}\; \bigg[\,\frac{1}{2} \pi
\left(7 \hat{r}_\psi^2-15\right) \hat{r}_\psi^2 \Big(\pmb{L}_{-1}(\hat{r}_\psi)
K_0(\hat{r}_\psi)+ \pmb{L}_0(\hat{r}_\psi) K_1(\hat{r}_\psi)\Big)-\frac{7 \pi
  \hat{r}_\psi^3}{2}+\frac{15 \pi  \hat{r}_\psi}{2}\\&\hspace{6em}+7
\hat{r}_\psi^3 K_1(\hat{r}_\psi)-7 \hat{r}_\psi^2 K_0(\hat{r}_\psi)-8 \hat{r}_\psi
K_1(\hat{r}_\psi)-30 K_0(\hat{r}_\psi)+ 24 K_2(\hat{r}_\psi)\,\bigg]\; .  
\end{align}
Finally, the branch-cut contribution arising from a $U(1)$ gauge boson
$A_\mu$ and its associate ghost field is given by
\begin{align}
\Delta V_A(r)\ & =\ -\frac{G m_A^2}{360 \pi}\; \bigg[\, \pi \left(7 \hat{r}_A^2
+ 195\right) \hat{r}_A^2 \Big(\pmb{L}_{-1}(\hat{r}_A) K_0(\hat{r}_A)+
\pmb{L}_0(\hat{r}_A) K_1(\hat{r}_A)\Big) - 7 \pi  \hat{r}_A^3-195 \pi
\hat{r}_A\\&\hspace{4em}+14 \hat{r}_A^3 K_1(\hat{r}_A)-14 \hat{r}_A^2
K_0(\hat{r}_A) + 404\hat{r}_A K_1(\hat{r}_A) - 240 K_0(\hat{r}_A) - 192
K_2(\hat{r}_A)\,\bigg]\; , 
\end{align}
with $\hat{r}_A  = 2  m_A r$.  We  have checked that  our perturbative
analytical  expressions  for  the  branch  cut  contributions  are  in
excellent agreement with numerical  results derived by using the fully
resummed graviton propagator to  less than 1 part in~$10^{-16}$. Plots
of  the different contributions  to the  potential for  different loop
masses are given in  Fig.~\ref{fig:plots}. These plots demonstrate the
exponential decay of loop effects due to particles with non-zero mass,
as a function of distance~$r$.

Using the above analytical results, it is not difficult to verify that 
the loop-corrected potential exhibits the desirable property:
\begin{equation}
\lim_{r \to \infty} V(r)\ =\ 0\; .
\end{equation}
To see this explicitly, we use the large-$r$ asymptotic formulae for
the modified Bessel and Struve functions:
\begin{equation}
  \label{eq:asymptoticexpansions}
K_\alpha(r)\ \sim\ \sqrt{\frac{\pi}{2 r}} e^{-r}\;, \qquad
\pmb{L}_\alpha(r)\ \sim\ \sqrt{\frac{1}{2 \pi r}} e^r\ -\ \frac{2^{1 -
    \alpha }r^{\alpha - 1}}{\sqrt{\pi}\,\Gamma \big( \alpha +
  \frac{1}{2} \big)}\; . 
\end{equation}
In particular, for the scalar case, the branch-cut term $\Delta
V_H(r)$ for $\hat{r}_H \gg 1$ simplifies  to
\begin{equation}
\Delta V_H(r)\ =\ -\: \frac{7 G m_H^2}{240 \sqrt{\pi
    m_H}}\para{\frac{e^{-2m_H r}}{\sqrt{r}}\ -\ \frac{2}{3}\,m_H\sqrt{r}e^{-2 m_H
    r}}\; . 
\end{equation}
For the fermion case, we have for $\hat{r}_\psi \gg 1$,
\begin{equation}
\Delta V_\psi (r)\ =\ \frac{7 G m_\psi^2}{60 \sqrt{\pi
    m_\psi}}\para{\frac{e^{-2m_\psi r}}{\sqrt{r}}\ -\
  \frac{2}{3}\,m_\psi\sqrt{r}e^{-2 m_\psi r}} 
\end{equation}
and for the gauge boson case for $\hat{r}_A \gg 1$, 
\begin{equation}
\Delta V_A(r)\ =\ -\: \frac{7 G m_A^2}{120 \sqrt{\pi
    m_A}}\para{\frac{e^{-2m_A r}}{\sqrt{r}}\ -\ \frac{2}{3}\,m_A\sqrt{r}e^{-2 m_A
    r}}\; . 
\end{equation}

In  the  opposite limit  where  $\hat{r}_{H,\psi,A}  \ll  1$, we  find
respectively   for   the  scalar,   Dirac   fermion  and   gauge-boson
contributions to $\Delta V (r)$ that
\begin{eqnarray}
\Delta V_H(r)\ &=&\ \frac{G}{20 \pi r^2}\ +\ \frac{G m^2_H}{6 \pi}\, \bigg(
\ln(m_H r)\: +\: \gamma_E\: +\: \frac{1}{3}\,\bigg)\ +\ {\cal O}(\hat{r})\; ,\\
\Delta V_\psi (r)\ &=&\ \frac{G}{15 \pi r^2}\ + \frac{G m^2_\psi}{3
  \pi}\, \bigg(  
\ln(m_\psi r)\: +\:  \gamma_E\: -\: \frac{2}{3}\, \bigg)\  
+\ {\cal O}(\hat{r})\; ,\\
\Delta V_A (r)\ &=&\ \frac{4 G}{15 \pi r^2}\ -\ \frac{2 G m^2_A}{3
  \pi}\, \bigg(
\ln(m_A r)\: +\: \gamma_E\: -\: \frac{25}{12}\,\bigg)\ +\ {\cal O}(\hat{r})\; . 
\end{eqnarray}
In the above  small-$\hat{r}$ expansion, the first term  on the RHS of
the  above  equations  represents  the  correction  to  the  potential
assuming that  the particle in  the loop is strictly  massless.  These
leading terms are consistent with the ones presented in the literature
\cite{Hamber1995,Bjerrum-Bohr2002a,Duff:1974ud,Duff:2000mt}.        For
massive fields, however, the relevant subleading correction to $\Delta
V(r)$ is logarithmically enhanced in $r$, as long as $r \ll 1/2m$.

\section{Conclusions \label{Concl}}

We have  revisited the  calculation of matter  quantum effects  on the
graviton self-energy, assuming a flat Minkowski background metric. One
of  the  central goals  of  our  study has  been  to  obtain a  deeper
understanding of  the underlying  mechanism that renders  the graviton
massless. To this end, we have first considered a gauged Abelian Higgs
model, which has been quantized within the framework of the background
field  method.  After  writing down  the  respective diffeomorphically
invariant path  integral, we have  derived a master Ward  identity for
the  path   integral  as  a   consequence  of  its   invariance  under
diffeomorphisms.   This Ward identity  does not  ensure by  itself the
transversality of  the graviton  self-energy.  The latter  property of
masslessness  of   the  graviton   is  only  obtained   upon  imposing
minimization conditions  to the effective action. In  this respect, we
have found that  the minimization of the effective  action is strongly
related  with   the  renormalization  of   the  cosmological  constant
$\Lambda$,  and  this  relation  can  be enforced  to  all  orders  in
perturbation theory, by means of a Graviton Low-Energy Theorem (GLET),
which we derived in this paper.

In the  context of  the Abelian Higgs  model mentioned above,  we have
also  calculated  the  matter  quantum corrections  to  the  Newtonian
potential.  As we have not considered graviton quantum loop effects in
our study,  it is  evident that matter  contributions to  the graviton
self-energy  are independent  of the  gauge fixing  parameters $\xi_D$
and~$\sigma$  of the diffeomorphisms.   In our  calculations, however,
the gauge  dependence due to diffeomorphisms does  formally enter when
considering  the  resummed  graviton propagator.   Nevertheless,  when
calculating the  ${\rm S}$-matrix amplitude for the  scattering of two
scalar fields,  this background gauge dependence is  removed by virtue
of the Ward identity derived in Section~\ref{TF} and the fact that the
gravitationally scattered  particles are on their  mass shell.  Hence,
the  analytic   results  we  have   presented  in  this   article  are
diffeomorphisms invariant.  On the  other hand, gauge-boson loops have
been  calculated in  the Feynman-'t  Hooft gauge  $\xi_G =  1$.  Since
${\rm  S}$-matrix   elements  are  independent   of  the  gauge-fixing
parameter  $\xi_G$,  the  graviton   self-energy  is  expected  to  be
independent  of  $\xi_G$  as  well, especially  when  considering  the
elastic gravitational  scattering of two  {\em gauge-singlet} scalars.
As a consequence,  we expect that the Newtonian  potential $V(r)$ will
not depend on the gauge-fixing parameter~$\xi_G$.

Treating  quantum  gravity  as  an  effective field  theory,  we  have
presented  analytical  formulae  for  matter quantum  effects  on  the
Newtonian  potential $V(r)$, in  terms of  modified Bessel  and Struve
functions which depend  on the particle masses in  the loop.  Thus, we
have  found that  the  corrections to  $V(r)$  exhibit an  exponential
fall-off  dependence on  the distance~$r$,  once  the non-relativistic
limit with respect to the  non-zero loop masses is properly taken into
account. In the  massless limit of scalars, fermions  and gauge bosons
in  the  loops, we  recover  the  well-known  results that  have  been
presented in the literature.

Like  the  well-known Higgs-boson  low-energy  theorem  that holds  in
particle-physics models,  such as  the Standard Model,  the GLET  is a
very powerful  theorem.  As was  explicitly shown in this  paper, both
the       GLET~\eqref{eq:GLET}      and       the      diffeomorphisms
WI~\eqref{eq:secondwardidentitymomentum}  are required  to  forbid the
appearance of a mass for the graviton field, which might be induced by
quantum-loop effects.  We  have derived the GLET for  a flat geometry,
where a  global shift symmetry  between the background  graviton field
and  the Minkowski metric  exists.  Given  the property  of background
independence of the  background field method, we expect  to be able to
extend this  theorem to  general curved background  metrics.  However,
such a  generalization is  beyond the scope  of the present  paper. We
hope to  be able  to report  progress on this  issue in  a forthcoming
communication.

\subsection*{Acknowledgements}

We would like to  thank Luis Alvarez--Gaume for discussions concerning
the topic  of the  graviton mass in  theories of quantum  gravity. The
work D.B.  is supported by  an STFC PhD  studentship, and the  work of
A.P.   is supported  by the  Lancaster-Manchester-Sheffield Consortium
for Fundamental Physics under STFC grant ST/L000520/1.

\newpage

\appendix

\section{Feynman Rules \label{FR}}

In this  appendix we list all  relevant Feynman rules  which have been
used in our calculations.  We  define all momenta as outgoing from the
vertex, obeying energy-momentum conservation.

\subsection{Graviton Propagator}

Since  our computations  pertain to  gauge-invariant  ${\rm S}$-matrix
amplitudes, we  employ the simplified form of  the graviton propagator
in    the   harmonic   gauge,    which   is    given   by    the   RHS
of~\eqref{eq:propagatorreplacement}.   For this  choice of  gauge, the
diffeomorphisms gauge-fixing  parameters $\xi_D$ and  $\sigma$ take on
the values: $\xi_D  = 1$, $\sigma = 1/2$.   For completeness, however,
we  present  the  general   expression  for  the  tree-level  graviton
propagator for arbitrary gauge parameters $\xi_D$ and $\sigma$:
\begin{equation} 
  \ba{
    \raisebox{-0.45\height}{\includegraphics[scale=1]{gravitonprop.1}
    } \hspace{1em}  =&\; \frac{1}{p^2  + i\epsilon} \Bigg[  P^{\mu \nu
        \rho  \sigma}  -  \bigg(4(1+\xi_D)  +  \frac{8}{\sigma  -1}  +
      \frac{3-\xi_D}{(\sigma-1)^2}   \bigg)\frac{p^\mu   p^\nu  p^\rho
        p^\sigma}{(p^2)^2}\\&  +   \bigg(  2  +   \frac{1}{\sigma  -1}
      \bigg)\bigg(\frac{p^\mu     p^\nu}{p^2}    \eta^{\rho    \sigma}
      +\frac{p^\rho p^\sigma}{p^2} \eta^{\mu \nu}\bigg) \\& + (\xi_D -
      1)\bigg(\frac{p^\mu  p^\rho}{p^2}  \eta^{\nu \sigma}+\frac{p^\mu
        p^\sigma}{p^2}    \eta^{\nu   \rho}+\frac{p^\nu   p^\rho}{p^2}
      \eta^{\mu     \sigma}+\frac{p^\nu    p^\sigma}{p^2}    \eta^{\mu
        \rho}\bigg)\Bigg]\;,  } 
\end{equation}  
with  $P^{\mu \nu \rho  \sigma} =  \eta^{\mu \rho}\eta^{\nu  \sigma} +
\eta^{\mu \rho}\eta^{\nu \sigma} - \eta^{\mu \nu}\eta^{\rho \sigma}$.

\subsection{Graviton-Scalar-Scalar Vertex}

The     coupling     for     the     scalar-scalar-graviton     vertex
$H(p)$-$H(q)$-$h^{\mu\nu}(l)$
reads:               \begin{equation} 
  \raisebox{-0.45\height}{\includegraphics[scale=1]{2scalar1graviton.1}
  } \equiv  V^{\mu \nu}_{HHh}(p,q) = \frac{i  \kappa}{2}(p^\mu q^\nu +
  q^\mu p^\nu - \eta^{\mu \nu}(p \cdot q + m_H^2))\; .  
\end{equation}

\subsection{Graviton-Graviton-Scalar-Scalar Vertex}

The quartic coupling $H(p)$-$H(q)$-$h^{\mu\nu}(l)$-$h^{\rho\sigma}(k)$
is             given              by             
\begin{equation}             
\ba{
    \raisebox{-0.45\height}{\includegraphics[scale=1]{2scalar2graviton.1}
    }  \equiv&\;  V_{HHhh}^{\mu  \nu , \rho  \sigma}(p,q) \\  =&  \;  i
    \kappa^2  \Bigg(\sqpara{\frac{1}{4}(\eta^{\mu \nu}  I^{\rho \sigma
        \alpha   \beta}  +  \eta^{\rho   \sigma}  I^{\mu   \nu  \alpha
        \beta})-I^{\mu \nu \alpha}_{\;\;\;\;\;\;\delta} I^{\rho \sigma
        \beta  \delta}}(p_\alpha q_\beta  + q_\alpha  p_\beta) \\  & +
    \frac{1}{2}\para{I^{\mu  \nu \rho  \sigma}  - \frac{1}{2}\eta^{\mu
        \nu} \eta^{\rho \sigma}}[(p  \cdot q + m_H^2)] \Bigg)\;  , } 
\end{equation}
where 
\begin{equation}
I^{\mu  \nu \rho \sigma}  \equiv \frac{1}{2}(\eta^{\mu \rho}
  \eta^{\nu \sigma}  + \eta^{\mu  \sigma} \eta^{\nu \rho})\;  .
\end{equation}
Note           that           the           quartic           coupling
$H(p)$-$H(q)$-$h^{\mu\nu}(l)$-$h^{\rho\sigma}(k)$ only  depends on the
four-momenta $p$ and $q$ of the scalar particles.

\subsection{Graviton-Fermion-Fermion Vertex}

The   fermion-fermion-graviton   interaction  $\bar{\psi}   (p)$-$\psi
(q)$-$h^{\mu\nu}(l)$         is          given         by         
\begin{equation}
  \raisebox{-0.45\height}{\includegraphics[scale=1]{2fermion1graviton.1}
  } \equiv V^{\mu \nu}_{\psi \psi h}(p,q) = -\frac{i \kappa}{8}\Big[(p
    - q)^\mu \gamma^\nu + (p-q)^\mu \gamma^\nu - 2\eta^{\mu
      \nu}(\slashed{p} - \slashed{q} - 2 m_\psi^2)\Big]\; .  
\end{equation}

\subsection{Graviton-Graviton-Fermion-Fermion Coupling}

The fermion-fermion-graviton-graviton interaction $\bar{\psi} (p)$-$\psi
(q)$-$h^{\mu\nu}(l)$-$h^{\rho\sigma}(k)$ reads:
\begin{equation}
	\ba{
	\raisebox{-0.45\height}{\includegraphics[scale=1]{2fermion2graviton.1}
}
	\equiv&\; V^{\mu \nu , \rho \sigma}_{\psi \psi hh}(p,q) \\ =&\;
        i \frac{\kappa^2}{8} \Bigg[ \frac{1}{4}\Big(\Big[\eta^{\mu
              \nu} \gamma^\rho(p^\sigma - q^\sigma) + \eta^{\rho
              \sigma} \gamma^\mu (p^\nu - q^\nu) + \frac{3}{2}
            \eta^{\nu \rho}\gamma^\mu (p^\sigma - q^\sigma) 
	\\ & + \frac{3}{2} \eta^{\nu \rho}\gamma^\sigma (p^\mu -
        q^\mu) + (\mu \leftrightarrow \nu)\Big] + (\rho
          \leftrightarrow \sigma)\Big) \\ & -\frac{1}{2}\para{I^{\mu
              \nu \rho \sigma} - \frac{1}{2}\eta^{\mu \nu} \eta^{\rho
              \sigma}}[(\slashed{p} - \slashed{q} -
            2m_\psi^2)]\Bigg]\; .
	}
\end{equation}
Like in the scalar case,         the           quartic           coupling
$\bar{\psi} (p)$-$\psi
(q)$-$h^{\mu\nu}(l)$-$h^{\rho\sigma}(k)$ only depends  on the
four-momenta $p$ and $q$ of the fermion particles.

\subsection{Graviton-Gauge-Gauge Vertex}

The  interaction vertex involving  two gauge  bosons $A^\rho  (p)$ and
$A^\sigma  (q)$   and  one   graviton  $h^{\mu\nu}(l)$  is   given  by
\begin{equation}
\hspace{-0.4cm}                                                
\ba{
    \raisebox{-0.45\height}{\includegraphics[scale=1]{2gauge1graviton.1}
    }  \equiv&\;  V^{\rho,\sigma,\mu\nu}_{AAh}(p,q) \\=&\;-  \frac{i
      \kappa}{2} \Bigg [(p \cdot q  + m_A^2)(2 I^{\mu \nu \rho \sigma}
      - \eta^{\mu  \nu}  \eta^{\rho  \sigma})  \\&  +  \eta^{\mu  \nu}
      p^\sigma  q^\rho   -  \Big(\eta^{\mu  \sigma}   p^\nu  q^\rho  +
      \eta^{\mu \sigma} p^\nu q^\rho  - \eta^{\rho \sigma} p^\mu q^\nu
      +  (\mu   \leftrightarrow  \nu)  \Big)   \\&  +  \frac{1}{\xi_G}
      \Big(\eta^{\mu \nu}(p^\rho  p^\sigma + p^\rho  q^\sigma + q^\rho
      q^\sigma)  - \Big[  \eta^{\nu \sigma}  p^\mu p^\rho  + \eta^{\nu
          \rho}  q^\mu  q^\sigma  +  (\mu \leftrightarrow  \nu)  \Big]
      \Big)\Bigg ]\; .  } 
\end{equation}

\subsection{Graviton-Graviton-Gauge-Gauge Quartic Coupling}

The  quartic coupling involving  two gauge  bosons $A^\alpha  (p)$ and
$A^\beta(q)$ and two gravitons $h^{\mu\nu}(l)$ and $h^{\rho\sigma}(k)$
is found to be
\begin{equation}         
 \ba{
    \raisebox{-0.45\height}{\includegraphics[scale=1]{2gauge2graviton.1}
    } &  \equiv V^{\alpha, \beta, \mu  \nu, \rho \sigma}_{AAhh}(p,q,l)
    \\  & =  \frac{i \kappa^2}{4}\bigg[K^{\mu  \nu \rho  \sigma \alpha
        \beta}(p  \cdot q  + m_A^2)  + L^{\mu  \nu \rho  \sigma \alpha
        \beta}(p,q)  + \frac{1}{\xi_G} M^{\mu  \nu \rho  \sigma \alpha
        \beta}(p,q,l)\bigg]\;, }  
\end{equation}
where 
\begin{equation} 
\ba{ &  L^{\mu \nu \rho
      \sigma \alpha \beta}(p,q) = \\&\eta^{\nu \rho } \eta^{\mu \sigma
    }  p^{\beta }  q^{\alpha  }+\eta^{\mu \rho  }  \eta^{\nu \sigma  }
    p^{\beta  }  q^{\alpha }+\eta^{\beta  \nu  }  \eta^{\rho \sigma  }
    p^{\mu } q^{\alpha }-\eta^{\beta  \sigma } \eta^{\mu \rho } p^{\nu
    }  q^{\alpha }-\eta^{\beta  \rho  } \eta^{\mu  \sigma  } p^{\nu  }
    q^{\alpha  }+\eta^{\beta  \mu  }  \eta^{\rho  \sigma  }  p^{\nu  }
    q^{\alpha  }  +\eta^{\mu \nu  }  \eta^{\beta  \sigma  } p^{\rho  }
    q^{\alpha  } \\&-\eta^{\beta \nu  } \eta^{\mu  \sigma }  p^{\rho }
    q^{\alpha  }-\eta^{\beta  \mu  }  \eta^{\nu  \sigma  }  p^{\rho  }
    q^{\alpha  }-\eta^{\beta  \sigma  }  p^{\mu  }  \eta^{\nu  \rho  }
    q^{\alpha  }-\eta^{\beta  \mu  }  p^{\sigma  }  \eta^{\nu  \rho  }
    q^{\alpha  }+\eta^{\mu  \nu  }  \eta^{\beta  \rho  }  p^{\sigma  }
    q^{\alpha  }-\eta^{\beta  \nu  }  \eta^{\mu  \rho  }  p^{\sigma  }
    q^{\alpha  }-\eta^{\beta  \rho  }  p^{\mu  }  \eta^{\nu  \sigma  }
    q^{\alpha  } \\&-p^{\beta }  \eta^{\mu \nu  } \eta^{\rho  \sigma }
    q^{\alpha  }+\eta^{\alpha \nu  }  \eta^{\rho \sigma  } p^{\beta  }
    q^{\mu }+\eta^{\beta \rho }  \eta^{\alpha \sigma } p^{\nu } q^{\mu
    }+\eta^{\alpha  \rho }  \eta^{\beta  \sigma }  p^{\nu  } q^{\mu  }
    -\eta^{\alpha  \nu  }  \eta^{\beta   \sigma  }  p^{\rho  }  q^{\mu
    }+\eta^{\alpha  \beta  }  \eta^{\nu  \sigma  }  p^{\rho  }  q^{\mu
    }-\eta^{\alpha  \sigma }  p^{\beta  } \eta^{\nu  \rho  } q^{\mu  }
    \\&-\eta^{\alpha  \nu  } \eta^{\beta  \rho  }  p^{\sigma }  q^{\mu
    }+\eta^{\alpha  \beta  }  \eta^{\nu  \rho  }  p^{\sigma  }  q^{\mu
    }-\eta^{\alpha  \rho }  p^{\beta  } \eta^{\nu  \sigma  } q^{\mu  }
    -\eta^{\alpha  \beta  }  p^{\nu   }  \eta^{\rho  \sigma  }  q^{\mu
    }+\eta^{\alpha  \mu  }  \eta^{\rho  \sigma  }  p^{\beta  }  q^{\nu
    }+\eta^{\beta  \rho  }  \eta^{\alpha  \sigma  }  p^{\mu  }  q^{\nu
    }+\eta^{\alpha  \rho }  \eta^{\beta  \sigma }  p^{\mu  } q^{\nu  }
    \\&-\eta^{\alpha  \mu  } \eta^{\beta  \sigma  }  p^{\rho }  q^{\nu
    }+\eta^{\alpha  \beta }  \eta^{\mu  \sigma }  p^{\rho  } q^{\nu  }
    -\eta^{\alpha  \sigma  }  p^{\beta   }  \eta^{\mu  \rho  }  q^{\nu
    }-\eta^{\alpha  \mu  }  \eta^{\beta  \rho  }  p^{\sigma  }  q^{\nu
    }+\eta^{\alpha  \beta  }  \eta^{\mu  \rho  }  p^{\sigma  }  q^{\nu
    }-\eta^{\alpha  \rho  }  p^{\beta  }  \eta^{\mu  \sigma  }  q^{\nu
    }-\eta^{\alpha  \beta }  p^{\mu  } \eta^{\rho  \sigma  } q^{\nu  }
    \\&+\eta^{\mu  \nu } \eta^{\alpha  \sigma }  p^{\beta }  q^{\rho }
    -\eta^{\alpha  \sigma  }  \eta^{\beta   \nu  }  p^{\mu  }  q^{\rho
    }+\eta^{\alpha  \beta  }  \eta^{\nu  \sigma  }  p^{\mu  }  q^{\rho
    }-\eta^{\alpha  \sigma  }  \eta^{\beta  \mu  }  p^{\nu  }  q^{\rho
    }+\eta^{\alpha  \beta  }  \eta^{\mu  \sigma  }  p^{\nu  }  q^{\rho
    }+\eta^{\beta  \mu  }  \eta^{\alpha  \nu  }  p^{\sigma  }  q^{\rho
    }+\eta^{\alpha  \mu }  \eta^{\beta  \nu }  p^{\sigma  } q^{\rho  }
    \\&-\eta^{\alpha  \beta  } \eta^{\mu  \nu  }  p^{\sigma }  q^{\rho
    }-\eta^{\alpha  \nu  }  p^{\beta  }  \eta^{\mu  \sigma  }  q^{\rho
    }-\eta^{\alpha  \mu }  p^{\beta  } \eta^{\nu  \sigma  } q^{\rho  }
    +\eta^{\mu  \nu  }  \eta^{\alpha   \rho  }  p^{\beta  }  q^{\sigma
    }-\eta^{\alpha  \rho  }  \eta^{\beta  \nu  }  p^{\mu  }  q^{\sigma
    }+\eta^{\alpha  \beta }  \eta^{\nu  \rho }  p^{\mu  } q^{\sigma  }
    -\eta^{\alpha  \rho  } \eta^{\beta  \mu  }  p^{\nu  } q^{\sigma  }
    \\&+\eta^{\alpha  \beta  } \eta^{\mu  \rho  }  p^{\nu }  q^{\sigma
    }+\eta^{\beta  \mu  }  \eta^{\alpha  \nu  }  p^{\rho  }  q^{\sigma
    }+\eta^{\alpha  \mu  }  \eta^{\beta  \nu  }  p^{\rho  }  q^{\sigma
    }-\eta^{\alpha  \beta  }  \eta^{\mu  \nu  }  p^{\rho  }  q^{\sigma
    }-\eta^{\alpha  \nu  }  p^{\beta  }  \eta^{\mu  \rho  }  q^{\sigma
    }-\eta^{\alpha \mu } p^{\beta }  \eta^{\nu \rho } q^{\sigma}\; , }
\end{equation}
\begin{equation}
	\ba{
	& K^{\mu \nu \rho \sigma \alpha \beta} = 
	\\ & \eta^{\alpha \mu} \eta^{\beta \rho} \eta^{\nu \sigma}+
        \eta^{\alpha \beta} \eta^{\mu \nu} \eta^{\rho \sigma} +
        \eta^{\alpha \sigma} \eta^{\beta \nu} \eta^{\mu \rho} +
        \eta^{\alpha \nu} \eta^{\beta \sigma} \eta^{\mu
          \rho}+\eta^{\alpha \rho} \eta^{\beta \nu} \eta^{\mu \sigma}
        +\eta^{\alpha \nu} \eta^{\beta \rho} \eta^{\mu \sigma}+
        \eta^{\alpha \sigma} \eta^{\beta \mu} \eta^{\nu
          \rho}+\eta^{\alpha \mu} \eta^{\beta \sigma} \eta^{\nu
          \rho}\\& 
	+\eta^{\alpha \rho} \eta^{\beta \mu} \eta^{\nu
          \sigma}-\eta^{\alpha \beta} \eta^{\sigma \mu} \eta^{\nu
          \rho}- \eta^{\alpha \beta} \eta^{\mu \rho} \eta^{\nu \sigma}
        - \eta^{\alpha \nu} \eta^{\beta \mu} \eta^{\rho \sigma}-
        \eta^{\alpha \mu} \eta^{\beta \nu} \eta^{\rho \sigma}
        -\eta^{\alpha \sigma} \eta^{\beta \rho} \eta^{\mu \nu} -
        \eta^{\alpha \rho} \eta^{\beta \sigma} \eta^{\mu \nu}\;, 
	}
\end{equation}
\begin{equation}
	\ba{
	&M^{\mu \nu \rho \sigma \alpha \beta}(p,q,l) = \\&  \eta^{\nu
            \rho } \eta^{\mu  \sigma } p^{\beta } p^{\alpha
          }+\eta^{\mu  \rho } \eta^{\nu  \sigma } p^{\beta } p^{\alpha
          }+\eta^{\nu  \rho } \eta^{\mu  \sigma } q^{\beta } p^{\alpha
          }+\eta^{\mu  \rho } \eta^{\nu  \sigma } q^{\beta } p^{\alpha
          }+\eta^{\beta  \nu } \eta^{\rho  \sigma } p^{\mu } p^{\alpha
          }+\eta^{\beta  \nu } \eta^{\rho  \sigma } q^{\mu } p^{\alpha
          }+\eta^{\beta  \nu } \eta^{\rho  \sigma } l^{\mu } p^{\alpha
          } 
	\\ &-\eta^{\beta  \sigma } \eta^{\mu  \rho } p^{\nu }
        p^{\alpha }-\eta^{\beta  \rho } \eta^{\mu  \sigma } p^{\nu }
        p^{\alpha }+\eta^{\beta  \mu } \eta^{\rho  \sigma } p^{\nu }
        p^{\alpha }+\eta^{\beta  \mu } \eta^{\rho  \sigma } q^{\nu }
        p^{\alpha }+\eta^{\beta  \mu } \eta^{\rho  \sigma } l^{\nu }
        p^{\alpha }+\eta^{\mu  \nu } \eta^{\beta  \sigma } p^{\rho }
        p^{\alpha }-\eta^{\beta  \nu } \eta^{\mu  \sigma } p^{\rho }
        p^{\alpha } 
	\\ &-\eta^{\beta  \mu } \eta^{\nu  \sigma } p^{\rho }
        p^{\alpha }-\eta^{\beta  \sigma } p^{\mu } \eta^{\nu  \rho }
        p^{\alpha }-\eta^{\beta  \mu } p^{\sigma } \eta^{\nu  \rho }
        p^{\alpha }+\eta^{\mu  \nu } \eta^{\beta  \rho } p^{\sigma }
        p^{\alpha }-\eta^{\beta  \nu } \eta^{\mu  \rho } p^{\sigma }
        p^{\alpha }-\eta^{\beta  \rho } p^{\mu } \eta^{\nu  \sigma }
        p^{\alpha }-p^{\beta } \eta^{\mu  \nu } \eta^{\rho  \sigma }
        p^{\alpha } 
	\\ &-q^{\beta } \eta^{\mu  \nu } \eta^{\rho  \sigma }
        p^{\alpha }-l^{\beta } \eta^{\mu  \nu } \eta^{\rho  \sigma }
        p^{\alpha }+\eta^{\alpha  \nu } \eta^{\rho  \sigma } p^{\mu }
        p^{\beta }+\eta^{\alpha  \nu } \eta^{\rho  \sigma } l^{\mu }
        p^{\beta }+\eta^{\alpha  \mu } \eta^{\rho  \sigma } p^{\nu }
        p^{\beta }+\eta^{\alpha  \mu } \eta^{\rho  \sigma } l^{\nu }
        p^{\beta }-l^{\alpha } \eta^{\mu  \nu } \eta^{\rho  \sigma }
        p^{\beta } 
	\\ &+\eta^{\alpha  \nu } \eta^{\rho  \sigma } q^{\beta }
        p^{\mu }+\eta^{\alpha  \nu } \eta^{\rho  \sigma } l^{\beta }
        p^{\mu }-\eta^{\alpha  \nu } \eta^{\beta  \sigma } p^{\rho }
        p^{\mu }-\eta^{\alpha  \nu } \eta^{\beta  \sigma } l^{\rho }
        p^{\mu }-\eta^{\alpha  \nu } \eta^{\beta  \rho } p^{\sigma }
        p^{\mu }-\eta^{\alpha  \nu } \eta^{\beta  \rho } l^{\sigma }
        p^{\mu }+\eta^{\alpha  \mu } \eta^{\rho  \sigma } q^{\beta }
        p^{\nu } 
	\\ &+\eta^{\alpha  \mu } \eta^{\rho  \sigma } l^{\beta }
        p^{\nu }-\eta^{\alpha  \mu } \eta^{\beta  \sigma } p^{\rho }
        p^{\nu }-\eta^{\alpha  \mu } \eta^{\beta  \sigma } l^{\rho }
        p^{\nu }-\eta^{\alpha  \mu } \eta^{\beta  \rho } p^{\sigma }
        p^{\nu }-\eta^{\alpha  \mu } \eta^{\beta  \rho } l^{\sigma }
        p^{\nu }+\eta^{\mu  \nu } \eta^{\beta  \sigma } l^{\alpha }
        p^{\rho }-\eta^{\alpha  \nu } \eta^{\beta  \sigma } l^{\mu }
        p^{\rho } 
	\\&-\eta^{\alpha  \mu } \eta^{\beta  \sigma } l^{\nu } p^{\rho
        }+\eta^{\mu  \nu } \eta^{\beta  \rho } l^{\alpha } p^{\sigma
        }-\eta^{\alpha  \nu } \eta^{\beta  \rho } l^{\mu } p^{\sigma
        }-\eta^{\alpha  \mu } \eta^{\beta  \rho } l^{\nu } p^{\sigma
        }+\eta^{\beta  \nu } \eta^{\rho  \sigma } q^{\mu } l^{\alpha
        }+\eta^{\beta  \mu } \eta^{\rho  \sigma } q^{\nu } l^{\alpha
        }+\eta^{\nu  \rho } \eta^{\mu  \sigma } q^{\alpha } q^{\beta } 
	\\&+\eta^{\mu  \rho } \eta^{\nu  \sigma } q^{\alpha } q^{\beta
        }+\eta^{\mu  \nu } \eta^{\alpha  \sigma } q^{\rho } l^{\beta
        }+\eta^{\mu  \nu } \eta^{\alpha  \rho } q^{\sigma } l^{\beta
        }+\eta^{\beta  \nu } \eta^{\rho  \sigma } q^{\alpha } q^{\mu
        }+\eta^{\alpha  \nu } \eta^{\rho  \sigma } q^{\beta } q^{\mu
        }+\eta^{\beta  \nu } \eta^{\rho  \sigma } q^{\alpha } l^{\mu
        }+\eta^{\beta  \nu } \eta^{\rho  \sigma } l^{\alpha } l^{\mu } 
	\\&+\eta^{\alpha  \nu } \eta^{\rho  \sigma } q^{\beta } l^{\mu
        }+\eta^{\alpha  \nu } \eta^{\rho  \sigma } l^{\beta } l^{\mu
        }+\eta^{\beta  \mu } \eta^{\rho  \sigma } q^{\alpha } q^{\nu
        }+\eta^{\alpha  \mu } \eta^{\rho  \sigma } q^{\beta } q^{\nu
        }-\eta^{\alpha  \sigma } q^{\beta } \eta^{\mu  \rho } q^{\nu
        }-\eta^{\alpha  \rho } q^{\beta } \eta^{\mu  \sigma } q^{\nu
        }+\eta^{\beta  \mu } \eta^{\rho  \sigma } q^{\alpha } l^{\nu } 
	\\&+\eta^{\beta  \mu } \eta^{\rho  \sigma } l^{\alpha } l^{\nu
        }+\eta^{\alpha  \mu } \eta^{\rho  \sigma } q^{\beta } l^{\nu
        }+\eta^{\alpha  \mu } \eta^{\rho  \sigma } l^{\beta } l^{\nu
        }+\eta^{\mu  \nu } \eta^{\alpha  \sigma } q^{\beta } q^{\rho
        }-\eta^{\alpha  \sigma } \eta^{\beta  \nu } q^{\mu } q^{\rho
        }-\eta^{\alpha  \sigma } \eta^{\beta  \nu } l^{\mu } q^{\rho
        }-\eta^{\alpha  \sigma } \eta^{\beta  \mu } q^{\nu } q^{\rho } 
	\\&-\eta^{\alpha  \sigma } \eta^{\beta  \mu } l^{\nu } q^{\rho
        }-\eta^{\alpha  \nu } q^{\beta } \eta^{\mu  \sigma } q^{\rho
        }-\eta^{\alpha  \mu } q^{\beta } \eta^{\nu  \sigma } q^{\rho
        }+\eta^{\mu  \nu } \eta^{\beta  \sigma } l^{\alpha } l^{\rho
        }+\eta^{\mu  \nu } \eta^{\alpha  \sigma } l^{\beta } l^{\rho
        }-\eta^{\alpha  \sigma } \eta^{\beta  \nu } q^{\mu } l^{\rho
        }-\eta^{\alpha  \sigma } \eta^{\beta  \nu } l^{\mu } l^{\rho } 
	\\&-\eta^{\alpha  \nu } \eta^{\beta  \sigma } l^{\mu } l^{\rho
        } -\eta^{\alpha  \sigma } \eta^{\beta  \mu } q^{\nu } l^{\rho
        }-\eta^{\alpha  \sigma } \eta^{\beta  \mu } l^{\nu } l^{\rho
        }-\eta^{\alpha  \mu } \eta^{\beta  \sigma } l^{\nu } l^{\rho
        }-\eta^{\alpha  \sigma } q^{\beta } q^{\mu } \eta^{\nu  \rho
        }-\eta^{\alpha  \mu } q^{\beta } q^{\sigma } \eta^{\nu  \rho
        }+\eta^{\mu  \nu } \eta^{\alpha  \rho } q^{\beta } q^{\sigma } 
	\\&-\eta^{\alpha  \rho } \eta^{\beta  \nu } q^{\mu } q^{\sigma
        }-\eta^{\alpha  \rho } \eta^{\beta  \nu } l^{\mu } q^{\sigma
        }-\eta^{\alpha  \rho } \eta^{\beta  \mu } q^{\nu } q^{\sigma
        }-\eta^{\alpha  \rho } \eta^{\beta  \mu } l^{\nu } q^{\sigma
        }-\eta^{\alpha  \nu } q^{\beta } \eta^{\mu  \rho } q^{\sigma
        }+\eta^{\mu  \nu } \eta^{\beta  \rho } l^{\alpha } l^{\sigma }
        +\eta^{\mu  \nu } \eta^{\alpha  \rho } l^{\beta } l^{\sigma } 
	\\&-\eta^{\alpha  \rho } \eta^{\beta  \nu } q^{\mu } l^{\sigma
        } -\eta^{\alpha  \rho } \eta^{\beta  \nu } l^{\mu } l^{\sigma
        }-\eta^{\alpha  \nu } \eta^{\beta  \rho } l^{\mu } l^{\sigma
        }-\eta^{\alpha  \rho } \eta^{\beta  \mu } q^{\nu } l^{\sigma
        }-\eta^{\alpha  \rho } \eta^{\beta  \mu } l^{\nu } l^{\sigma
        }-\eta^{\alpha  \mu } \eta^{\beta  \rho } l^{\nu } l^{\sigma
        }-\eta^{\alpha  \rho } q^{\beta } q^{\mu } \eta^{\nu  \sigma } 
	\\& -q^{\alpha } q^{\beta } \eta^{\mu  \nu } \eta^{\rho
          \sigma }-l^{\alpha } q^{\beta } \eta^{\mu  \nu } \eta^{\rho
          \sigma }-q^{\alpha } l^{\beta } \eta^{\mu  \nu } \eta^{\rho
          \sigma }-2 l^{\alpha } l^{\beta } \eta^{\mu  \nu }
        \eta^{\rho  \sigma }\; . 
	}\hspace{-1em}
\end{equation}

\end{document}